\documentclass[twocolumn]{svjour3}

\usepackage[utf8]{inputenc}
\usepackage[T1]{fontenc}

\usepackage{cuted}

\usepackage{graphicx}
\usepackage{scalefnt}

\usepackage{amsmath}
\usepackage{amssymb}
\usepackage{amsfonts}

\usepackage{lmodern}
\usepackage{microtype}

\newcommand{\ra}{\rangle}
\newcommand{\pr}{\prime}

\newcommand{\spc}[1]{\ensuremath{c^\dagger_{#1}}}
\newcommand{\spa}[1]{\ensuremath{c_{#1}}}

\newcommand{\tr}[1]{\ensuremath{\text{Tr}#1}}

\newcommand{\qnooneb}{\ensuremath{\text{PQNO1B}}}

\newcommand{\kongno}[1]{\ensuremath{\text{N}[#1]}}
\newcommand{\no}[1]{\ensuremath{:#1:}}

\newcommand{\antisym}[1]{\ensuremath{\mathcal{A}(#1)}}

\newcommand{\Sum}[2]{\the\numexpr #1 + #2 \relax}

\newcommand{\bogoexpect}[1]{\ensuremath{\langle\Phi|#1|\Phi\rangle}}

\begin{document}

\title{Normal-ordered $k$-body approximation in particle-number-breaking theories}

\author{J.~Ripoche \and A.~Tichai \and T.~Duguet}
\institute{
J. Ripoche \at
CEA, DAM, DIF, F-91297 Arpajon, France \\
\email{julien.ripoche@cea.fr} \and
A. Tichai \at ESNT, CEA-Saclay, DRF, IRFU, D\'epartement de Physique Nucl\'eaire, Universit\'e de Paris Saclay, F-91191 Gif-sur-Yvette \\ 
\email{alexander.tichai@cea.fr} 
\and
T. Duguet \at
IRFU, CEA, Universit\'e Paris-Saclay, 91191 Gif-sur-Yvette, France, \and 
KU Leuven, Instituut voor Kern- en Stralingsfysica, 3001 Leuven, Belgium \\ \email{thomas.duguet@cea.fr}
}

\maketitle

\abstract{
The reach of \textit{ab initio} many-body theories is rapidly extending over the nuclear chart. However, dealing fully with three-nucleon, possibly four-nucleon, interactions makes the solving of the A-body Schr\"odinger equation particularly cumbersome, if not impossible beyond a certain nuclear mass. Consequently, \textit{ab initio} calculations of mid-mass nuclei are typically performed on the basis of the so-called \textit{normal-ordered two-body} (NO2B) \textit{approximation} that captures dominant effects of three-nucleon forces while effectively working with two-nucleon operators.
A powerful idea currently employed to extend \textit{ab initio} calculations to open-shell nuclei consists of expanding the exact solution of the A-body Schr\"odinger equation while authorizing the approximate solution to break symmetries of the Hamiltonian. In this context, operators are normal ordered with respect to a symmetry-breaking reference state such that proceeding to a naive truncation may lead to symmetry-breaking approximate operators. The purpose of the present work is to design a normal-ordering approximation of operators that is consistent with the symmetries of the Hamiltonian while working in the context of symmetry broken (and potentially restored) methods.
Focusing on many-body formalisms in which U(1) global-gauge symmetry associated with particle number conservation is broken (and potentially restored), a \textit{particle-number-conserving normal-ordered $k$-body} (PNOkB) \textit{approximation} of an arbitrary N-body operator is designed on the basis of Bogoliubov reference states. A numerical test based on particle-number projected Hartree-Fock-Bogoliubov calculations permits to check the particle-number conserving/violating character of a given approximation to a particle-number conserving operator.  
The PNOkB approximation of an arbitrary N-body operator is formulated. Based on this systematic approach, it is demonstrated that naive extensions of the normal-ordered two-body (NO2B) approximation employed so far on the basis of symmetry-conserving reference states lead to particle non-conserving operators. Alternatively, the PNOkB procedure is now available to generate particle-number-conserving approximate operators. The formal analysis is validated numerically. 
Using the presently proposed PNOkB approximation, \textit{ab initio} calculations based on symmetry-breaking and restored formalisms can be safely performed. The future formulation of an angular-momentum-conserving normal-ordered $k$-body approximation based on deformed Slater determinant or Bogoliubov reference states is envisioned.
}

\allowdisplaybreaks

\section{Introduction}

The reach of \textit{ab initio} many-body theories is rapidly extending over the nuclear chart. In particular, the development of many-body methods based on a systematic expansion of the exact solution of the A-body Schroedinger equation around a conveniently chosen reference state allows an efficient description of medium-mass (semi-)magic nuclei. Examples of such methods are many-body perturbation theory (MBPT)~\cite{Langhammer:2012jx,Hu:2016txm,Tichai:2017rqe,Tichai:2018mll,Arthuis:2018yoo,Hu:2018dza}, self-consistent Green's function (SCGF)~\cite{Dickhoff:2004xx,Soma:2011aj,Soma:2013xha,Carbone:2013eqa,Lapoux:2016exf,Duguet:2016wwr,Raimondi:2018aa,Raimondi:2018mtv}, coupled-cluster (CC)~\cite{Hagen:2013nca,Ja14,Signoracci:2014dia,Morris:2017vxi} or in-medium similarity renormalization group (IMSRG)~\cite{Hergert:2015awm,Hergert:2016iju,Parzuchowski:2017wcq,Morris:2017vxi} approaches. To extend their reach to open-shell nuclei and capture noticeably challenging {\it static} correlations, these single-reference methods employ a symmetry-breaking reference state~\cite{Soma:2011aj,Signoracci:2014dia,Tichai:2018mll} and possibly restore the broken symmetry~\cite{Duguet:2014jja,Duguet:2015yle,Qiu:2018edx}. 

The computational load is particularly acute due to the relevance of three-nucleon interactions, knowing that including even more demanding four-body forces could eventually be mandatory~\cite{Bazak:2018qnu}. As a matter of fact, dealing with three-nucleon interactions in full makes the solving of the A-body Schr\"odinger equation rapidly too costly as the mass of the system grows. To circumvent the explicit treatment of three-body operators, \textit{ab initio} calculations of mid-mass nuclei have  been performed on the basis of the so-called \emph{normal-ordered two-body} (NO2B) \emph{approximation}. This approximation captures dominant effects of three-nucleon forces, while effectively working with two-body operators. In large-scale no-core shell model (NCSM) calculations, the error induced by the NO2B approximation to the Hamiltonian was estimated to be of the order of $1$-$3\%$~\cite{RoBi12,Geb16} up to the oxygen region.

The NO2B approximation consists of normal ordering the operator with respect to a many-body reference state and discarding the normal-ordered three-body component. While typically formulated with respect to an uncorrelated reference state, i.e. a Slater determinant, normal-ordering techniques and associated NO2B approximations can be also be devised with respect to a correlated reference state~\cite{KuMu97}, as is employed in the multi-reference IMSRG approach~\cite{Hergert:2015awm,Hergert:2016iju}, or even with respect to the fully correlated solution of the problem as is done in SCGF theory~\cite{Carbone:2013eqa}. In any case, the NO2B approximation has been employed so far on the basis of {\it symmetry-conserving} states. Only recently such an approximation has been employed in Bogoliubov MBPT (BMBPT)~\cite{Tichai:2018mll,Tichai:2019ksh} in which U(1) symmetry associated with particle-number conservation is spontaneously broken by the (approximate) many-body state. In this context, the normal ordering of operators at play is performed with respect to a particle-number-breaking Bogoliubov reference state such that proceeding to a naive truncation may lead to approximating a particle-number-conserving operator by a particle-number-breaking one. A similar situation shall occur when using a SU(2)-breaking, i.e. deformed, reference state such that proceeding to a naive normal-ordered truncation of a rotationally invariant operator may lead to an angular-momentum non-conserving approximation.

The purpose of the present work is thus to design a general normal-ordering approximation of operators that is consistent with symmetries of the Hamiltonian while working with symmetry broken (and restored) reference state\footnote{The symmetry-breaking nature of the many-body methods of present interest concerns the approximate {\it many-body state} while making use of a {\it symmetry-conserving Hamiltonian}. Approximating the Hamiltonian in a symmetry-violating way is conceptually different and more problematic as one wishes to eventually employ a symmetry-restored approximate many-body state~\cite{Duguet:2014jja,Duguet:2015yle,Qiu:2018edx}.}. Focusing on many-body formalisms in which U(1) symmetry associated with particle number conservation is broken (and potentially restored), a \emph{particle-number-conserving normal-ordered $k$-body} (PNOkB) \emph{approximation} of an arbitrary N-body operator is designed on the basis of Bogoliubov reference states. A numerical test based on particle-number projected Hartree-Fock-Bogoliubov calculations is designed and employed to check the particle-number conserving/violating character of the approximate operator.

\section{Formalism}

\subsection{Bogoliubov reference state}

Many-body formalisms of present interest, e.g. BMBPT~\cite{Duguet:2015yle,Tichai:2018mll,Arthuis:2018yoo}, Bogoliubov coupled cluster (BCC)~\cite{Signoracci:2014dia}, Gorkov SCGF~\cite{Soma:2011aj,Soma:2013xha}, Projected BMBPT (PBMBPT)~\cite{Duguet:2015yle} and Projected BCC (PBCC)~\cite{Duguet:2015yle,Qiu:2018edx}, rely on the use of a particle-number breaking Bogoliubov reference state $| \Phi \rangle$. 

The state $| \Phi \rangle$ is a vacuum for a complete set of quasi-particle operators $\{\beta_k,\beta^{\dagger}_k\}$ obtained from an arbitrary single-particle basis $\{c_l,c^\dagger_l\}$ of the one-body Hilbert space through a unitary Bogoliubov transformation
\begin{subequations}
\begin{align}
\beta_k &\equiv \sum_{l_1} U^*_{l_1k} c_p + V^*_{l_1k} c^\dagger_{l_1}\, , \\
\beta_k^\dagger &\equiv \sum_{l_1} U_{l_1k} c^\dagger_{l_1} + V_{l_1k} c_{l_1} \, ,
\end{align}
\end{subequations}
such that $\beta_k | \Phi \rangle =0 $ for all $k$. Although not presently concerned with these technical details, note that the columns of the transformation matrices $(U,V)$ are typically obtained as eigenvectors of the Hartree-Fock-Bogoliubov (HFB) eigenvalue problem~\cite{RiSc80}. More importantly for the discussion given below, $| \Phi \rangle$ reduces to a Slater determinant, presently denoted as $|\text{SD}\rangle$, whenever U(1) symmetry is not spontaneously broken, e.g. in closed shell systems where the HFB eigenproblem typically reduces to the Hartree-Fock (HF) one.

Normal and anomalous elementary contractions of single-particle creation and annihilation operators are defined with respect to $|\Phi\ra$ by
\begin{subequations}
  \label{eq:diagonalcontractions}
  \begin{align}
    \rho_{l_1l_2}
    &\equiv \frac{\langle\Phi|c^\dagger_{l_2} c_{l_1}|\Phi\rangle}{\langle\Phi|\Phi\rangle} = +(V^{\ast}V^{T})_{l_1l_2} \, , \\*
    \kappa_{l_1l_2}
    &\equiv \frac{\langle\Phi|c_{l_2} c_{l_1}|\Phi\rangle}{\langle\Phi|\Phi\rangle} = +(V^{\ast}U^{T})_{l_1l_2} \, ,
  \end{align}
\end{subequations}
such that
\begin{subequations}
\label{densitymatrices}
\begin{align}
\rho_{l_2l_1} &= +\rho^{\ast}_{l_1l_2}  \ , \label{densitymatrices1} \\*
\kappa_{l_2l_1} &= - \kappa_{l_1l_2} \ . \label{densitymatrices2}
\end{align}
\end{subequations}

\subsection{Particle-number conserving operator}
\label{sec:normalorderedO}

We are typically concerned with the treatment of a N-body operator $O$ that commutes with the particle-number operator
\begin{align}
A &\equiv \sum_{l_1} c^\dagger_{l_1} c_{l_1} \, , \label{eq:A}
\end{align}
and whose general form in the basis $\{c_l,c^\dagger_l\}$ is
\begin{align}
  \label{eq:genericodef}
  O
  &\equiv o^{00}\nonumber \\
  & + \frac{1}{1!1!} \sum_{l_1l_2} o^{11}_{l_1l_2} c^\dagger_{l_1} c_{l_2} \nonumber \\
  &+ \frac{1}{2!2!} \sum_{l_1l_2l_3l_4} o^{22}_{l_1l_2l_3l_4} c^\dagger_{l_1} c^\dagger_{l_2} c_{l_4} c_{l_3}  \nonumber \\
  &+ \frac{1}{3!3!} \sum_{l_1l_2l_3l_4l_5l_6} o^{33}_{l_1l_2l_3l_4l_5l_6} c^\dagger_{l_1} c^\dagger_{l_2} c^\dagger_{l_3} c_{l_6} c_{l_5} c_{l_4} + \ldots \notag \\
  &\equiv o^{00} + o^{11} + o^{22} + o^{33} + \ldots \notag \\
  &= \sum_{n=0}^N o^{nn} \, ,
\end{align}
where matrix elements $o^{nn}_{l_1 \ldots l_n l_{n+1} \ldots l_{2n}}$ are fully antisymmetric with respect to the permutation of the $n$ first, resp. $n$ last, indices
\begin{equation}
  o^{nn}_{l_1 \ldots l_{n} l_{n+1} \ldots l_{2n}} = \epsilon(\sigma) \, o^{nn}_{\sigma(l_1 \ldots l_n | l_{n+1} \ldots l_{2n})} \, ,
\end{equation}
where $\epsilon(\sigma)$ refers to the signature of the permutation $\sigma$. The notation $\sigma(\ldots | \ldots)$ denotes a separation between the $n$ first and the $n$ last indices such that permutations are only considered between members of the same group. If $O$ is hermitian, the n-body matrix elements $o^{nn}_{l_1 \ldots l_n l_{n+1} \ldots l_{2n}}$ satisfy
\begin{align}
  o^{nn \ast}_{l_1 \ldots l_n l_{n+1} \ldots l_{2n}} &= o^{nn}_{l_{n+1} \ldots l_{2n} l_1 \ldots l_n} \, .
\end{align}
A graphical representation of the contributions to a three-body operator $O$ is given in Tab.~\ref{fig:no_sp_pv_grid} where the $o^{ij}$ contributions are sorted horizontally according to $i-j$ and vertically according to $i+j$. As $O$ commutes with $A$, there is no contribution with $i\neq j$ in the present case. This graphical representation will become useful below when normal-ordering
the operator with respect to the quasi-particle vacuum $|\Phi\rangle$.


\begin{table}[t!]
\centering
\setlength{\tabcolsep}{8.3pt}
\renewcommand{\arraystretch}{1.7}
\begin{tabular}{| c | c | c | c | c | c | c | c |}
\hline 
$\{c,c^\dagger\}, \, |0\ra $	&	-6	&	-4 	& 	-2	&	0	&	+2	&	+4	&	+6 \\
\hline 
\hline 
$o^{[0]}$	&	&	&	&	$o^{00}$	&	&	&	\\
\hline 
$o^{[2]}$	&	&	&	&	$o^{11}$	&	&	&	\\
\hline 
$o^{[4]}$	&	&	&	&	$o^{22}$	&	&	&	\\
\hline 
$o^{[6]}$	&	&	&	&	$o^{33}$	&	&	&	\\
\hline
\end{tabular}
\caption{Contributions to the three-body operator $O$ in normal-ordered form with respect to the particle vacuum $|0\rangle$ and expressed in $\{c,c^\dagger\}$. The $o^{ij}$ contributions are sorted horizontally according to $i-j$ and vertically according to $i+j$.}
\label{fig:no_sp_pv_grid}
\end{table}

\subsection{Normal ordering}

In order to address (in)appropriate approximations to the operator $O$, its normal-ordered forms with respect to $|\Phi\rangle$ expressed in two different bases are needed.

\subsubsection{Single-particle basis}
\label{sec:spnormalorder}

The normal ordering of $O$ with respect to the Bogoliubov vacuum $|\Phi\rangle$ leads to re-expressing the operator under the form
\begin{align}
  \label{eq:spnormalorder}
  O
  &\equiv \sum_{n=0}^N \sum_{\substack{i,j=0/\\i+j=2n}}^{N} \frac{1}{i!j!} \sum_{l_1 \ldots l_{i+j}} \Lambda^{ij}_{l_1 \ldots l_{i+j}} \notag \\ 
  &\times : \spc{l_1} \ldots \spc{l_i} \spa{l_{i+j}} \ldots \spa{l_{i+1}} : \notag \\
  &\equiv \sum_{n=0}^N O^{[2n]}
\end{align}
where $:\ldots:$ denotes the normal-ordered product with respect to $|\Phi\rangle$. Matrix elements $\Lambda^{ij}_{l_1 \ldots l_{i+j}}$ are fully antisymmetric with respect to the permutation of the $i$ first, resp. $j$ last, indices
\begin{align}
  \Lambda^{ij}_{l_1 \ldots l_{i} l_{i+1} \ldots l_{i+j}}
  &= \epsilon(\sigma) \, \Lambda^{ij}_{\sigma(l_1 \ldots l_i | l_{i+1} \ldots l_{i+j})} \, .
\end{align}
If $O$ is hermitian, field matrix elements $\Lambda^{ij}_{l_1 \ldots l_i l_{i+1} \ldots l_{i+j}}$ satisfy
\begin{align}
  \Lambda^{ij \ast}_{l_1 \ldots l_i l_{i+1} \ldots l_{i+j}} &= \Lambda^{ji}_{l_{i+1} \ldots l_{i+j} l_1 \ldots l_i} \, .
\end{align}
The class $O^{[2n]}$ groups all the terms containing a normal-ordered product of $2n$ single-particle operators, i.e. terms possibly containing different numbers of single-particle creation and annihilation operators according to
\begin{align}
  \label{eq:o2nclass}
  O^{[2n]}
  &\equiv \sum_{\substack{i,j=0/\\i+j=2n}}^{N} \Lambda^{ij} \, ,
\end{align}
where $\Lambda^{ij}$ is a $n$-body normal, resp. anomalous, field if $i=j$, resp. $i\neq j$, containing all the terms with a normal-ordered product of $i$, resp. $j$, single-particle creation, resp. annihilation, operators and reading as
\begin{align}
  \Lambda^{ij}
  &\equiv \frac{1}{i!j!} \sum_{l_1 \ldots l_{i+j}} \Lambda^{ij}_{l_1 \ldots l_i l_{i+1} \ldots l_{i+j}} \nonumber \\
  & \hspace{2cm} \times : \spc{l_1} \ldots \spc{l_i} \spa{l_{i+j}} \ldots \spa{l_{i+1}} :  \, . \label{eq:nofields1}
\end{align}
Applying standard Wick's theorem~\cite{Wi50} to $O$, $\Lambda^{ij}$ receives contributions from all $n$-body terms $o^{nn}$ with $n\ge \max(i,j)$, i.e.
\begin{align}
  \Lambda^{ij}_{l_1 \ldots l_{i+j}}
  &\equiv \sum_{n=\max(i,j)}^N \Lambda^{ij(nn)}_{l_1 \ldots l_{i+j}} \, ,
\end{align}
such that $\Lambda^{ij(nn)}_{l_1 \ldots l_{i+j}}$ accounts for all appropriate contraction patterns and is given by
\begin{align}
  \label{eq:explicitlambdaijk}
  \Lambda^{ij(nn)}_{l_1 \ldots l_{i+j}}
  &\equiv \sum_{\substack{(n_\rho, n_{\kappa^\ast}, n_\kappa)}}^{\substack{n_\rho+2n_{\kappa^\ast}=n-i\\n_\rho+2n_{\kappa}=n-j}} \frac{1}{n_\rho!n_{\kappa^\ast}!n_{\kappa}!} \left(\frac{1}{2}\right)^{n_{\kappa^\ast}} \left(\frac{1}{2}\right)^{n_{\kappa}}\notag \\
  &\hphantom{= \sum_{l_{i+j+1}}\,\,\,\,} 
  \times   \tr[o^{nn}\rho\kappa^\ast\kappa]^{(n_\rho, n_{\kappa^\ast}, n_\kappa)}_{l_1 \ldots l_{i+j}} \, ,
\end{align}
where the notation $\tr[o^{nn}\rho\kappa^\ast\kappa]^{(n_\rho, n_{\kappa^\ast}, n_\kappa)}_{l_1 \ldots l_{i+j}}$ denotes a trace over the indices of the $n_\rho$ normal contractions $\rho$, the $n_{\kappa^\ast}$ anomalous contractions $\kappa^\ast$ and the $n_\kappa$ anomalous contractions $\kappa$, i.e.
\begin{strip}
\begin{align}
  \tr[o^{nn}\rho\kappa^\ast\kappa]^{(n_\rho, n_{\kappa^\ast}, n_\kappa)}_{l_1 \ldots l_{i+j}}
  &\equiv \sum_{l_{i+j+1} \ldots l_{2n}} o^{nn}_{l_1 \ldots l_{i} l_{i+j+1} \ldots l_{j+n} l_{i+1} \ldots l_{i+j} l_{j+n+1} \dots l_{2n}}
  \rho_{l_{2n-n_\rho+1} l_{j+n-n_\rho+1}} \ldots \rho_{l_{2n} l_{j+n}} 
  \notag \\
  &\hphantom{= \sum_{l_{i+j+1} \ldots l_{2n}}} 
  \times 
  \kappa^\ast_{l_{i+j+1}l_{i+j+2}} \ldots \kappa^\ast_{l_{j+n-n_\rho-1} l_{j+n-n_\rho}} 
  \kappa_{l_{j+n+1}l_{j+n+2}} \ldots \kappa_{l_{2n-n_\rho-1} l_{2n-n_\rho}} \, . \label{contractpattern}
\end{align}
\end{strip}
Whenever the Bogoliubov vacuum reduces to a Slater determinant, all anomalous contractions are null, i.e. $\kappa=\kappa^{\ast}=0$, such that $\Lambda^{ij}=0$ for $i\neq j$. The proof leading to Eqs.~(\ref{eq:explicitlambdaijk}-\ref{contractpattern}), together with the explicit form of the $\Lambda^{ij}$ matrix elements associated with a three-body operator, are given in App.~\ref{proof1}.  

\begin{figure*}[t!]
  \centering
  \includegraphics[width=0.7\textwidth]{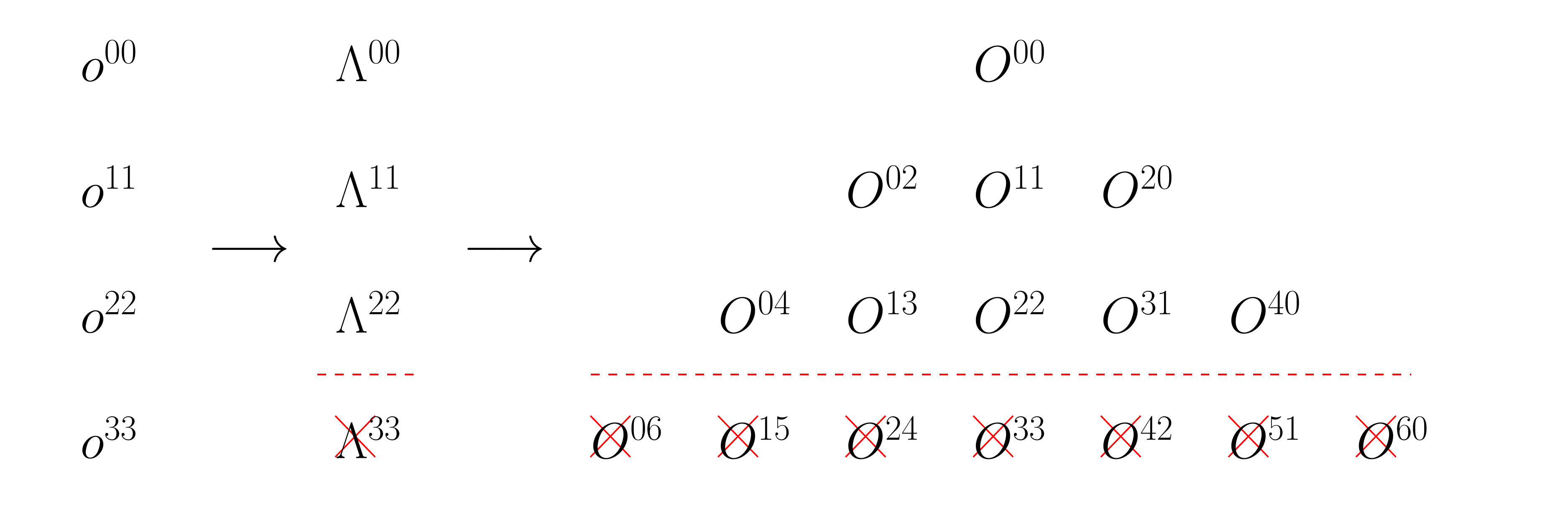}
  \caption{Representation of the NO2B approximation for a three-body operator $O$ normal-ordered with respect to a Slater Determinant $|\text{SD}\rangle$.
  Left column: normal-ordered form with respect to $|0\rangle$ expressed in $\{c,c^\dagger\}$.
  Middle column: normal-ordered form with respect to $|\text{SD}\rangle$ expressed in $\{c,c^\dagger\}$.
  Right column: normal-ordered form with respect to $|\text{SD}\rangle$ expressed in $\{\beta,\beta^\dagger\}$.  The contributions $o^{ij}/\Lambda^{ij}/O^{ij}$ are sorted horizontally according to $i-j$ and vertically according to $i+j$.
  Red crosses indicate terms that are suppressed in the NO2B approximation such that red dashed lines separate suppressed terms from retained ones.}
  \label{fig:nokb_sd}
\end{figure*}

\subsubsection{Quasi-particle basis}

One can rather choose to express the normal-ordered form of $O$ in the quasi-particle basis $\{\beta_k,\beta^\dagger_k\}$. Applying standard Wick's theorem leads to
\begin{align}
  O
  &\equiv \sum_{n=0}^N \sum_{\substack{i,j=0\\i+j=2n}}^{2N} \frac{1}{i!j!} \sum_{l_1 \ldots l_{i+j}} O^{ij}_{l_1 \ldots l_{i+j}} \nonumber \\
  & \hspace{2cm} \times\beta^{\dagger}_{k_1} \ldots \beta^{\dagger}_{k_i} \beta_{k_{i+j}} \ldots \beta_{k_{i+1}} \notag \\*
  &\equiv \sum_{n=0}^N O^{[2n]} \, ,
\end{align}
where matrix elements $O^{ij}_{k_1 \ldots k_{i+j}}$ are fully antisymmetric with respect to the permutation of the $i$ first, resp. $j$ last, indices
\begin{align}
  O^{ij}_{k_1 \ldots k_{i} k_{i+1} \ldots k_{i+j}}
  &= \epsilon(\sigma) \, O^{ij}_{\sigma(k_1 \ldots k_i | k_{i+1} \ldots k_{i+j})} \, .
\end{align}
If $O$ is hermitian, matrix elements $O^{ij}_{k_1 \ldots k_{i+j}}$ satisfy
\begin{align}
  O^{ij \ast}_{k_1 \ldots k_i k_{i+1} \ldots k_{i+j}} &= O^{ji}_{k_{i+1} \ldots k_{i+j} k_1 \ldots k_i} \, .
\end{align}

The class $O^{[2n]}$ groups all terms containing a normal-ordered product of $2n$ quasiparticle operators that can be further  differentiated according to the number of quasiparticle creation and annihilation operators they contained, i.e.
\begin{align}
  O^{[2n]}
  &\equiv \sum_{\substack{i,j=0\\i+j=2n}}^{2N} O^{ij} \, ,
\end{align}
where $O^{ij}$ gathers all terms with $i$, resp. $j$, quasiparticle creation, resp. annihilation, operators and reads
\begin{align}
  O^{ij}
  &\equiv \frac{1}{i!j!} \sum_{k_1 \ldots k_{i+j}} O^{ij}_{k_1 \ldots k_{i} k_{i+1} \ldots k_{i+j}} \nonumber \\
  &\hspace{2cm} \times \beta^{\dagger}_{k_1} \ldots \beta^{\dagger}_{k_i} \beta_{k_{i+j}} \ldots \beta_{k_{i+1}} \, .
\end{align}
Extending to $O^{[6]}$ the results provided in Ref.~\cite{Signoracci:2014dia}, the explicit form of the  $O^{ij}$ matrix elements associated with a three-body operator $O$ is given in App.~\ref{proof2} as an example.

\begin{figure*}[t!]
  \centering
  \includegraphics[width=\textwidth]{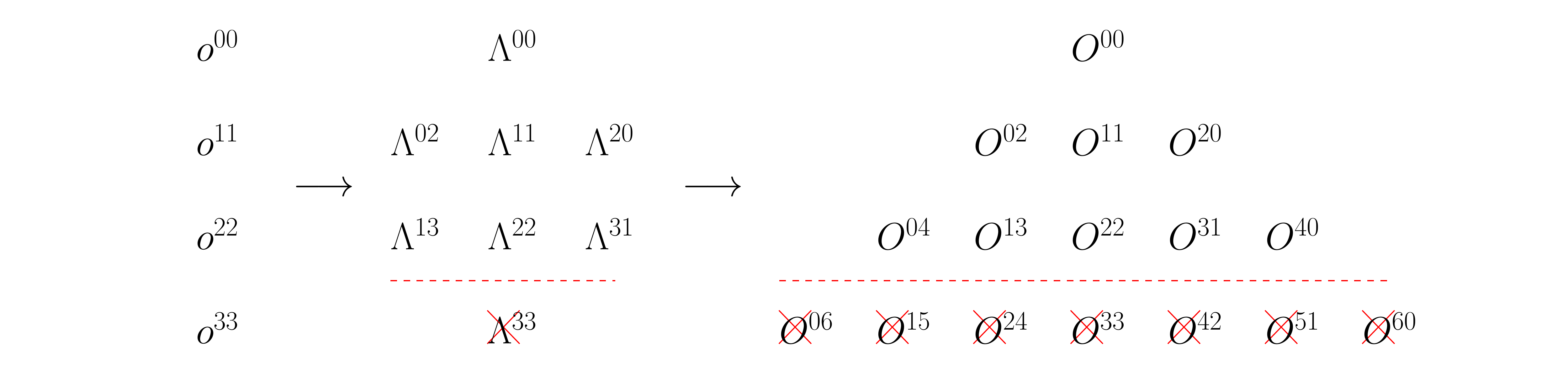}
  \caption{Representation of the naive extension of NO2B approximation for a three-body operator $O$ normal-ordered with respect to the quasi-particle vacuum $|\Phi\rangle$.
  Left column: normal-ordered form with respect to $|0\rangle$ expressed in $\{c,c^\dagger\}$.
  Middle column: normal-ordered form with respect to $|\Phi\rangle$ expressed in $\{c,c^\dagger\}$.
  Right column: normal-ordered form with respect to $|\Phi\rangle$ expressed in $\{\beta,\beta^\dagger\}$.
  Red crosses indicate terms that are suppressed in the naive extension of the NO2B approximation such that red dashed lines separate suppressed terms from retained ones.}
  \label{fig:nokb_naive}
\end{figure*}

\subsection{Approximation}

The need to handle (at least) three-nucleon interactions makes the solving of the A-body Schr\"odinger equation particularly cumbersome beyond the lightest nuclei. At the mean-field level one is able to handle the full Hamiltonian. Beyond the mean-field, however, three-body interactions prevent the implementation of many-body methods in their full glory. 

To tackle this difficulty, the so-called \textit{normal-ordered two-body} (NO2B) \textit{approximation} of three-body operators has been designed and used with success~\cite{RoBi12,Geb16}. This approximation amounts to employing the full Hamiltonian to generate the mean-field Slater determinant reference state $|\text{SD}\rangle$ and to truncating the normal-ordered form of $H$ with respect to $|\text{SD}\rangle$ at the effective two-body level for the beyond mean-field step. Thus, the NO2B approximation of a three-body operator $O$ commuting with $A$ reads as
\begin{align}
  \label{eq:no2bapprox}
  O^{\text{NO2B}}
  &\equiv O - O^{[6]} \notag \\*
  &= O^{[0]} + O^{[2]} + O^{[4]} \, ,
\end{align}
where all terms containing a normal-ordered product of six single-particle operators have been discarded. A graphical representation of this approximation is given in Fig.~\ref{fig:nokb_sd}.

\subsubsection{Naive extension}

Whereas this truncation is transparent and appropriate in the symmetry-conserving context, it has not been discussed in connection with symmetry-breaking reference states. It happens that, even if $O$ commutes with $A$, the approximate operator
$O^{\text{NO2B}}$ obtained via a naive generalization of Eq.~\eqref{eq:no2bapprox} on the basis of a Bogoliubov reference state $|\Phi\rangle$, does not commute with $A$, i.e.
\begin{align}
  \left[ O^{\text{nNO2B}} , A \right] &\neq 0 \, .
\end{align}
A graphical representation of this naive extension of the NO2B approximation is given in Fig.~\ref{fig:nokb_naive}.

For $O=H$ the naive extension of the NO2B approximation would imply working with a Hamiltonian $H^{\text{nNO2B}}$ whose exact eigenstates are not eigenstates of $A$. While this can be considered as part of a systematic error induced by the approximation, it eventually leads, when restoring the symmetry of the approximate wave function, to energies that problematically depend on the particular way the symmetry restoration is formulated. This key feature is discussed and illustrated later on. 

In this context, it is desirable to generalize the NOkB approximation of an arbitrary N-body operator in such a way that the truncated operator is particle-number conserving even if the reference state breaks $U(1)$ symmetry.

\subsubsection{Particle-number-conserving NOkB approximation}
\label{PNOkBprocedure}

In the following, $O$ denotes a normal-ordered $N$-body operator such that its naive NOkB (nNOkB) approximation reads as
\begin{align}
  O^{\text{nNOkB}}
  &\equiv \sum_{n=0}^{k} O^{[2n]} \, . \label{naiveNOkB}
\end{align}
The nNOkB approximation suffers from the same pathology as the naive NO2B approximation regarding particle-number violation. One now wishes to design an approximation to $O$ fulfilling the following three requirements
\begin{enumerate}
\item All normal-ordered terms of ranks higher than $k$ must be discarded as the practical goal of normal-ordered approximations is to eventually work with an effective operator characterized by a maximum rank $k<N$.
\item The approximate operator must commute with $A$.
\item The error generated by the approximation must be minimal.
\end{enumerate}
While the nNOkB approximation fulfills the first condition, it violates the second by \textit{fully retaining} the normal-ordered contributions with ranks lower or equal than $k$. One can thus anticipate that fulfilling the second condition in addition to the first one requires to further approximate specific parts of the operator displayed in Eq.~\ref{naiveNOkB}.  In the following, a systematic procedure to achieve this goal is devised for any $N$ and any $k<N$. It must be noted that there is no unique way to do so\footnote{In App.~\ref{alternativePNOkB}, an alternative particle-number conserving approximation based on the quasi-normal ordering proposed in Ref.~\cite{kong10a} is investigated.}. Eventually, while the third condition may be anticipated based on reasonable arguments, it can solely be validated through benchmark calculations. 

The {\it particle-number-conserving NOkB} (PNOkB) {\it approximation} to $O$ is given by
\begin{align}
  O^{\text{PNOkB}}
    &\equiv \tilde{o}^{00}\nonumber \\*
  & + \frac{1}{1!1!} \sum_{l_1l_2} \tilde{o}^{11}_{l_1l_2} c^\dagger_{l_1} c_{l_2} \nonumber \\*
  &+ \frac{1}{2!2!} \sum_{l_1l_2l_3l_4} \tilde{o}^{22}_{l_1l_2l_3l_4} c^\dagger_{l_1} c^\dagger_{l_2} c_{l_4} c_{l_3}  \nonumber \\*
  &+ \hspace{1.3cm} \vdots \nonumber \\*
  &+ \frac{1}{k!k!} \sum_{l_1 \ldots l_{2k}} \tilde{o}^{kk}_{l_1 \ldots l_{2k}} \spc{l_1} \ldots \spc{l_k} \spa{l_{2k}} \ldots \spa{l_{k+1}} \notag \\*
  &\equiv \sum_{n=0}^{k} \tilde{o}^{nn} \, , 
  \label{defPNOkB}
\end{align}
which is manifestly particle-number conserving. The $n$-body matrix elements $\tilde{o}^{nn}_{l_1 \ldots l_{2n}}$, $n\le k$, are recursively defined in decreasing order, from $n=k$ down to $n=0$, by
\begin{align}
  \label{eq:otildedef}
  \tilde{o}^{kk}_{l_1 \ldots l_{2k}}
  &\equiv \Lambda^{kk}_{l_1 \ldots l_{2k}} \notag \\
  \tilde{o}^{nn}_{l_1 \ldots l_{2n}}
  &\equiv \Lambda^{nn}_{l_1 \ldots l_{2n}} - \sum_{m=n+1}^{k} \tilde{\Lambda}^{nn(mm)}_{l_1 \ldots l_{2n}} \text{~for~} n<k \, ,
\end{align}
where $\tilde{\Lambda}^{nn(mm)}$, $m>n$, is the $m$-body contribution to the $n$-body normal field associated with $O^{\text{PNOkB}}$ defined in the same way as the $m$-body contribution to the $n$-body normal field $\Lambda^{nn(mm)}$ associated with the full operator $O$ was introduced in Eq.~\eqref{eq:explicitlambdaijk}.

It is clear that $O^{\text{PNOkB}}$ contains, through the fields $\Lambda^{nn}$, $n\leq k$, contributions from $o^{nn}$ up to $n=N$. It is the usual benefit of a normal-ordered approximation to capture the dominant part of all contributions to the original operator while working with effective operators of lower ranks.

To better appreciate the content of $O^{\text{PNOkB}}$, let us consider its normal-ordered form in the single-particle basis. As proven in App.~\ref{AppPNOkB}, $O^{\text{PNOkB}}$ is obtained from $O$ via a two-step process, i.e. by
\begin{enumerate}
\item removing all $\Lambda^{ij}$ fields with $\max(i,j)>k$,
\item adding $\breve{\Lambda}^{ij}$ defined below for $\max(i,j) \leq k$. 
\end{enumerate}
This leads to rewriting the approximate operator as
\begin{align}
  O^{\text{PNOkB}}
  &= \sum_{i,j=0}^{\max(i,j)\le k} \tilde{\Lambda}^{ij} \, ,
\end{align}
with
\begin{align}
  \tilde{\Lambda}^{ij}
  &\equiv \Lambda^{ij} + \breve{\Lambda}^{ij} \, ,
\end{align}
where the extra term $\breve{\Lambda}^{ij}$ is given by
\begin{strip}
\begin{align}
  \label{eq:lambdabrevedef1}
  \breve{\Lambda}^{ij}_{l_1 \ldots l_{i+j}}
  &\equiv \frac{1}{\left(\frac{i-j}{2}\right)!}\left(\frac{1}{2}\right)^{\frac{i-j}{2}} \sum_{n=i}^{N} \sum_{n_\kappa=0}^{\left\lfloor \frac{n-i}{2} \right\rfloor} \left(1 - \frac{1}{\binom{n_\kappa + \frac{i-j}{2}}{n_\kappa}}\right) \sum_{l_{i+j+1}\ldots l_{2i}} \left[ \Lambda^{ii(nn)(n_\kappa)}_{l_1 \ldots l_{2i}} - \tilde{\Lambda}^{ii(nn)(n_\kappa)}_{l_1 \ldots l_{2i}} \right] \kappa_{l_{i+j+1}l_{i+j+2}} \ldots \kappa_{l_{2i-1} l_{2i}}
\end{align}
for $i>j$, by
\begin{align}
  \label{eq:lambdabrevedef2}
  \breve{\Lambda}^{ij}_{l_1 \ldots l_{i+j}}
  &\equiv \frac{1}{\left(\frac{j-i}{2}\right)!}\left(\frac{1}{2}\right)^{\frac{j-i}{2}} \sum_{n=j}^{N} \sum_{n_{\kappa^\ast}=0}^{\left\lfloor \frac{n-j}{2} \right\rfloor} \left(1 - \frac{1}{\binom{n_\kappa + \frac{j-i}{2}}{n_\kappa}}\right) \sum_{l_{i+j+1}\ldots l_{2j}} \left[ \Lambda^{jj(nn)(n_{\kappa^\ast})}_{l_1 \ldots l_i l_{i+j+1} \ldots l_{2j} l_{i+1} \ldots l_{i+j}} - \tilde{\Lambda}^{jj(nn)(n_{\kappa^\ast})}_{l_1 \ldots l_i l_{i+j+1} \ldots l_{2j} l_{i+1} \ldots l_{i+j}} \right] \nonumber \\
 & \hphantom{\equiv \frac{1}{\left(\frac{j-i}{2}\right)!}\left(\frac{1}{2}\right)^{\frac{j-i}{2}} \sum_{n=j}^{N} \sum_{n_{\kappa^\ast}=0}^{\left\lfloor \frac{n-j}{2} \right\rfloor} \left(1 - \frac{1}{\binom{n_\kappa + \frac{j-i}{2}}{n_\kappa}}\right) \sum_{l_{i+j+1}\ldots l_{2j}} \,\,\, } 
 \times \kappa^\ast_{l_{i+j+1}l_{i+j+2}} \ldots \kappa^\ast_{l_{2j-1} l_{2j}}
\end{align}
\end{strip}
for $i<j$ and by
\begin{align}
  \label{eq:lambdabrevedef3}
  \breve{\Lambda}^{ij}_{l_1 \ldots l_{i+j}}
  &\equiv 0
\end{align}
for $i=j$. In the end, $\breve{\Lambda}^{ij}$ is non-zero only if $i-j\ge 2$ and $n_\kappa \ge 1$, or $j-i\ge 2$ and $n_{\kappa^\ast} \ge 1$. This is only possible for an initial $N$-body operator $O$ with $N\ge 4$, i.e. up to an initial three-body operator, no extra term is to be considered and the retained fields $\tilde{\Lambda}^{ij}$ are trivially given by the original fields $\Lambda^{ij}$.

The leading criterion behind the definition provided in Eqs.~(\ref{defPNOkB}-\ref{eq:otildedef}) is to ensure that $ O^{\text{PNOkB}}$ shares the same normal fields as $O$ for $n\le k$, i.e.
\begin{align}
  \tilde{\Lambda}^{nn}
  &= \Lambda^{nn} \, ,
\end{align}
as those are believed to be the dominant contributions to the normal-ordered operator. This requirement possibly induces extra contributions $\breve{\Lambda}^{ij}$ to the anomalous fields. This feature relates to the fact that $\tilde{\Lambda}^{ij}$ is obtained from the original operator through two successive normal orderings steps rather than one for $\Lambda^{ij}$. When proceeding to anomalous contractions in the application of Wick's theorem, it possibly leads to the non-equivalent combinatorial prefactors
\begin{align}
  (n_{\kappa_1}+n_{\kappa_2})!
  &\neq n_{\kappa_1}!n_{\kappa_2}! \text{~~~if~} n_{\kappa_1} \ge 1 \text{~and~} n_{\kappa_2} \ge 1 \, , \nonumber
\end{align}
in both procedures.

Eventually, and as best illustrated through the examples worked out in App.~\ref{AppPNOkB}, the effect of the correction terms defined through Eqs.~(\ref{eq:lambdabrevedef1}-\ref{eq:lambdabrevedef3}) is nothing but to modify the numerical prefactors of specific contributions to the normal-ordered anomalous fields expressed in the single-particle basis. Noticing that plainly omitting $\Lambda^{ij}$ with $\max(i,j)>k$ can also be viewed as the mere replacement of its original prefactor by an approximate one (0 in such cases), the function of the extra terms $\breve{\Lambda}^{ij}$ is in fact not different, i.e. it corresponds to modifying the prefactors of a specific set of fields such that the initial objectives are fulfilled. The quantitative performance of the procedure can only be judged \textit{a posteriori} by comparing many-body results obtained using the full operator and its PNOkB approximation.

\subsubsection{PNO1B approximation of a two-body operator}
\label{sec:PNO1B}

Let us exemplify the PNOkB procedure by building the PNO1B approximation of the two-body operator
\begin{align}
  O &= o^{00} + o^{11} + o^{22} \, .
\end{align}
This constitutes the simplest possible case. While it is not of interest for realistic \textit{ab initio} calculations, it is used throughout the present paper to illustrate the difference between a nNOkB and the newly designed PNOkB.

The normal ordering of $O$ with respect to $| \Phi \rangle$ reads in the single-particle basis as
\begin{align}
  O &= 
 \Lambda^{00}  +\Lambda^{20}+ \Lambda^{11}+ \Lambda^{02}  + \Lambda^{22}  
 \label{NO2body} 
\end{align}
and involves the following normal and anomalous fields
  \begin{subequations}
  \label{lambdaijtwobody}
    \begin{align}
      \Lambda^{22}
      &= o^{22} \, , \\
      \Lambda^{20}
      &= \frac{1}{2}\tr[o^{22}\kappa] \, , \\
      \Lambda^{11} 
      &= o^{11} + \tr[o^{22}\rho] \, , \\
      \Lambda^{00}
      &= o^{00} + \tr[o^{11}\rho]  + \frac{1}{2}\tr[o^{22}\rho\rho] + \frac{1}{4}\tr[o^{22}\kappa^\ast\kappa] \, ,
    \end{align}
  \end{subequations}
where each contribution, in fact denoting {\it matrix elements} of the fields, is written in a compact form without indices such that the key informations, i.e. the original $o^{nn}$ term, the set of elementary contractions involved and the overall prefactor, can be easily extracted. 

The nNO1B approximation leads to just dropping $O^{[2]}=\Lambda^{22}$ in Eq.~\eqref{NO2body}. Using properties of elementary contractions
\begin{subequations}
  \begin{align}
    c^\dagger_{l_1} c^\dagger_{l_2}
    &=  :c^\dagger_{l_1} c^\dagger_{l_2}: + \kappa^\ast_{l_1l_2} \, , \\
    c^\dagger_{l_1} c_{l_2}
    &=  :c^\dagger_{l_1} c_{l_2}: + \rho_{l_2l_1} \, , \\
    c_{l_2} c_{l_1}
    &=  :c_{l_2} c_{l_1}: + \kappa_{l_1l_2} \, ,
  \end{align}
\end{subequations}
one can write
\begin{align}
  O^{\text{nNO1B}}
  &\equiv \Lambda^{00} + \Lambda^{20} + \Lambda^{11} + \Lambda^{02} \notag \\
  &= \Lambda^{00}  \notag \\
    & \phantom{=}  + \frac{1}{2!0!} \sum_{l_1l_2} \Lambda^{20}_{l_1l_2} :c^\dagger_{l_1} c^\dagger_{l_2}:  \notag \\
    & \phantom{=} + \frac{1}{1!1!} \sum_{l_1l_2} \Lambda^{11}_{l_1l_2} :c^\dagger_{l_1} c_{l_2}:   \notag \\
    & \phantom{=} + \frac{1}{0!2!} \sum_{l_1l_2} \Lambda^{02}_{l_1l_2} :c_{l_2} c_{l_1}: \notag \\
  &\equiv \tilde{o}^{00}\notag \\
  & \phantom{=} + \frac{1}{2!0!} \sum_{l_1l_2} \tilde{o}^{20}_{l_1l_2} c^\dagger_{l_1} c^\dagger_{l_2}  \notag \\
    & \phantom{=}  + \frac{1}{1!1!} \sum_{l_1l_2} \tilde{o}^{11}_{l_1l_2} c^\dagger_{l_1} c_{l_2}  \notag \\
    & \phantom{=}  + \frac{1}{0!2!} \sum_{l_1l_2} \tilde{o}^{02}_{l_1l_2} c_{l_2} c_{l_1} \notag \\
  &= \tilde{o}^{00} + \tilde{o}^{20} + \tilde{o}^{11} + \tilde{o}^{02} \, , \label{no1bpartbasis}
\end{align}
where the matrix elements are given by
\begin{subequations}
\label{no1bpartbasisME}
  \begin{align}
    \tilde{o}^{00}
    &\equiv \Lambda^{00} - \frac{1}{2} \tr[\Lambda^{20}\kappa^\ast] - \tr[\Lambda^{11} \rho] - \frac{1}{2} \tr[\Lambda^{02} \kappa] \notag \\
    &= o^{00} - \frac{1}{2}\tr[o^{22}\rho\rho] - \frac{1}{4}\tr[o^{22}\kappa^\ast\kappa] \, , \\
    \tilde{o}^{20}
    &\equiv \Lambda^{20} \notag \\
    &= \frac{1}{2}\tr[o^{22}\kappa] \, , \\
    \tilde{o}^{11}
    &\equiv \Lambda^{11} \notag \\
    &= o^{11} + \tr[o^{22}\rho] \, , \\
    \tilde{o}^{02}
    &\equiv \Lambda^{02} \notag \\
    &= \frac{1}{2}\tr[o^{22}\kappa^\ast] \, . 
  \end{align}
\end{subequations}
Equations~(\ref{no1bpartbasis}-\ref{no1bpartbasisME}) explicitly demonstrate that the nNO1B approximation leads to an operator that does not conserve particle number as it contains non-zero two-particle-addition $\tilde{o}^{20}$ and two-particle removal $\tilde{o}^{02}$ contributions. 

The application of the PNO1B approximation is more involved and we now proceed to the construction of the corresponding operator
\begin{align}
  O^{\text{PNO1B}}
  &\equiv \tilde{o}^{00}+\tilde{o}^{11} \, , \label{PNOkBexample0}
\end{align}
where the terms are to be obtained recursively on the basis of Eq.~\eqref{eq:otildedef}. The procedure starts with  $\tilde{o}^{kk}=\Lambda^{kk}$, which is normal-ordered in the single-particle basis to generate $\tilde{o}^{(k-1)(k-1)}$ etc. until $\tilde{o}^{00}$ is reached. In the present case, the procedure is trivial since it starts with $k=1$ and leads directly to $\tilde{o}^{00}$ in the first step. Eventually, the result reads as
  \begin{subequations}
  \label{PNOkBexample0ME}
    \begin{align}
      \tilde{o}^{11}
      &= \Lambda^{11} \notag \\
      &= o^{11} + \tr[o^{22}\rho] \, , \\
      \tilde{o}^{00} 
      &= \Lambda^{00} - \tilde{\Lambda}^{00(11)} \, \notag \\
      &= o^{00} - \frac{1}{2} \tr[o^{22}\rho\rho] + \frac{1}{4} \tr[o^{22}\kappa^\ast\kappa] \, ,
    \end{align}
  \end{subequations}
where the only needed intermediate quantity is
  \begin{subequations}
    \begin{align}
      \tilde{\Lambda}^{00(11)}
      &= \tr[\tilde{o}^{11}\rho] \notag \\
      &= \tr[o^{11}\rho] + \tr[o^{22}\rho\rho] \, .
    \end{align}
  \end{subequations}
One first observes that, even if $O$ has no constant term $o^{00}$ to begin with, $O^{\text{PNO1B}}$ does acquire one. The corresponding contributions originate from the two-body operator whose normal-ordered two-body part is omitted. When normal ordering $O^{\text{PNO1B}}$ in the single-particle basis, this term combines with the fully contracted part obtained from $\tilde{o}^{11}$ to generate the complete fully contracted part of the original two-body operator. One further notes that $\tilde{o}^{00}$ entering $O^{\text{PNO1B}}$ is different from the one appearing in Eq.~\eqref{no1bpartbasisME}, which underlines the fact that the procedure to obtain the PNO1B does {\it not} correspond to performing the nNO1B before dropping particle-number non-conserving terms. 

Equations~(\ref{PNOkBexample0}-\ref{PNOkBexample0ME}) fully define the PNO1B approximation of the two-body operator. It is particle-number conserving by construction. It is useful to further characterize the operator by closely inspecting its normal-ordered contributions. Focusing for example on the normal field $\tilde{\Lambda}^{11}$, one obtains
  \begin{align}
    \tilde{\Lambda}^{11}
    &= \tilde{o}^{11} \notag \\
    &= o^{11} + \tr[o^{22}\rho] \notag \\
    &= \Lambda^{11} \, ,
  \end{align}
which indeed satisfies the systematic property $\tilde{\Lambda}^{ii}=\Lambda^{ii}$. Eventually, the complete set of normal-ordered contributions to $O^{\text{PNO1B}}$ expressed in the single-particle basis relates to those of the original two-body operator through
  \begin{subequations}
    \begin{align}
      \tilde{\Lambda}^{22}
      &= 0 \, , \\
      \tilde{\Lambda}^{20}
      &= 0 \, ,\\
      \tilde{\Lambda}^{11}
      &= \Lambda^{11} \, ,\\
      \tilde{\Lambda}^{00}
      &= \Lambda^{00} \, .
    \end{align}
  \end{subequations}
One thus observes that the non-zero fields are strictly equal to those associated with the original operator, i.e. there is no extra term when approximating a two-body operator.
  
\begin{figure*}[t!]
  \centering
  \includegraphics[width=\textwidth]{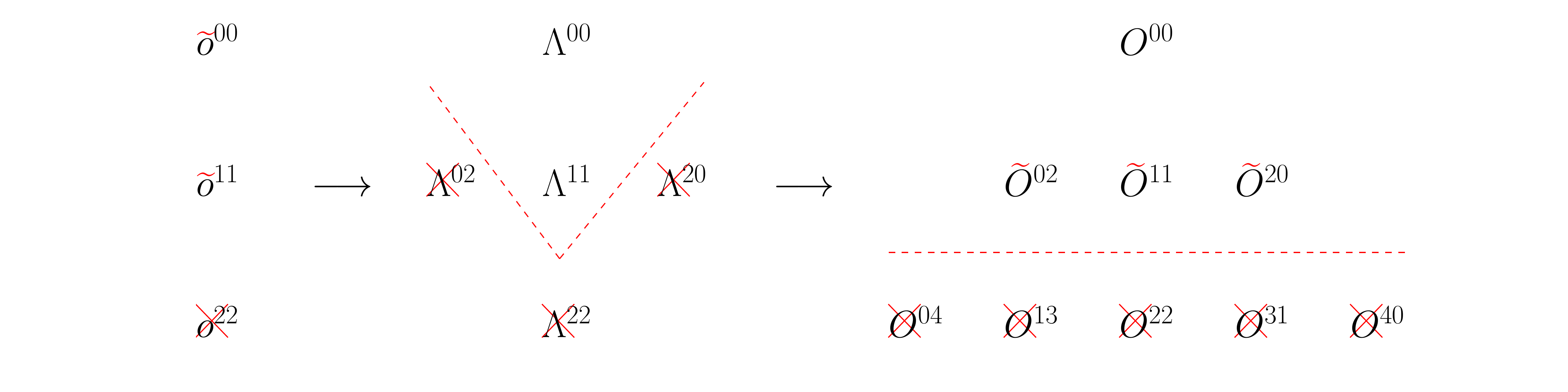}
  \caption{Representation of the PNO1B approximation of a two-body operator $O$.
  Left column: Normal-ordered form with respect to the particle vacuum $|0\rangle$ expressed in $\{c,c^\dagger\}$.
  Middle column: Normal-ordered form with respect to the Bogoliubov vacuum $|\Phi\rangle$ expressed in $\{c,c^\dagger\}$.
  Right column: Normal-ordered form with respect to $|\Phi\rangle$ expressed in $\{\beta,\beta^\dagger\}$.
  The red tildes, crosses and dashed lines embody the effect of the PNO1B approximation, i.e. (1) red crosses indicate the suppressed terms, (2) red dashed lines separate suppressed terms from retained ones and (3) red tildes represent the retained terms that are modified. }
  \label{fig:pno1b}
\end{figure*}

A graphical representation of the PNO1B approximation of a two-body operator $O$ is given in Fig.~\ref{fig:pno1b} .
 
\subsubsection{Other applications}

The PNOkB approximation procedure is worked out in details for three other cases in App.~\ref{ExamplesapplicationPNOkB}, i.e. 
\begin{enumerate}
\item the PNO2B approximation of a 3-body operator,
\item the PNO2B approximation of a 4-body operator,
\item the PNO3B approximation of a 4-body operator,
\end{enumerate}
The first case applies to state-of-the-art \textit{ab initio} calculations~\cite{Tichai:2018mll,Tichai:2019ksh}. The second case illustrates how the extra terms $\breve{\Lambda}^{ij}$ come into play for the first time and may become of practical interest in the future in case four-nucleon interactions have to be accounted for at the two-body effective level. If one becomes capable of working at the effective three-body level, the third case provides the PNO3B approximation of a 4-body operator. 

Finally, the main outcomes of the PNO3B and PNO4B approximations of a 5-body operator are briefly compared in App.~\ref{ExamplesapplicationPNOkB} to illustrate how the extra terms depend on the rank $k$ of the approximate operator.  

\subsection{Testing $[O^{\text{(P)NOkB}},A]=0$}

A method is now introduced to test (formally and numerically) whether or not a given operator $F$ is commuting with the particle-number operator $A$. Starting from a particle-number-conserving operator $O$, the test is meant to be applied to both\footnote{The test constitutes a necessary but not sufficient condition to prove that $F$ commutes with $A$, i.e. the test can thus be used to prove that $O^{\text{nNOkB}}$ does {\it not} commute with $A$ but can  only indicate that $O^{\text{PNOkB}}$ {\it probably} commutes with $A$. This limitation is not problematic in the present case given that $O^{\text{PNOkB}}$ {\it does} commute with $A$ by construction.} $F\equiv O^{\text{nNOkB}}$ and $O^{\text{PNOkB}}$.

\subsubsection{Particle-number projection}

The test is based on the fact that, if $F$ commutes with $A$, it must also commute with the particle-number projection operator $P^{\text{A}}$ defined as
\begin{equation}
P^{\text{A}} \equiv \frac{1}{2\pi} \int_0^{2\pi} \!d\varphi e^{-i\varphi \text{A}} R(\varphi) \, , \label{dePA}
\end{equation}
where $R(\varphi)\equiv e^{iA \varphi}$ is the rotation operator in gauge space.  Given that $P^{\text{A}}$ is idempotent, i.e. $(P^{\text{A}})^2=P^{\text{A}}$, the commutation of $F$ with $P^{\text{A}}$ can be re-expressed as
\begin{align}
P^{\text{A}} F P^{\text{A}}  &= P^{\text{A}} F  \, .
\end{align}
Based on the above, the ratio of singly over doubly projected mean-field matrix elements
\begin{align}
  \label{defQAF}
  Q_{F}^{\text{A}}
  &\equiv \frac{\langle\Phi|P^{\text{A}}F|\Phi\rangle}{\langle\Phi|P^{\text{A}}FP^{\text{A}}|\Phi\rangle} \, ,
\end{align}
is formed such that the particle-number conserving (violating) character of $F$ corresponds to $Q_{F}^{\text{A}} = 1 (\neq 1)$.

\subsubsection{Computation of $Q_{F}^{\text{A}}$}

The computation of $Q_{F}^{\text{A}}$ relies on the representation of $P^{\text{A}}$ given in Eq.~\eqref{dePA} such that
\begin{align}
  \label{eq:quotientcommutation}
  Q_{F}^{\text{A}}
  &= \frac{ \int_0^{2\pi} \!\frac{d\varphi}{2\pi} e^{-i\varphi \text{A}} f^{(0)}(\varphi) \mathcal{N}^{(0)}(\varphi)}{ \int_0^{2\pi} \!\!\!\int_0^{2\pi} \!\frac{d\varphi}{2\pi}\frac{d\varphi^\pr}{2\pi}  e^{-i(\varphi-\varphi^\pr) \text{A}} f^{(0)}(\varphi,\varphi^\pr) \mathcal{N}^{(0)}(\varphi,\varphi^\pr)} \nonumber \, ,
\end{align}
where singly- and doubly-rotated mean-field off-diagonal norm and operator kernels are given by
\begin{subequations}
  \begin{align}
    \mathcal{N}^{(0)}(\varphi)
    &\equiv \langle\Phi(\varphi)|\Phi\rangle \, , \\
    \mathcal{N}^{(0)}(\varphi,\varphi^\pr)
    &\equiv \langle\Phi(\varphi)|\Phi(\varphi^\pr)\rangle = \mathcal{N}^{(0)}(\varphi-\varphi^\pr) \, ,
  \end{align}
\end{subequations}
and by
\begin{subequations}
  \begin{align}
    f^{(0)}(\varphi)
    &\equiv \frac{\langle\Phi(\varphi)|F|\Phi\rangle}{\langle\Phi(\varphi)|\Phi\rangle} \, , \\
    f^{(0)}(\varphi,\varphi^\pr)
    &\equiv \frac{\langle\Phi(\varphi)|F|\Phi(\varphi^\pr)\rangle}{\langle\Phi(\varphi)|\Phi(\varphi^\pr)\rangle} \, ,
  \end{align}
\end{subequations}
where the gauge-rotated Bogoliubov state is defined through $\langle \Phi(\varphi) \vert \equiv \langle \Phi \vert R(\varphi)$.

\subsubsection{One-body operator}

The numerical illustrations provided in the present paper rely on the (P)NO1B of a two-body nuclear Hamiltonian, i.e. $O\equiv H$. Let us thus characterize $Q_{F}^{\text{A}}$ for a generic, i.e. possibly particle-number violating, one-body operator
\begin{align}
  F
  &\equiv f^{00} + f^{20} + f^{11} + f^{02} \\
  &= f^{00} \notag \\
  & \phantom{=}  + \frac{1}{2} \sum_{l_1l_2} f^{20}_{l_1l_2} c^\dagger_{l_1} c^\dagger_{l_2} \notag \\
  & \phantom{=}  +  \phantom{\frac{1}{2}} \sum_{l_1l_2} f^{11}_{l_1l_2} c^\dagger_{l_1} c_{l_2} \notag \\
  & \phantom{=}  + \frac{1}{2} \sum_{l_1l_2} f^{02}_{l_1l_2} c_{l_2} c_{l_1} \, .  \notag
\end{align}
A graphical representation of the operator $F$ is given in Fig.~\ref{fig:no_sp_pv_grid_o11o20o02}. Starting from a two-body operator $O$, the explicit expression of the matrix elements $(f^{00}, f^{20}_{l_1l_2}, f^{11}_{l_1l_2},f^{02}_{l_1l_2})$ associated to $O^{\text{nNO1B}}$ were provided in Sec.~\ref{sec:PNO1B}, where it was formally proven that the particle-number non-conserving terms $f^{20}$ and $f^{02}$ are indeed non zero in this case. For $O^{\text{PNO1B}}$, the sole non-zero terms are $f^{00}$ and $f^{11}$.

\begin{table}[t!]
\centering
\setlength{\tabcolsep}{7.5pt}
\renewcommand{\arraystretch}{1.7}
\begin{tabular}{| c | c | c | c | c | c | c | c |}
\hline 
$\{c,c^\dagger\}, \, |0\ra $	&	-6	&	-4 	& 	-2	&	0	&	+2	&	+4	&	+6 \\
\hline 
\hline 
  $f^{[0]}$	&	&	&	&	$f^{00}$	&	&	&	\\
\hline 
$f^{[2]}$	&	&	&  $f^{02}$ &	$f^{11}$	& $f^{20}$	&	&	\\
\hline 
$f^{[4]}$	&	&	&	&		&	&	&	\\
\hline 
$f^{[6]}$	&	&	&	&		&	&	&	\\
\hline
\end{tabular}
\caption{Contributions to the one-body operator $F$ in normal-ordered form with respect to the particle vacuum $|0\rangle$ and expressed in single-particle basis $\{c,c^\dagger\}$. The $f^{ij}$ contributions are sorted horizontally according to $i-j$ and vertically according to $i+j$.}
\label{fig:no_sp_pv_grid_o11o20o02}
\end{table}

By virtue of the off-diagonal Wick's theorem~\cite{BaBr69}, singly- and doubly-rotated connected operator kernels write respectively as
\begin{subequations}
  \label{eq:sdgaugerotatedoperatorkernels}
  \begin{align}
    f^{(0)}(\varphi)
    &= f^{00} \\
    & \phantom{=} + \frac{1}{2} \sum_{l_1l_2} f^{20}_{l_1l_2} \bar{\kappa}^\ast_{l_1l_2}(\varphi)  \notag \\
    & \phantom{=} +  \phantom{\frac{1}{2}}\sum_{l_1l_2} f^{11}_{l_1l_2} \rho_{l_2l_1}(\varphi)\notag \\
    & \phantom{=} + \frac{1}{2} \sum_{l_1l_2} f^{02}_{l_1l_2} \kappa_{l_1l_2}(\varphi) \, , \notag
  \end{align}
  and as
  \begin{align}
    f^{(0)}(\varphi,\varphi^\pr)
    &= f^{00} \\
    & \phantom{=} + \frac{1}{2} \sum_{l_1l_2}f^{20}_{l_1l_2} \bar{\kappa}^\ast_{l_1l_2}(\varphi,\varphi^\pr)  \notag \\
    & \phantom{=} +  \phantom{\frac{1}{2}}\sum_{l_1l_2} f^{11}_{l_1l_2} \rho_{l_2l_1}(\varphi,\varphi^\pr) \notag \\
    & \phantom{=} + \frac{1}{2} \sum_{l_1l_2} f^{02}_{l_1l_2} \kappa_{l_1l_2}(\varphi,\varphi^\pr) \notag \\
    &= f^{00}  \\
    & \phantom{=} + \left[\frac{1}{2} \sum_{l_1l_2} f^{20}_{l_1l_2}  \bar{\kappa}^\ast_{l_1l_2}(\varphi-\varphi^\pr) \right]  e^{+2i\varphi^\pr} \notag \\
    & \phantom{=} +  \left[\sum_{l_1l_2} f^{11}_{l_1l_2} \rho_{l_2l_1}(\varphi-\varphi^\pr) \right] \notag \\
    & \phantom{=} + \left[\frac{1}{2}\sum_{l_1l_2} f^{02}_{l_1l_2}  \kappa_{l_1l_2}(\varphi-\varphi^\pr)  \right] e^{-2i\varphi^\pr} \, , \notag
  \end{align}
\end{subequations}
where singly and doubly gauge-rotated contractions are defined and calculated in App.~\ref{Approtcontract}. Using the change of variables $\phi\equiv \varphi-\varphi^\pr$, $f^{(0)}(\varphi,\varphi^\pr)$ can be expressed
  \begin{align}
    f^{(0)}(\varphi,\varphi^\pr)
    &= f^{00} \notag \\
    & \phantom{=}  + f^{20(0)}(\phi) \, e^{+i 2\varphi^\pr}   \notag  \\
    & \phantom{=} + f^{11(0)}(\phi)  \notag  \\
    & \phantom{=}  +   f^{02(0)}(\phi) \, e^{-i 2\varphi^\pr} \, , \label{fourier}
  \end{align}
i.e. it displays a Fourier decomposition in $\varphi^\pr$ whose components are nothing but the singly gauge-rotated kernels associated with the various contributions to $F$. The appearance of non-trivial Fourier components, i.e. irreducible representation of $U(1)$, constitute a fingerprint of the particle-number non-conserving character of $F$. In the present case, two such non-trivial modes appear in connection with $f^{20}$ and $f^{02}$. This result can obviously be extended to higher-body operators. Considering a general operator $F$ containing arbitrary combinations of (an even number of) single-particle creation and annihilation operators,  its doubly gauge-rotated mean-field connected kernel takes the form
  \begin{align}
    f^{(0)}(\varphi,\varphi^\pr)
    &\equiv f^{00} \label{fouriergeneral} \\
    & \phantom{=} + \sum_{(m-n)/2 \in \mathbb{Z}} f^{mn(0)}(\phi) \, e^{i(m-n)\varphi^\pr} \, ,  \notag
  \end{align}
such that the Fourier component labelled by $k=m-n$ receives contributions from all $k$-particle addition operators contributing to $F$.

With Eqs.~(\ref{eq:sdgaugerotatedoperatorkernels}-\ref{fourier}) at hand, and as demonstrated in App.~\ref{Appprojsingdoub}, one obtains
\begin{align}
  \label{eq:quotientcommutation2}
  Q_{F}^{\text{A}}
  &= 1 + \frac{\langle\Phi|P^{\text{A}}f^{20}|\Phi\rangle}{\langle\Phi|P^{\text{A}}f^{11}|\Phi\rangle} + \frac{\langle\Phi|P^{\text{A}}f^{02}|\Phi\rangle}{\langle\Phi|P^{\text{A}}f^{11}|\Phi\rangle} \, ,
\end{align}
which is (\textit{a priori} not) equal to $1$ when the particle non-conserving parts of $F$ are (non) vanishing.

\section{Results}
\label{results}

\subsection{Calculations set up}

The nuclear Hamiltonian used in this work includes a chiral two-nucleon (2N) interaction at next-to-next-to-next-to leading order with a cutoff of $\Lambda_{2N}=500 \,\text{MeV}$~\cite{EnMa03}. The three-nucleon interaction is omitted to be able to compare results obtained with the full Hamiltonian to those obtained via its nNO1B and PNO1B approximations. The Hamiltonian is further softened using a Similarity Renormalization Group (SRG) transformation with a flow parameter $\alpha=0.08\,\text{fm}^4$~\cite{BoFu07,HeRo07,RoRe08,RoLa11,JuMa13} such that only up to transformed two-body operators are retained. 

Calculations are performed using the one-body eigenbasis of the spherical harmonic oscillator (HO) Hamiltonian with frequency $\hbar \Omega=20$ MeV. One- and two-body operators are represented using all single-particle states up to $e_\text{max} = (2n +l ) _\text{max} =10$. Spherical Hartree-Fock-Bogoliubov (HFB) calculations are performed in J-coupled scheme while the (single and double) particle-number projection (PHFB) is performed after the variation (PAV) on the basis of $n_{{\text int}}=500$ equally-spaced integration points over the interval\footnote{Working with a Bogoliubov state carrying good, e.g. even, number-parity quantum number, the integration over the gauge angles can indeed be reduced to the interval $[0,\pi]$ in Eq.~\eqref{eq:quotientcommutation}.} $\varphi \in [0,\pi]$. 

\subsection{Energetics in O isotopes}
\label{energetics}

Before coming to normal-ordered approximations, let us discuss the ground-state energetics obtained with the full Hamiltonian to set the orders of magnitude at play. Since three-body forces are presently discarded for the sake of the demonstration, computed energies are not meant to reproduce experimental data.

\begin{figure}
  \centering
  \includegraphics[width=0.35\textwidth]{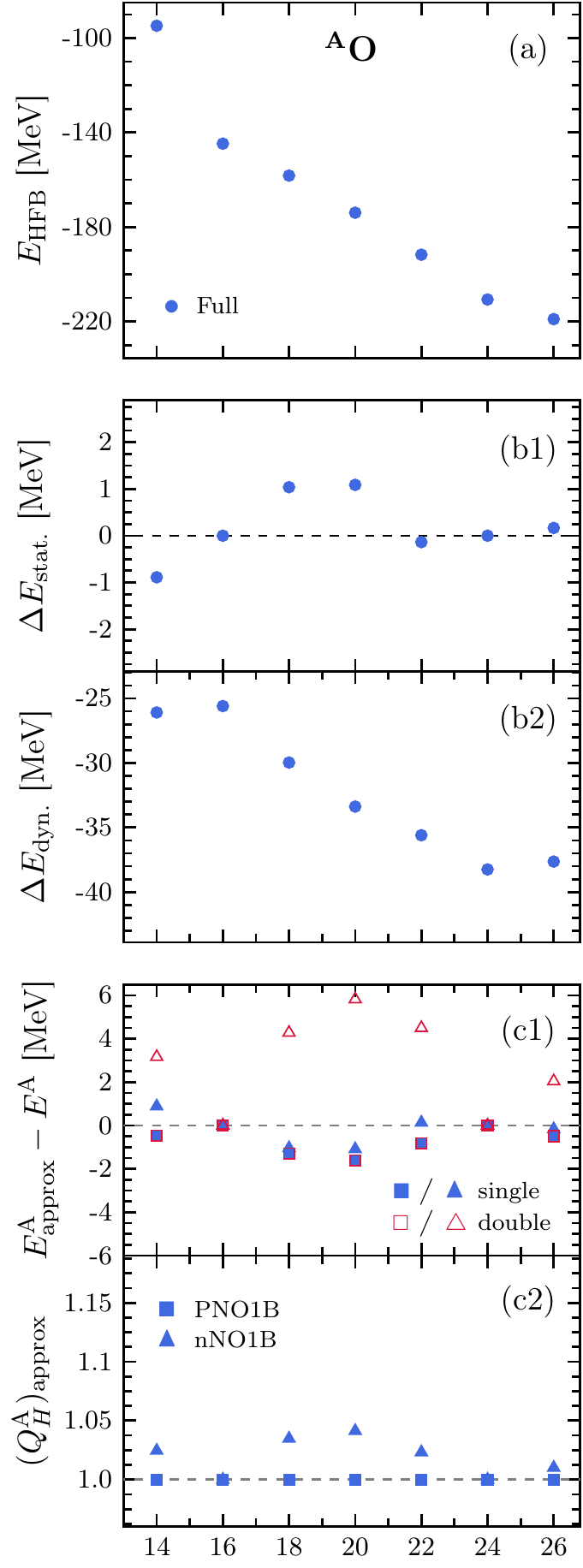}
  \caption{Energetics along oxygen isotopes. Panel (a): HFB energy. Panel (b1): static correlation energy brought about by the particle number projection (after variation). Panel (b2): dynamical correlation energy computed via BMBPT(2). Panel (c1): difference between PHFB energies obtained with the full two-body Hamiltonian and with its nNO1B or PNO1B approximations. Energies are computed twice, i.e. using a single or a double particle-number projection. Panel (c2): Ratios $Q_{H^{\text{nNO1B}}}^{\text{A}}$ and $Q_{H^{\text{PNO1B}}}^{\text{A}}$ of singly- over doubly-projected energies obtained on the basis of $H^{\text{nNO1B}}$ and $H^{\text{PNO1B}}$.}
  \label{Fig:energeticsO}
\end{figure}

In Fig.~\ref{Fig:energeticsO}, ground-state energetics along the Oxygen isotopic chain are displayed. Panel (a) provides HFB energies ranging from about $-100$\,MeV in $^{14}$O to about $-220$\,MeV in $^{26}$O. Panel (b1) displays static correlations associated with particle number projection that are of the order of\footnote{In energy density functional (EDF) calculations, particle-number projection typically lowers the energy, which is not the case with the presently used chiral 2N interaction. Because PHFB calculations with realistic nuclear interactions are rather novel, this feature is briefly analyzed in App.~\ref{sumrulesproj} by looking at the decomposition of the HFB vacuum into its particle-number projected components. } $\pm 1$MeV. This order of magnitude is to be compared with dynamical correlations displayed in panel (b2). Calculated via, e.g., BMBPT(2)~\cite{Duguet:2015yle,Tichai:2018mll} on the basis of the same Hamiltonian and model space, dynamical correlations range from $-26$ to $-38$ MeV in O isotopes. Eventually, both types of correlations can be captured consistently via PBMBPT~\cite{Duguet:2015yle,ripoche19b} or PBCC~\cite{Duguet:2015yle,Qiu:2018edx}.

With results from the full Hamiltonian at hands, one can now analyze the effect of approximating it. Panel (c1) of Fig.~\ref{Fig:energeticsO} displays the difference of PHFB energies obtained with the two-body Hamiltonian and with its nNO1B or PNO1B approximations. While results are the same for single or double particle-number projection when using $H^{\text{PNO1B}}$, it is not the case for  $H^{\text{nNO1B}}$. This observation is confirmed in Panel(c2) where the ratios $Q_{H^{\text{nNO1B}}}^{\text{A}}$ and $Q_{H^{\text{PNO1B}}}^{\text{A}}$ are displayed. While $Q_{H^{\text{PNO1B}}}^{\text{A}}=1$ for all isotopes, $Q_{H^{\text{nNO1B}}}^{\text{A}}\neq 1$ except in doubly closed-shell isotopes. This key result proves that $H^{\text{nNO1B}}$ is particle-number violating, thus making PHFB energies dependent on the way the particle-number projection is defined. Contrarily, $H^{\text{PNO1B}}$ does behave as a particle-number conserving operator. 

One further observes that the effect of the PNO1B approximation is significant, i.e. it is of the same order as the effect of the projection itself displayed in Panel (b1). This is not surprising given that ignoring the residual part of the two-body interaction has never believed to be an appropriate approximation. In order to quantitatively gauge the quality of the PNOkB approximations, one must at least test the PNO2B approximation of a three-nucleon interaction, which is beyond the scope of the present paper that is rather focusing on the symmetry violating/conserving character of a given approximation method.

\subsection{Doubly gauge-rotated operator kernel}

\begin{figure}[t!]
  \centering
  \includegraphics[width=1.\columnwidth]{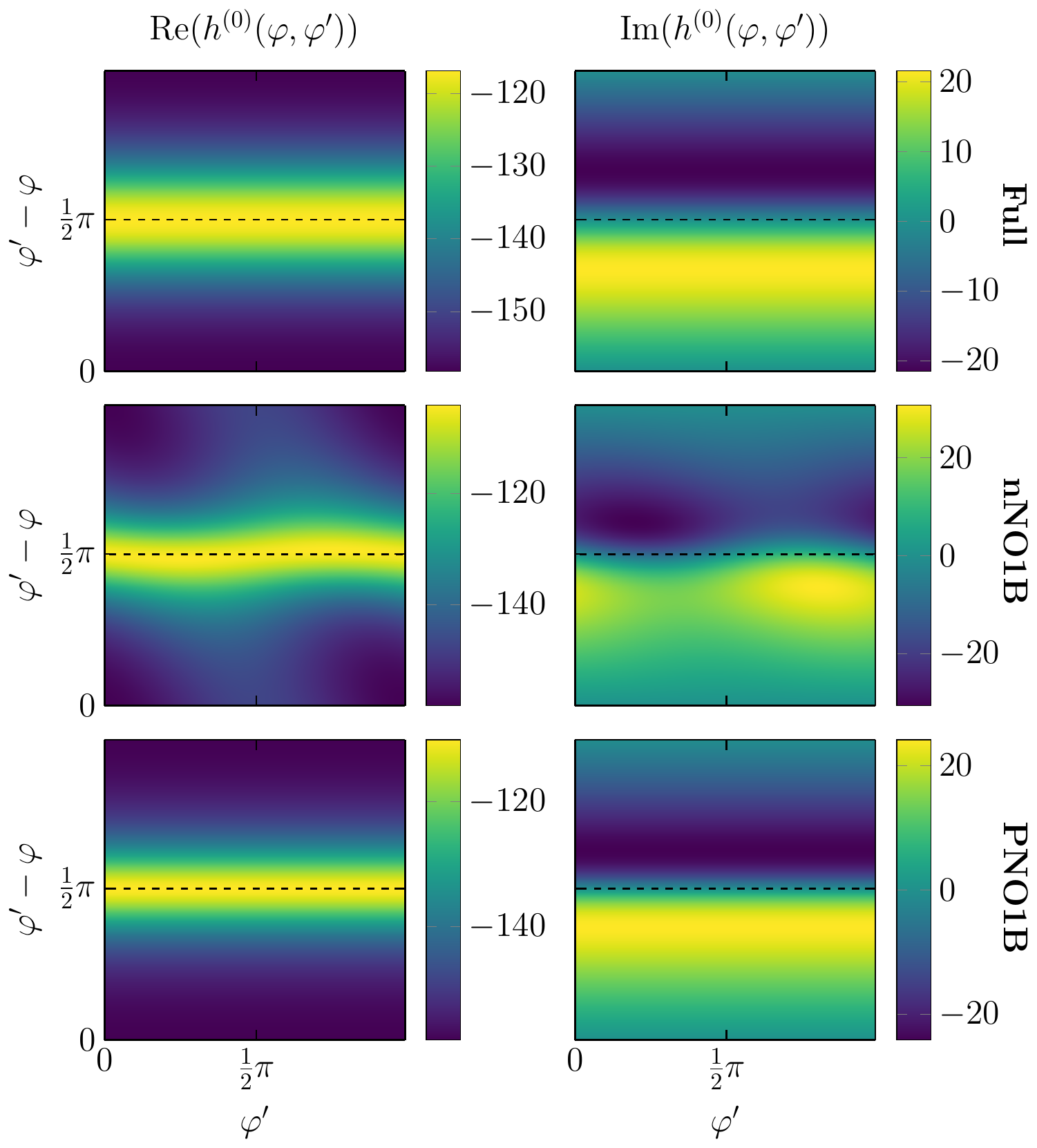}
  \caption{Real (left) and imaginary (right) parts of the doubly gauge-rotated mean-field connected Hamiltonian kernel $h^{(0)}(\varphi,\varphi^\pr)$ in $^{18}$O. Upper panel:  full Hamiltonian. Middle panel: nNO1B. Lower panel: PNO1B.}
  \label{kernels}
\end{figure}

To further characterize the normal-ordered approximations, the doubly gauge-rotated mean-field connected Hamiltonian kernel $h^{(0)}(\varphi,\varphi^\pr)$ is analyzed along with those associated with $H^{\text{nNO1B}}$ and $H^{\text{PNO1B}}$. The real and imaginary parts of the kernels are shown in Fig.~\ref{kernels} for $^{18}$O. They are displayed as contour plots with respect to the variables $(\phi=\varphi-\varphi^\pr,\varphi^\pr)$. 

\begin{figure}
  \centering
  \includegraphics[width=0.5\textwidth]{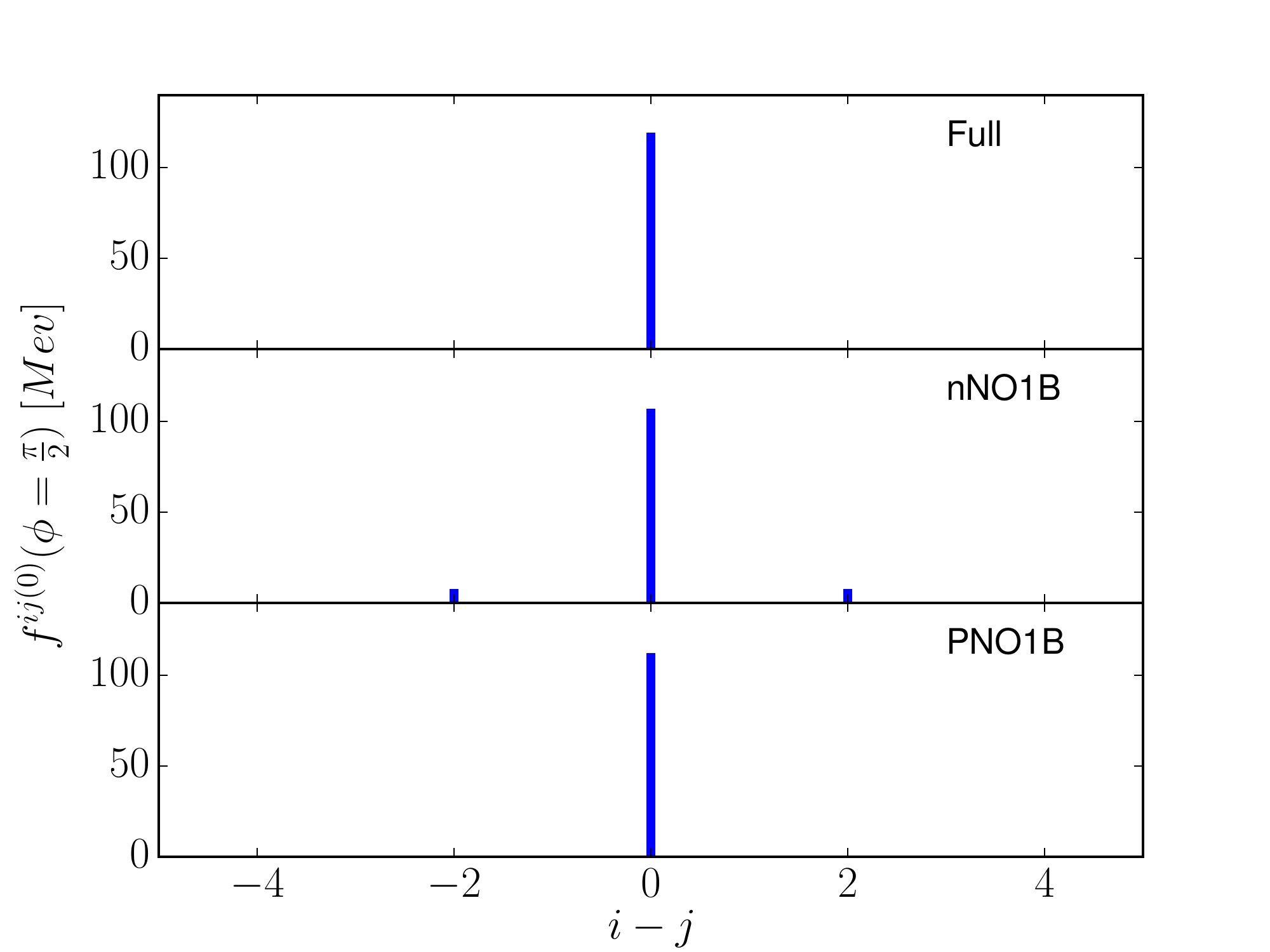}
  \caption{Fourier decomposition of the doubly gauge-rotated mean-field connected Hamiltonian kernel $h^{(0)}(\varphi,\varphi^\pr)$ with respect to $\varphi^\pr$ for fixed $\phi=\pi/2$ in $^{18}$O. Upper panel:  full Hamiltonian. Middle panel: nNO1B. Lower panel: PNO1B.}
  \label{decompofourier}
\end{figure}

Because $H$ is particle-number conserving, its kernel solely depends on $\phi$ and is independent of $\varphi^\pr$, i.e. the Fourier expansion of $h^{(0)}(\varphi,\varphi^\pr)$ with respect to $\varphi^\pr$ in Eq.~\eqref{fouriergeneral} only contains the trivial component, i.e. the irreducible representation of $U(1)$, characterized by $k=0$. While this feature is already manifest in the upper panels of Fig.~\ref{kernels}, it is confirmed in Fig.~\ref{decompofourier} where the Fourier components are numerically extracted. 

Performing the nNO1B approximation, the kernel displayed in the middle panels of Fig.~\ref{kernels} is obtained. It is clear that both the real and imaginary parts now vary with $\varphi^\pr$. This indicates that non-trivial components are present in the Fourier decomposition of Eq.~\eqref{fourier} due to particle-number non-conserving contributions to $H^{\text{nNO1B}}$. As visible from the middle panel of Fig.~\ref{decompofourier} non-zero Fourier components confirm the presence of $f^{20(0)}(\phi)$ and $f^{02(0)}(\phi)$, in agreement with the analytical derivation provided in Eqs.~(\ref{no1bpartbasis}-\ref{no1bpartbasisME}).

Moving from $H^{\text{nNO1B}}$ to $H^{\text{PNO1B}}$, the kernel displayed in the bottom panels of Fig.~\ref{kernels} are obtained. The independence of the kernel with respect to $\varphi^\pr$ is recovered, thus testifying of the particle-number conserving nature of the approximate Hamiltonian $H^{\text{PNO1B}}$. This is confirmed in the lower panel of Fig.~\ref{decompofourier}.

\subsection{Systematics}
\begin{figure}[t!]
\centering
  \includegraphics[width=1.0\columnwidth]{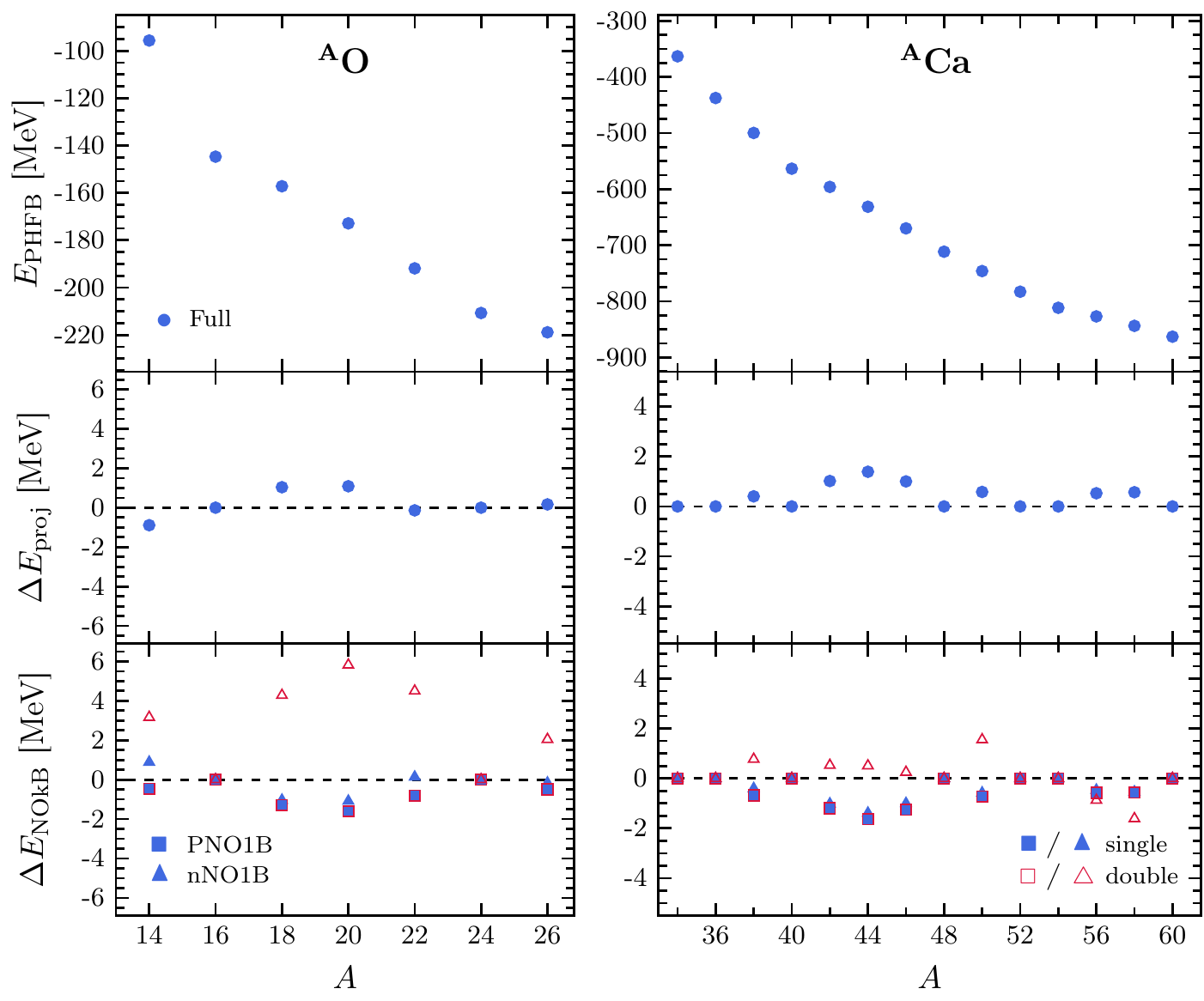}
  \caption{Energetics along oxygen (left) and calcium (right) isotopes. Upper panel: HFB energy. Middle panel: static correlation energy brought about by the particle number projection (after variation). Lower panel: difference between PHFB energies obtained with the full two-body Hamiltonian and with its nNO1B or PNO1B approximations. Energies are computed twice, i.e. using a single or a double particle-number projection.}
  \label{test1}
\end{figure}

The analysis provided in Sec.~\ref{energetics} is now extended to Ca isotopes and to the variance operator. The energetics provided in Fig.~\ref{test1} confirms the conclusions drawn earlier, i.e. while PHFB energies in O and Ca isotopes originating from $H^{\text{PNO1B}}$ are identical when using single or double projection, it is not the case for $H^{\text{nNO1B}}$. It confirms that $H^{\text{nNO1B}}$ ($H^{\text{PNO1B}}$) is particle-number violating (conserving). 

\begin{figure}[t!]
\centering
  \includegraphics[width=1.0\columnwidth]{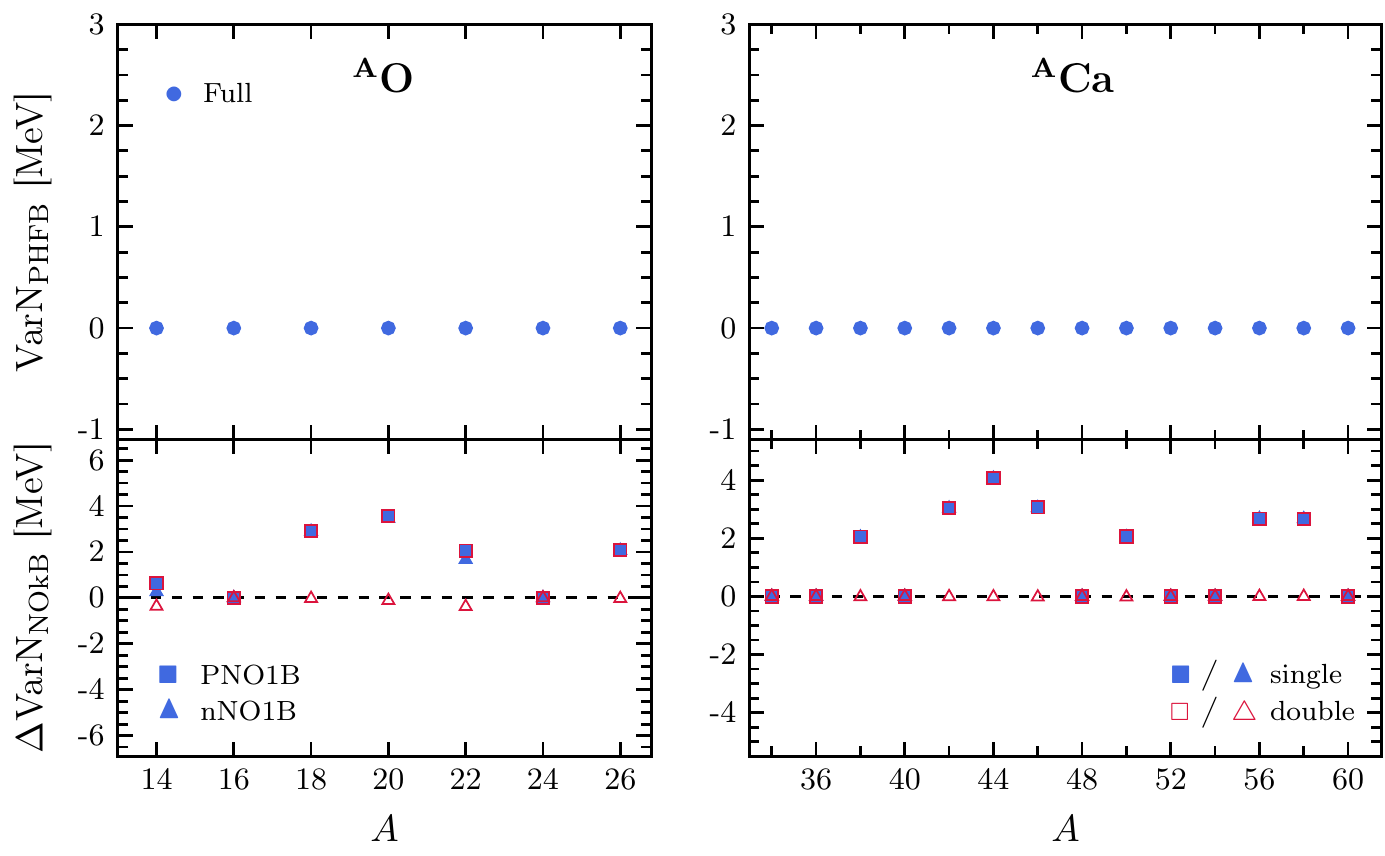}
  \caption{Particle-number variance obtained via PHFB calculations along oxygen (left) and calcium (right) isotopes. Upper panel: results obtained with the full two-body operator. Lower panel: results obtained via the nNO1B and PNO1B approximation to the full operator. Results are computed twice, i.e. using a single or a double particle-number projection.}
  \label{test2}
\end{figure}

As a curiosity, nNO1B and PNO1B approximations are further applied to the particle-number variance operator. Corresponding results are displayed in Fig.~\ref{test2}. When using the full operator, PHFB calculations obviously deliver a null variance, independently of whether the single or double projection is employed. Next, results obtained via the approximate nNO1B (PNO1B) one-body operator  do (not) depend on the way the projection is performed, thus confirming that the approximate operator is particle-number violating (conserving). 
Furthermore, results do depart significantly from the correct one in the PNO1B approximation, which is again not surprising given the expected crudeness of the one-body approximation to a two-body operator in general. Interestingly, nNO1B delivers essentially identical results to PNO1B when the single projection is employed. More surprisingly, nNO1B does provide essentially exact (i.e. null) results when the double projection is used. The reasons for this unexpected result  are analyzed in App.~\ref{approxVARPART} where they are shown to be specific to the one-body approximation of the particle-number variance operator and thus to be accidental.

\section{Conclusion}

In view of dealing efficiently with three-, possibly four-, nucleon interactions in \textit{ab initio} calculations of open-shell nuclei, the present paper addresses approximations based on normal-ordering techniques in the context of many-body methods in which the exact solution of the A-body Schr\"odinger equation is expanded around a symmetry-breaking reference state. Because a naive extension of the standard normal-ordered approximation, originally designed on the basis of symmetry-conserving states, may lead to symmetry-breaking approximate operators, a systematic approximation techniques delivering symmetry-conserving operators is necessary in this context. 

Focusing here on many-body formalisms in which U(1) global-gauge symmetry associated with particle number conservation is broken (and potentially restored), a particle-number-conserving normal-ordered $k$-body (PNOkB) approximation of an arbitrary N-body operator has been designed on the basis of Bogoliubov reference states. After laying down the general formalism, the explicit form of the approximate operator has been provided for various relevant combinations of N and $k$. Furthermore, numerical tests based on particle-number projected Hartree-Fock-Bogoliubov calculations have allowed to check (i) the particle-number violating character of a naive extension of the standard normal-ordered approximation and (ii) the particle-number conserving character of the newly designed PNOkB approximation. Using the PNOkB approximation, \textit{ab initio} calculations based on formalisms exploiting the breaking and restoration of particle-number can thus be safely performed. The future formulation of an angular-momentum-conserving normal-ordered $k$-body approximation based on deformed Slater determinant or Bogoliubov reference states is envisioned.

\section{Acknowledgements}

The authors thank H. Hergert for pointing out to them the existence of the quasi-normal ordering of Ref.~\cite{kong10a}.

\begin{appendix}
\setcounter{equation}{0}
\renewcommand\theequation{A.\arabic{equation}}
\allowdisplaybreaks

\section{Normal-ordered operator}

\subsection{Particle basis}
\label{proof1}

\subsubsection{Derivation}

Applying Wick's theorem~\cite{Wi50} to the n-body part of $O$ gives
  \begin{align}
    o^{nn}
    &= \frac{1}{n!n!} \sum_{l_1 \ldots l_{2n}} o^{nn}_{l_1 \ldots l_{2n}} c^\dagger_{l_1} \ldots c^\dagger_{l_n} c_{l_{2n}} \ldots c_{l_{n+1}} \notag \\
    &= \frac{1}{n!n!} \sum_{l_1 \ldots l_{2n}} o^{nn}_{l_1 \ldots l_{2n}} \left[ :c^\dagger_{l_1} \ldots c^\dagger_{l_n} c_{l_{2n}} \ldots c_{l_{n+1}}: \right. \notag \\
    &\phantom{=}+ \rho_{l_{2n} l_n} :c^\dagger_{l_1} \ldots c^\dagger_{l_{n-1}} c_{l_{2n-1}} \ldots c_{l_{n+1}}: + \ldots \notag \\
    &\phantom{=}+ \kappa^\ast_{l_{n-1} l_{n}} :c^\dagger_{l_1} \ldots c^\dagger_{l_{n-2}} c_{l_{2n}} \ldots c_{l_{n+1}}: + \ldots \notag \\
    &\phantom{=}+ \kappa_{l_{2n-1} l_{2n}} :c^\dagger_{l_1} \ldots c^\dagger_{l_{n}} c_{l_{2n-2}} \ldots c_{l_{n+1}}: + \ldots \notag \\
    &\hphantom{= \frac{1}{n!n!} \sum_{l_1 \ldots l_{2n}} o^{nn}_{l_1 \ldots l_{2n}} \left[\right.} + \ldots \left. \vphantom{c^\dagger_{l_1}} \right] \, .
  \end{align}
The n-body matrix elements $o^{nn}_{l_1 \ldots l_n l_{n+1} \ldots l_{2n}}$ are fully antisymmetric with respect to the permutation of the n first, resp. n last, indices, so that all terms coming from $n_\rho$ normal $\rho$ contractions can be recasted into a single term with an appropriate numerical factor, and similarly for all terms coming from $n_{\kappa^\ast}$ anomalous $\kappa^\ast$ contractions or $n_\kappa$ anomalous $\kappa$ contractions. For the term with $n_\rho$ normal $\rho$ contractions, there are $n_\rho$ creation, resp. annihilitation, operators among $n$ to be considered. Furthermore there are $n_\rho!$ ways to pair $n_\rho$ creation and $n_\rho$ annihilation operators. Thus, the numerical factor resulting from performing $n_\rho$ normal $\rho$ contractions and no anomalous contraction of an initial n-body operator is
  \begin{align}
    \text{Prefactor}(n_\rho,0,0)_n
    &= \frac{n_\rho!\binom{n}{n_\rho}\binom{n}{n_\rho}}{n!n!} \notag \\
    &= \frac{1}{n_\rho!} \frac{1}{(n-n_\rho)!(n-n_\rho)!} \, .
  \end{align}
  Adding $n_{\kappa^\ast}$, resp. $n_{\kappa}$, anomalous $\kappa^\ast$, resp. $\kappa$ contractions, the numerical factor becomes
\begin{strip}
  \begin{align}
    \label{eq:noprefact}
    \text{Prefactor}(n_\rho,n_{\kappa^\ast},n_\kappa)_n
    &= \frac{1}{n_\rho!} \frac{1}{n_{\kappa^\ast}!} \frac{\binom{n-n_\rho}{2} \binom{n-n_\rho-2}{2} \ldots \binom{n-n_\rho-2(n_{\kappa^\ast}-1)}{2}}{(n-n_\rho)!} \frac{1}{n_{\kappa}!} \frac{\binom{n-n_\rho}{2} \binom{n-n_\rho-2}{2} \ldots \binom{n-n_\rho-2(n_{\kappa}-1)}{2}}{(n-n_\rho)!} \notag \\
    &= \frac{1}{n_\rho!n_{\kappa^\ast}!n_{\kappa}!} \left(\frac{1}{2}\right)^{n_{\kappa^\ast}} \left(\frac{1}{2}\right)^{n_\kappa} \frac{1}{(n-n_\rho-2n_{\kappa^\ast})!(n-n_\rho-2n_\kappa)!} \, .
  \end{align}
\end{strip}
The term whose prefactor is given by Eq.~\eqref{eq:noprefact} contributes to $\Lambda^{ij}$ with $i=n-n_\rho-2n_{\kappa^\ast}$ and $j=n-n_\rho-2n_\kappa$. This prefactor accounts for the factor $1/i!j!$ involved in the definition of the operator $\Lambda^{ij}$ in Eq.~\eqref{eq:nofields1} and for the factor appearing in Eq.~\eqref{eq:explicitlambdaijk}.

\subsubsection{Example}
Taking a three-body operator $O$ commuting with $A$ as an example, the matrix elements $\Lambda^{ij}_{l_1\ldots l_i l_{i+1}\ldots l_{i+j}}$ are given by
\begin{subequations}
  \begin{align}
    \Lambda^{00} &\equiv \sum_{n=1}^3 \Lambda^{00(nn)} \notag \\
    &= o^{00} + \sum_{l_1l_2} o^{11}_{l_1l_2} \rho_{l_2l_1} \notag \\*
    &\hphantom{=} + \frac{1}{2} \sum_{l_1l_2l_3l_4} o^{22}_{l_1l_2l_3l_4} \rho_{l_3l_1} \rho_{l_4l_2}  \notag \\
    &\hphantom{=} + \frac{1}{4} \sum_{l_1l_2l_3l_4} o^{22}_{l_1l_2l_3l_4} \kappa^\ast_{l_1l_2} \kappa_{l_3l_4} \notag \\*
    &\hphantom{=} + \frac{1}{6} \sum_{l_1l_2l_3l_4l_5l_6} o^{33}_{l_1l_2l_3l_4l_5l_6} \rho_{l_4l_1} \rho_{l_5l_2} \rho_{l_6l_3} \notag \\ 
    &\hphantom{=} + \frac{1}{4} \sum_{l_1l_2l_3l_4l_5l_6} o^{33}_{l_1l_2l_3l_4l_5l_6} \kappa^\ast_{l_1l_2} \kappa_{l_4l_5} \rho_{l_6l_3} \, , \notag \\*
    \Lambda^{11}_{l_1l_2}&\equiv \sum_{n=1}^3 \Lambda^{11(nn)}_{l_1l_2} \notag \\
    &= o^{11}_{l_1l_2} \notag \\*
    &\hphantom{=} + \sum_{l_3l_4} o^{22}_{l_1l_3l_2l_4} \rho_{l_4l_3} \notag \\*
    &\hphantom{=} + \frac{1}{2} \sum_{l_3l_4l_5l_6} o^{33}_{l_1l_3l_4l_2l_5l_6} \rho_{l_5l_3}\rho_{l_6l_4} \notag \\ 
    &\hphantom{=} + \frac{1}{4} \sum_{l_3l_4l_5l_6} o^{33}_{l_1l_3l_4l_2l_5l_6} \kappa^\ast_{l_3l_4} \kappa_{l_5l_6} \, , \notag \\*
    \Lambda^{20}_{l_1l_2}\equiv \sum_{n=2}^3 \Lambda^{20(nn)}_{l_1l_2}
    &= \frac{1}{2} \sum_{l_3l_4} o^{22}_{l_1l_2l_3l_4} \kappa_{l_3l_4} \notag \\*
    &\hphantom{=} + \frac{1}{2} \sum_{l_3l_4l_5l_6} o^{33}_{l_1l_2l_3l_4l_5l_6} \kappa_{l_4l_5} \rho_{l_6l_3} \, , \\*
    \Lambda^{02}_{l_1l_2} &\equiv \sum_{n=2}^3 \Lambda^{02(nn)}_{l_1l_2} \notag \\
    &= \frac{1}{2} \sum_{l_3l_4} o^{22}_{l_3l_4l_1l_2} \kappa^\ast_{l_3l_4} \notag \\*
    &\hphantom{=} + \frac{1}{2} \sum_{l_3l_4l_5l_6} o^{33}_{l_4l_5l_6l_1l_2l_3} \kappa^\ast_{l_4l_5} \rho_{l_3l_6} \, , \\*
    \Lambda^{22}_{l_1l_2l_3l_4}&\equiv \sum_{n=2}^3 \Lambda^{22(nn)}_{l_1l_2l_3l_4} \notag \\
    &= o^{22}_{l_1l_2l_3l_4}  + \sum_{l_5l_6} o^{33}_{l_1l_2l_5l_3l_4l_6} \rho_{l_6l_5} \, , \\*
    \Lambda^{31}_{l_1l_2l_3l_4} &\equiv \Lambda^{31(33)}_{l_1l_2l_3l_4} \notag \\
    &= \frac{1}{2} \sum_{l_5l_6} o^{33}_{l_1l_2l_3l_4l_5l_6} \kappa_{l_5l_6} \, , \\*
    \Lambda^{13}_{l_1l_2l_3l_4} &\equiv \Lambda^{13(33)}_{l_1l_2l_3l_4} \notag \\
    &= \frac{1}{2} \sum_{l_5l_6} o^{33}_{l_1l_5l_6l_2l_3l_4} \kappa^\ast_{l_5l_6} \, , \\* 
    \Lambda^{33}_{l_1l_2l_3l_4l_5l_6} &\equiv \Lambda^{33(33)}_{l_1l_2l_3l_4l_5l_6} \notag \\
    &= o^{33}_{l_1l_2l_3l_4l_5l_6} \, .
  \end{align}
\end{subequations}
The graphical representation of $O$ expressed in normal-ordered form with respect to $|\Phi\rangle$ is given in Fig.~\ref{fig:no_sp_qpv_grid}. Contrarily to the case displayed in Fig.~\ref{fig:no_sp_pv_grid}, non-diagonal contributions $\Lambda^{20}$, $\Lambda^{02}$, $\Lambda^{31}$ and $\Lambda^{13}$ appear.

\begin{table}[t!]
\centering
\setlength{\tabcolsep}{7.5pt}
\renewcommand{\arraystretch}{1.7}
\begin{tabular}{| c | c | c | c | c | c | c | c |}
\hline 
$\{c,c^\dagger\}, \, |\Phi \ra $	&	-6	&	-4 	& 	-2	&	0	&	+2	&	+4	&	+6 \\
\hline 
\hline 
$\Lambda^{[0]}$	&	&	&	&	$\Lambda^{00}$	&	&	&	\\
\hline 
$\Lambda^{[2]}$	&	&	&	$\Lambda^{02}$ &	$\Lambda^{11}$	&	$\Lambda^{20}$ &	&	\\
\hline 
$\Lambda^{[4]}$	&	&	&	$\Lambda^{13}$ &	$\Lambda^{22}$	&	$\Lambda^{31}$ &	&	\\
\hline 
$\Lambda^{[6]}$	&	&	&	&	$\Lambda^{33}$	&	&	&	\\
\hline
\end{tabular}
\caption{Contributions to the three-body operator $O$ in normal-ordered form with respect to the Bogoliubov vacuum $|\Phi\rangle$ and expressed in the single-particle basis $\{c,c^\dagger\}$. The $\Lambda^{ij}$ contributions are sorted horizontally according to $i-j$ and vertically according to $i+j$.}
\label{fig:no_sp_qpv_grid}
\end{table}

In the Slater determinant limit, for which $\rho_{ip}=\delta_{ip}$, $\rho_{ap}=0$ and $\kappa_{pq}=0$
\footnote{Labels $i,j,\ldots$ denotes hole indices, i.e. occupied single-particle states in the Slater determinant, whereas $a,b,\ldots$ denote particle indices, i.e. occupied single-particle states in the Slater determinant. Indices $p,q,\ldots$ denotes either of those.}, the set of matrix elements reduce to
\begin{subequations}
  \begin{align}
    \Lambda^{00} &= \sum_{i} o^{11}_{ii} + \frac{1}{2} \sum_{ij} o^{22}_{ijij} + \frac{1}{6} \sum_{ijk} o^{33}_{ijkijk} \, , \\
    \Lambda^{11}_{pq} &= o^{11}_{pq} + \sum_{i} o^{22}_{piqi} + \frac{1}{2} \sum_{ij} o^{33}_{pijqij} \, , \\
    \Lambda^{22}_{pqrs} &= o^{22}_{pqrs} + \sum_{i} o^{33}_{pqirsi} \, , \\
    \Lambda^{33}_{pqrstu} &= o^{33}_{pqrstu} \, .
  \end{align}
\end{subequations}
The graphical representation of $O$ in normal-ordered form with respect to $|\text{SD}\rangle$ is given in Fig.~\ref{fig:no_sp_sd_grid}. In this case, no non-diagonal contribution appears.

\begin{table}[h!]
\centering
\setlength{\tabcolsep}{7.8pt}
\renewcommand{\arraystretch}{1.7}
\begin{tabular}{| c | c | c | c | c | c | c | c |}
\hline 
$\{c,c^\dagger\}, \, |\text{SD} \ra $	&	-6	&	-4 	& 	-2	&	0	&	+2	&	+4	&	+6 \\
\hline 
\hline 
$\Lambda^{[0]}$	&	&	&	&	$\Lambda^{00}$	&	&	&	\\
\hline 
$\Lambda^{[2]}$	&	&	&	&	$\Lambda^{11}$	&	&	&	\\
\hline 
$\Lambda^{[4]}$	&	&	&	&	$\Lambda^{22}$	&	&	&	\\
\hline 
$\Lambda^{[6]}$	&	&	&	&	$\Lambda^{33}$	&	&	&	\\
\hline
\end{tabular}
\caption{Contributions to the three-body operator $O$ in normal-ordered form with respect to the slater determinant $|\text{SD}\rangle$ and expressed in the single-particle basis $\{c,c^\dagger\}$. $\Lambda^{ij}$ contributions are sorted horizontally according to $i-j$ and vertically according to $i+j$.}
\label{fig:no_sp_sd_grid}
\end{table}

\subsection{Quasi-particle basis}
\label{proof2}

The derivation relates to the application of Wick's theorem with respect to $| \Phi \rangle$, which is particularly straighforward using the set of quasi-particle operators of which $| \Phi \rangle$ is a vacuum. This procedure has already been discussed and exemplified in Ref.~\cite{Signoracci:2014dia} for a three-body operator $O$ commuting with $A$. However, the expressions of the matrix elements $O^{ij}_{k_1\ldots k_i k_{i+1}\ldots k_{i+j}}$ were only provided for $O^{[0]}$, $O^{[2]}$ and $O^{[4]}$. Extending for completeness the results of Ref.~\cite{Signoracci:2014dia}, matrix elements up to $O^{[6]}$ are given by
\begin{strip}
\begin{align}
  O^{00}
  &= \Lambda^{00} \, , \\
  O^{20}_{k_{1}k_{2}}
  &= \sum_{l_{1}l_{2}}
    \Lambda^{11}_{l_{1}l_{2}}U^\ast_{l_{1}k_{1}}V^\ast_{l_{2}k_{2}}
  - \Lambda^{11}_{l_{1}l_{2}}V^\ast_{l_{2}k_{1}}U^\ast_{l_{1}k_{2}}
  + \Lambda^{20}_{l_{1}l_{2}}U^\ast_{l_{1}k_{1}}U^\ast_{l_{2}k_{2}}
  - \Lambda^{02}_{l_{1}l_{2}}V^\ast_{l_{1}k_{1}}V^\ast_{l_{2}k_{2}} \, , \\
  O^{11}_{k_{1}k_{2}}
  &= \sum_{l_{1}l_{2}}
    \Lambda^{11}_{l_{1}l_{2}}U^\ast_{l_{1}k_{1}}U_{l_{2}k_{2}}
  - \Lambda^{11}_{l_{1}l_{2}}V^\ast_{l_{2}k_{1}}V_{l_{1}k_{2}}
  + \Lambda^{20}_{l_{1}l_{2}}U^\ast_{l_{1}k_{1}}V_{l_{2}k_{2}}
  - \Lambda^{02}_{l_{1}l_{2}}V^\ast_{l_{1}k_{1}}U_{l_{2}k_{2}} \, , \\
  O^{02}_{k_{1}k_{2}}
  &= \sum_{l_{1}l_{2}}
    \Lambda^{11}_{l_{1}l_{2}}U_{l_{2}k_{1}}V_{l_{1}k_{2}}
  - \Lambda^{11}_{l_{1}l_{2}}V_{l_{1}k_{1}}U_{l_{2}k_{2}}
  - \Lambda^{20}_{l_{1}l_{2}}V_{l_{1}k_{1}}V_{l_{2}k_{2}}
  + \Lambda^{02}_{l_{1}l_{2}}U_{l_{1}k_{1}}U_{l_{2}k_{2}} \, , \\
  O^{40}_{k_{1}k_{2}k_{3}k_{4}}
  &= \sum_{l_{1}l_{2}l_{3}l_{4}} \Lambda^{22}_{l_{1}l_{2}l_{3}l_{4}} \left( \vphantom{U^\ast_{l_1k_1}} \right.
  - U^\ast_{l_{1}k_{1}}U^\ast_{l_{2}k_{2}}V^\ast_{l_{3}k_{3}}V^\ast_{l_{4}k_{4}}
  + U^\ast_{l_{1}k_{1}}V^\ast_{l_{3}k_{2}}U^\ast_{l_{2}k_{3}}V^\ast_{l_{4}k_{4}}
  - U^\ast_{l_{1}k_{1}}V^\ast_{l_{3}k_{2}}V^\ast_{l_{4}k_{3}}U^\ast_{l_{2}k_{4}} \notag \\*[-10pt]
  &\hphantom{= \sum_{l_{1}l_{2}l_{3}l_{4}} \Lambda^{22}_{l_{1}l_{2}l_{3}l_{4}} \left( \vphantom{U^\ast_{l_1k_1}} \right.}
  - V^\ast_{l_{3}k_{1}}U^\ast_{l_{1}k_{2}}U^\ast_{l_{2}k_{3}}V^\ast_{l_{4}k_{4}}
  + V^\ast_{l_{3}k_{1}}U^\ast_{l_{1}k_{2}}V^\ast_{l_{4}k_{3}}U^\ast_{l_{2}k_{4}}
  - V^\ast_{l_{3}k_{1}}V^\ast_{l_{4}k_{2}}U^\ast_{l_{1}k_{3}}U^\ast_{l_{2}k_{4}} \left. \vphantom{U^\ast_{l_1k_1}} \right) \\*
  &\hphantom{= \sum_{l_{1}l_{2}l_{3}l_{4}}} \hspace*{-7pt} + \Lambda^{31}_{l_{1}l_{2}l_{3}l_{4}} \left( \vphantom{U^\ast_{l_1k_1}} \right.
  + U^\ast_{l_{1}k_{1}}U^\ast_{l_{2}k_{2}}U^\ast_{l_{3}k_{3}}V^\ast_{l_{4}k_{4}}
  - U^\ast_{l_{1}k_{1}}U^\ast_{l_{2}k_{2}}V^\ast_{l_{4}k_{3}}U^\ast_{l_{3}k_{4}} \notag \\*
  &\hphantom{= \sum_{l_{1}l_{2}l_{3}l_{4}} \hspace*{-7pt} + \Lambda^{31}_{l_{1}l_{2}l_{3}l_{4}} \left( \vphantom{U^\ast_{l_1k_1}} \right.}
  + U^\ast_{l_{1}k_{1}}V^\ast_{l_{4}k_{2}}U^\ast_{l_{2}k_{3}}U^\ast_{l_{3}k_{4}}
  - V^\ast_{l_{4}k_{1}}U^\ast_{l_{1}k_{2}}U^\ast_{l_{2}k_{3}}U^\ast_{l_{3}k_{4}} \left. \vphantom{U^\ast_{l_1k_1}} \right) \\*
  &\hphantom{= \sum_{l_{1}l_{2}l_{3}l_{4}}} \hspace*{-7pt} + \Lambda^{13}_{l_{1}l_{2}l_{3}l_{4}} \left( \vphantom{U^\ast_{l_1k_1}} \right.
  - U^\ast_{l_{1}k_{1}}V^\ast_{l_{5}k_{2}}V^\ast_{l_{6}k_{3}}V^\ast_{l_{2}k_{4}}
  + V^\ast_{l_{5}k_{1}}U^\ast_{l_{1}k_{2}}V^\ast_{l_{6}k_{3}}V^\ast_{l_{2}k_{4}} \notag \\*
  &\hphantom{= \sum_{l_{1}l_{2}l_{3}l_{4}} \hspace*{-7pt} + \Lambda^{13}_{l_{1}l_{2}l_{3}l_{4}} \left( \vphantom{U^\ast_{l_1k_1}} \right.}
  - V^\ast_{l_{5}k_{1}}V^\ast_{l_{6}k_{2}}U^\ast_{l_{1}k_{3}}V^\ast_{l_{2}k_{4}}
  + V^\ast_{l_{5}k_{1}}V^\ast_{l_{6}k_{2}}V^\ast_{l_{2}k_{3}}U^\ast_{l_{1}k_{4}} \left. \vphantom{U^\ast_{l_1k_1}} \right) \, , \\
  O^{31}_{k_{1}k_{2}k_{3}k_{4}}
  &= \sum_{l_{1}l_{2}l_{3}l_{4}} \Lambda^{22}_{l_{1}l_{2}l_{3}l_{4}} \left( \vphantom{U^\ast_{l_1k_1}} \right.
  - U^\ast_{l_{1}k_{1}}V^\ast_{l_{3}k_{2}}V^\ast_{l_{4}k_{3}}V_{l_{2}k_{4}}
  + V^\ast_{l_{3}k_{1}}U^\ast_{l_{1}k_{2}}V^\ast_{l_{4}k_{3}}V_{l_{2}k_{4}}
  - V^\ast_{l_{3}k_{1}}V^\ast_{l_{4}k_{2}}U^\ast_{l_{1}k_{3}}V_{l_{2}k_{4}} \notag \\*[-10pt]
  &\hphantom{= \sum_{l_{1}l_{2}l_{3}l_{4}} \Lambda^{22}_{l_{1}l_{2}l_{3}l_{4}} \left( \vphantom{U^\ast_{l_1k_1}} \right.}
  - U^\ast_{l_{1}k_{1}}U^\ast_{l_{2}k_{2}}V^\ast_{l_{3}k_{3}}U_{l_{4}k_{4}}
  + U^\ast_{l_{1}k_{1}}V^\ast_{l_{3}k_{2}}U^\ast_{l_{2}k_{3}}U_{l_{4}k_{4}}
  - V^\ast_{l_{3}k_{1}}U^\ast_{l_{1}k_{2}}U^\ast_{l_{2}k_{3}}U_{l_{4}k_{4}} \left. \vphantom{U^\ast_{l_1k_1}} \right) \\*
  &\hphantom{= \sum_{l_{1}l_{2}l_{3}l_{4}}} \hspace*{-7pt} + \Lambda^{31}_{l_{1}l_{2}l_{3}l_{4}} \left( \vphantom{U^\ast_{l_1k_1}} \right.
  - U^\ast_{l_{1}k_{1}}U^\ast_{l_{2}k_{2}}V^\ast_{l_{4}k_{3}}V_{l_{3}k_{4}}
  + U^\ast_{l_{1}k_{1}}V^\ast_{l_{4}k_{2}}U^\ast_{l_{2}k_{3}}V_{l_{3}k_{4}} \notag \\*
  &\hphantom{= \sum_{l_{1}l_{2}l_{3}l_{4}} \hspace*{-7pt} + \Lambda^{31}_{l_{1}l_{2}l_{3}l_{4}} \left( \vphantom{U^\ast_{l_1k_1}} \right.}
  - V^\ast_{l_{4}k_{1}}U^\ast_{l_{1}k_{2}}U^\ast_{l_{2}k_{3}}V_{l_{3}k_{4}}
  + U^\ast_{l_{1}k_{1}}U^\ast_{l_{2}k_{2}}U^\ast_{l_{3}k_{3}}U_{l_{4}k_{4}} \left. \vphantom{U^\ast_{l_1k_1}} \right) \\*
  &\hphantom{= \sum_{l_{1}l_{2}l_{3}l_{4}}} \hspace*{-7pt} + \Lambda^{13}_{l_{1}l_{2}l_{3}l_{4}} \left( \vphantom{U^\ast_{l_1k_1}} \right.
  + V^\ast_{l_{5}k_{1}}V^\ast_{l_{6}k_{2}}V^\ast_{l_{2}k_{3}}V_{l_{1}k_{4}}
  - U^\ast_{l_{1}k_{1}}V^\ast_{l_{5}k_{2}}V^\ast_{l_{6}k_{3}}U_{l_{2}k_{4}} \notag \\*
  &\hphantom{= \sum_{l_{1}l_{2}l_{3}l_{4}} \hspace*{-7pt} + \Lambda^{13}_{l_{1}l_{2}l_{3}l_{4}} \left( \vphantom{U^\ast_{l_1k_1}} \right.}
  + V^\ast_{l_{5}k_{1}}U^\ast_{l_{1}k_{2}}V^\ast_{l_{6}k_{3}}U_{l_{2}k_{4}}
  - V^\ast_{l_{5}k_{1}}V^\ast_{l_{6}k_{2}}U^\ast_{l_{1}k_{3}}U_{l_{2}k_{4}} \left. \vphantom{U^\ast_{l_1k_1}} \right) \, , \\
  O^{22}_{k_{1}k_{2}k_{3}k_{4}}
  &= \sum_{l_{1}l_{2}l_{3}l_{4}} \Lambda^{22}_{l_{1}l_{2}l_{3}l_{4}} \left( \vphantom{U^\ast_{l_1k_1}} \right.
  + V^\ast_{l_{3}k_{1}}V^\ast_{l_{4}k_{2}}V_{l_{1}k_{3}}V_{l_{2}k_{4}}
  + U^\ast_{l_{1}k_{1}}V^\ast_{l_{3}k_{2}}U_{l_{4}k_{3}}V_{l_{2}k_{4}}
  - U^\ast_{l_{1}k_{1}}V^\ast_{l_{3}k_{2}}V_{l_{2}k_{3}}U_{l_{4}k_{4}} \notag \\*[-10pt]
  &\hphantom{= \sum_{l_{1}l_{2}l_{3}l_{4}} \Lambda^{22}_{l_{1}l_{2}l_{3}l_{4}} \left( \vphantom{U^\ast_{l_1k_1}} \right.}
  - V^\ast_{l_{3}k_{1}}U^\ast_{l_{1}k_{2}}U_{l_{4}k_{3}}V_{l_{2}k_{4}}
  + V^\ast_{l_{3}k_{1}}U^\ast_{l_{1}k_{2}}V_{l_{2}k_{3}}U_{l_{4}k_{4}}
  + U^\ast_{l_{1}k_{1}}U^\ast_{l_{2}k_{2}}U_{l_{3}k_{3}}U_{l_{4}k_{4}} \left. \vphantom{U^\ast_{l_1k_1}} \right) \\*
  &\hphantom{= \sum_{l_{1}l_{2}l_{3}l_{4}}} \hspace*{-7pt} + \Lambda^{31}_{l_{1}l_{2}l_{3}l_{4}} \left( \vphantom{U^\ast_{l_1k_1}} \right.
  - U^\ast_{l_{1}k_{1}}V^\ast_{l_{4}k_{2}}V_{l_{2}k_{3}}V_{l_{3}k_{4}}
  + V^\ast_{l_{4}k_{1}}U^\ast_{l_{1}k_{2}}V_{l_{2}k_{3}}V_{l_{3}k_{4}} \notag \\*
  &\hphantom{= \sum_{l_{1}l_{2}l_{3}l_{4}} \hspace*{-7pt} + \Lambda^{31}_{l_{1}l_{2}l_{3}l_{4}} \left( \vphantom{U^\ast_{l_1k_1}} \right.}
  + U^\ast_{l_{1}k_{1}}U^\ast_{l_{2}k_{2}}U_{l_{4}k_{3}}V_{l_{3}k_{4}}
  - U^\ast_{l_{1}k_{1}}U^\ast_{l_{2}k_{2}}V_{l_{3}k_{3}}U_{l_{4}k_{4}} \left. \vphantom{U^\ast_{l_1k_1}} \right) \\*
  &\hphantom{= \sum_{l_{1}l_{2}l_{3}l_{4}}} \hspace*{-7pt} + \Lambda^{13}_{l_{1}l_{2}l_{3}l_{4}} \left( \vphantom{U^\ast_{l_1k_1}} \right.
  - V^\ast_{l_{5}k_{1}}V^\ast_{l_{6}k_{2}}U_{l_{2}k_{3}}V_{l_{1}k_{4}}
  + V^\ast_{l_{5}k_{1}}V^\ast_{l_{6}k_{2}}V_{l_{1}k_{3}}U_{l_{2}k_{4}} \notag \\*
  &\hphantom{= \sum_{l_{1}l_{2}l_{3}l_{4}} \hspace*{-7pt} + \Lambda^{13}_{l_{1}l_{2}l_{3}l_{4}} \left( \vphantom{U^\ast_{l_1k_1}} \right.}
  + U^\ast_{l_{1}k_{1}}V^\ast_{l_{5}k_{2}}U_{l_{6}k_{3}}U_{l_{2}k_{4}}
  - V^\ast_{l_{5}k_{1}}U^\ast_{l_{1}k_{2}}U_{l_{6}k_{3}}U_{l_{2}k_{4}} \left. \vphantom{U^\ast_{l_1k_1}} \right) \, , \\
  O^{13}_{k_{1}k_{2}k_{3}k_{4}}
  &= \sum_{l_{1}l_{2}l_{3}l_{4}} \Lambda^{22}_{l_{1}l_{2}l_{3}l_{4}} \left( \vphantom{U^\ast_{l_1k_1}} \right.
  + V^\ast_{l_{3}k_{1}}U_{l_{4}k_{2}}V_{l_{1}k_{3}}V_{l_{2}k_{4}}
  - V^\ast_{l_{3}k_{1}}V_{l_{1}k_{2}}U_{l_{4}k_{3}}V_{l_{2}k_{4}}
  + V^\ast_{l_{3}k_{1}}V_{l_{1}k_{2}}V_{l_{2}k_{3}}U_{l_{4}k_{4}} \notag \\*[-10pt]
  &\hphantom{= \sum_{l_{1}l_{2}l_{3}l_{4}} \Lambda^{22}_{l_{1}l_{2}l_{3}l_{4}} \left( \vphantom{U^\ast_{l_1k_1}} \right.}
  + U^\ast_{l_{1}k_{1}}U_{l_{3}k_{2}}U_{l_{4}k_{3}}V_{l_{2}k_{4}}
  - U^\ast_{l_{1}k_{1}}U_{l_{3}k_{2}}V_{l_{2}k_{3}}U_{l_{4}k_{4}}
  + U^\ast_{l_{1}k_{1}}V_{l_{2}k_{2}}U_{l_{3}k_{3}}U_{l_{4}k_{4}} \left. \vphantom{U^\ast_{l_1k_1}} \right) \\*
  &\hphantom{= \sum_{l_{1}l_{2}l_{3}l_{4}}} \hspace*{-7pt} + \Lambda^{31}_{l_{1}l_{2}l_{3}l_{4}} \left( \vphantom{U^\ast_{l_1k_1}} \right.
  + V^\ast_{l_{4}k_{1}}V_{l_{1}k_{2}}V_{l_{2}k_{3}}V_{l_{3}k_{4}}
  - U^\ast_{l_{1}k_{1}}U_{l_{4}k_{2}}V_{l_{2}k_{3}}V_{l_{3}k_{4}} \notag \\*
  &\hphantom{= \sum_{l_{1}l_{2}l_{3}l_{4}} \hspace*{-7pt} + \Lambda^{31}_{l_{1}l_{2}l_{3}l_{4}} \left( \vphantom{U^\ast_{l_1k_1}} \right.}
  + U^\ast_{l_{1}k_{1}}V_{l_{2}k_{2}}U_{l_{4}k_{3}}V_{l_{3}k_{4}}
  - U^\ast_{l_{1}k_{1}}V_{l_{2}k_{2}}V_{l_{3}k_{3}}U_{l_{4}k_{4}} \left. \vphantom{U^\ast_{l_1k_1}} \right) \\*
  &\hphantom{= \sum_{l_{1}l_{2}l_{3}l_{4}}} \hspace*{-7pt} + \Lambda^{13}_{l_{1}l_{2}l_{3}l_{4}} \left( \vphantom{U^\ast_{l_1k_1}} \right.
  - V^\ast_{l_{5}k_{1}}U_{l_{6}k_{2}}U_{l_{2}k_{3}}V_{l_{1}k_{4}}
  + V^\ast_{l_{5}k_{1}}U_{l_{6}k_{2}}V_{l_{1}k_{3}}U_{l_{2}k_{4}} \notag \\*
  &\hphantom{= \sum_{l_{1}l_{2}l_{3}l_{4}} \hspace*{-7pt} + \Lambda^{13}_{l_{1}l_{2}l_{3}l_{4}} \left( \vphantom{U^\ast_{l_1k_1}} \right.}
  - V^\ast_{l_{5}k_{1}}V_{l_{1}k_{2}}U_{l_{6}k_{3}}U_{l_{2}k_{4}}
  + U^\ast_{l_{1}k_{1}}U_{l_{5}k_{2}}U_{l_{6}k_{3}}U_{l_{2}k_{4}} \left. \vphantom{U^\ast_{l_1k_1}} \right) \, , \\
  O^{04}_{k_{1}k_{2}k_{3}k_{4}}
  &= \sum_{l_{1}l_{2}l_{3}l_{4}} \Lambda^{22}_{l_{1}l_{2}l_{3}l_{4}} \left( \vphantom{U^\ast_{l_1k_1}} \right.
  - U_{l_{3}k_{1}}U_{l_{4}k_{2}}V_{l_{1}k_{3}}V_{l_{2}k_{4}}
  + U_{l_{3}k_{1}}V_{l_{1}k_{2}}U_{l_{4}k_{3}}V_{l_{2}k_{4}}
  - U_{l_{3}k_{1}}V_{l_{1}k_{2}}V_{l_{2}k_{3}}U_{l_{4}k_{4}} \notag \\*[-10pt]
  &\hphantom{= \sum_{l_{1}l_{2}l_{3}l_{4}} \Lambda^{22}_{l_{1}l_{2}l_{3}l_{4}} \left( \vphantom{U^\ast_{l_1k_1}} \right.}
  - V_{l_{1}k_{1}}U_{l_{3}k_{2}}U_{l_{4}k_{3}}V_{l_{2}k_{4}}
  + V_{l_{1}k_{1}}U_{l_{3}k_{2}}V_{l_{2}k_{3}}U_{l_{4}k_{4}}
  - V_{l_{1}k_{1}}V_{l_{2}k_{2}}U_{l_{3}k_{3}}U_{l_{4}k_{4}} \left. \vphantom{U^\ast_{l_1k_1}} \right) \\*
  &\hphantom{= \sum_{l_{1}l_{2}l_{3}l_{4}}} \hspace*{-7pt} + \Lambda^{31}_{l_{1}l_{2}l_{3}l_{4}} \left( \vphantom{U^\ast_{l_1k_1}} \right.
  - U_{l_{4}k_{1}}V_{l_{1}k_{2}}V_{l_{2}k_{3}}V_{l_{3}k_{4}}
  + V_{l_{1}k_{1}}U_{l_{4}k_{2}}V_{l_{2}k_{3}}V_{l_{3}k_{4}} \notag \\*
  &\hphantom{= \sum_{l_{1}l_{2}l_{3}l_{4}} \hspace*{-7pt} + \Lambda^{31}_{l_{1}l_{2}l_{3}l_{4}} \left( \vphantom{U^\ast_{l_1k_1}} \right.}
  - V_{l_{1}k_{1}}V_{l_{2}k_{2}}U_{l_{4}k_{3}}V_{l_{3}k_{4}}
  + V_{l_{1}k_{1}}V_{l_{2}k_{2}}V_{l_{3}k_{3}}U_{l_{4}k_{4}} \left. \vphantom{U^\ast_{l_1k_1}} \right) \\*
  &\hphantom{= \sum_{l_{1}l_{2}l_{3}l_{4}}} \hspace*{-7pt} + \Lambda^{13}_{l_{1}l_{2}l_{3}l_{4}} \left( \vphantom{U^\ast_{l_1k_1}} \right.
  + U_{l_{5}k_{1}}U_{l_{6}k_{2}}U_{l_{2}k_{3}}V_{l_{1}k_{4}}
  - U_{l_{5}k_{1}}U_{l_{6}k_{2}}V_{l_{1}k_{3}}U_{l_{2}k_{4}} \notag \\*
  &\hphantom{= \sum_{l_{1}l_{2}l_{3}l_{4}} \hspace*{-7pt} + \Lambda^{13}_{l_{1}l_{2}l_{3}l_{4}} \left( \vphantom{U^\ast_{l_1k_1}} \right.}
  + U_{l_{5}k_{1}}V_{l_{1}k_{2}}U_{l_{6}k_{3}}U_{l_{2}k_{4}}
  - V_{l_{1}k_{1}}U_{l_{5}k_{2}}U_{l_{6}k_{3}}U_{l_{2}k_{4}} \left. \vphantom{U^\ast_{l_1k_1}} \right) \, , \\
  O^{60}_{k_1k_2k_3k_4k_5k_6}
  &= \sum_{l_1l_2l_3l_4l_5l_6} \Lambda^{33}_{l_1l_2l_3l_4l_5l_6} \left( \vphantom{U^\ast_{l_1k_1}} \right. \notag \\*
  &- U^\ast_{l_1k_1}U^\ast_{l_2k_2}U^\ast_{l_3k_3}V^\ast_{l_4k_4}V^\ast_{l_5k_5}V^\ast_{l_6k_6}
  + U^\ast_{l_1k_1}U^\ast_{l_2k_2}V^\ast_{l_4k_3}U^\ast_{l_3k_4}V^\ast_{l_5k_5}V^\ast_{l_6k_6}
  - U^\ast_{l_1k_1}U^\ast_{l_2k_2}V^\ast_{l_4k_3}V^\ast_{l_5k_4}U^\ast_{l_3k_5}V^\ast_{l_6k_6} \notag \\*
  &+ U^\ast_{l_1k_1}U^\ast_{l_2k_2}V^\ast_{l_4k_3}V^\ast_{l_5k_4}V^\ast_{l_6k_5}U^\ast_{l_3k_6}
  - U^\ast_{l_1k_1}V^\ast_{l_4k_2}U^\ast_{l_2k_3}U^\ast_{l_3k_4}V^\ast_{l_5k_5}V^\ast_{l_6k_6}
  + U^\ast_{l_1k_1}V^\ast_{l_4k_2}U^\ast_{l_2k_3}V^\ast_{l_5k_4}U^\ast_{l_3k_5}V^\ast_{l_6k_6} \notag \\*
  &- U^\ast_{l_1k_1}V^\ast_{l_4k_2}U^\ast_{l_2k_3}V^\ast_{l_5k_4}V^\ast_{l_6k_5}U^\ast_{l_3k_6}
  - U^\ast_{l_1k_1}V^\ast_{l_4k_2}V^\ast_{l_5k_3}U^\ast_{l_2k_4}U^\ast_{l_3k_5}V^\ast_{l_6k_6}
  + U^\ast_{l_1k_1}V^\ast_{l_4k_2}V^\ast_{l_5k_3}U^\ast_{l_2k_4}V^\ast_{l_6k_5}U^\ast_{l_3k_6} \notag \\*
  &- U^\ast_{l_1k_1}V^\ast_{l_4k_2}V^\ast_{l_5k_3}V^\ast_{l_6k_4}U^\ast_{l_2k_5}U^\ast_{l_3k_6}
  + V^\ast_{l_4k_1}U^\ast_{l_1k_2}U^\ast_{l_2k_3}U^\ast_{l_3k_4}V^\ast_{l_5k_5}V^\ast_{l_6k_6}
  - V^\ast_{l_4k_1}U^\ast_{l_1k_2}U^\ast_{l_2k_3}V^\ast_{l_5k_4}U^\ast_{l_3k_5}V^\ast_{l_6k_6} \notag \\*
  &+ V^\ast_{l_4k_1}U^\ast_{l_1k_2}U^\ast_{l_2k_3}V^\ast_{l_5k_4}V^\ast_{l_6k_5}U^\ast_{l_3k_6}
  + V^\ast_{l_4k_1}U^\ast_{l_1k_2}V^\ast_{l_5k_3}U^\ast_{l_2k_4}U^\ast_{l_3k_5}V^\ast_{l_6k_6}
  - V^\ast_{l_4k_1}U^\ast_{l_1k_2}V^\ast_{l_5k_3}U^\ast_{l_2k_4}V^\ast_{l_6k_5}U^\ast_{l_3k_6} \notag \\*
  &+ V^\ast_{l_4k_1}U^\ast_{l_1k_2}V^\ast_{l_5k_3}V^\ast_{l_6k_4}U^\ast_{l_2k_5}U^\ast_{l_3k_6}
  - V^\ast_{l_4k_1}V^\ast_{l_5k_2}U^\ast_{l_1k_3}U^\ast_{l_2k_4}U^\ast_{l_3k_5}V^\ast_{l_6k_6}
  + V^\ast_{l_4k_1}V^\ast_{l_5k_2}U^\ast_{l_1k_3}U^\ast_{l_2k_4}V^\ast_{l_6k_5}U^\ast_{l_3k_6} \notag \\*
  &- V^\ast_{l_4k_1}V^\ast_{l_5k_2}U^\ast_{l_1k_3}V^\ast_{l_6k_4}U^\ast_{l_2k_5}U^\ast_{l_3k_6}
  + V^\ast_{l_4k_1}V^\ast_{l_5k_2}V^\ast_{l_6k_3}U^\ast_{l_1k_4}U^\ast_{l_2k_5}U^\ast_{l_3k_6} \left. \vphantom{U^\ast_{l_1k_1}} \right) \, , \\
  O^{51}_{k_1k_2k_3k_4k_5k_6}
  &= \sum_{l_1l_2l_3l_4l_5l_6} \Lambda^{33}_{l_1l_2l_3l_4l_5l_6} \left( \vphantom{U^\ast_{l_1k_1}} \right. \notag \\*
  &+ U^\ast_{l_1k_1}U^\ast_{l_2k_2}V^\ast_{l_4k_3}V^\ast_{l_5k_4}V^\ast_{l_6k_5}V_{l_3k_6}
  - U^\ast_{l_1k_1}V^\ast_{l_4k_2}U^\ast_{l_2k_3}V^\ast_{l_5k_4}V^\ast_{l_6k_5}V_{l_3k_6}
  + U^\ast_{l_1k_1}V^\ast_{l_4k_2}V^\ast_{l_5k_3}U^\ast_{l_2k_4}V^\ast_{l_6k_5}V_{l_3k_6} \notag \\*
  &- U^\ast_{l_1k_1}V^\ast_{l_4k_2}V^\ast_{l_5k_3}V^\ast_{l_6k_4}U^\ast_{l_2k_5}V_{l_3k_6}
  + V^\ast_{l_4k_1}U^\ast_{l_1k_2}U^\ast_{l_2k_3}V^\ast_{l_5k_4}V^\ast_{l_6k_5}V_{l_3k_6}
  - V^\ast_{l_4k_1}U^\ast_{l_1k_2}V^\ast_{l_5k_3}U^\ast_{l_2k_4}V^\ast_{l_6k_5}V_{l_3k_6} \notag \\*
  &+ V^\ast_{l_4k_1}U^\ast_{l_1k_2}V^\ast_{l_5k_3}V^\ast_{l_6k_4}U^\ast_{l_2k_5}V_{l_3k_6}
  + V^\ast_{l_4k_1}V^\ast_{l_5k_2}U^\ast_{l_1k_3}U^\ast_{l_2k_4}V^\ast_{l_6k_5}V_{l_3k_6}
  - V^\ast_{l_4k_1}V^\ast_{l_5k_2}U^\ast_{l_1k_3}V^\ast_{l_6k_4}U^\ast_{l_2k_5}V_{l_3k_6} \notag \\*
  &+ V^\ast_{l_4k_1}V^\ast_{l_5k_2}V^\ast_{l_6k_3}U^\ast_{l_1k_4}U^\ast_{l_2k_5}V_{l_3k_6}
  - U^\ast_{l_1k_1}U^\ast_{l_2k_2}U^\ast_{l_3k_3}V^\ast_{l_4k_4}V^\ast_{l_5k_5}U_{l_6k_6}
  + U^\ast_{l_1k_1}U^\ast_{l_2k_2}V^\ast_{l_4k_3}U^\ast_{l_3k_4}V^\ast_{l_5k_5}U_{l_6k_6} \notag \\*
  &- U^\ast_{l_1k_1}U^\ast_{l_2k_2}V^\ast_{l_4k_3}V^\ast_{l_5k_4}U^\ast_{l_3k_5}U_{l_6k_6}
  - U^\ast_{l_1k_1}V^\ast_{l_4k_2}U^\ast_{l_2k_3}U^\ast_{l_3k_4}V^\ast_{l_5k_5}U_{l_6k_6}
  + U^\ast_{l_1k_1}V^\ast_{l_4k_2}U^\ast_{l_2k_3}V^\ast_{l_5k_4}U^\ast_{l_3k_5}U_{l_6k_6} \notag \\*
  &- U^\ast_{l_1k_1}V^\ast_{l_4k_2}V^\ast_{l_5k_3}U^\ast_{l_2k_4}U^\ast_{l_3k_5}U_{l_6k_6}
  + V^\ast_{l_4k_1}U^\ast_{l_1k_2}U^\ast_{l_2k_3}U^\ast_{l_3k_4}V^\ast_{l_5k_5}U_{l_6k_6}
  - V^\ast_{l_4k_1}U^\ast_{l_1k_2}U^\ast_{l_2k_3}V^\ast_{l_5k_4}U^\ast_{l_3k_5}U_{l_6k_6} \notag \\*
  &+ V^\ast_{l_4k_1}U^\ast_{l_1k_2}V^\ast_{l_5k_3}U^\ast_{l_2k_4}U^\ast_{l_3k_5}U_{l_6k_6}
  - V^\ast_{l_4k_1}V^\ast_{l_5k_2}U^\ast_{l_1k_3}U^\ast_{l_2k_4}U^\ast_{l_3k_5}U_{l_6k_6} \left. \vphantom{U^\ast_{l_1k_1}} \right) \, , \\
  O^{42}_{k_1k_2k_3k_4k_5k_6}
  &= \sum_{l_1l_2l_3l_4l_5l_6} \Lambda^{33}_{l_1l_2l_3l_4l_5l_6} \left( \vphantom{U^\ast_{l_1k_1}} \right. \notag \\*
  &+ U^\ast_{l_1k_1}V^\ast_{l_4k_2}V^\ast_{l_5k_3}V^\ast_{l_6k_4}V_{l_2k_5}V_{l_3k_6}
  - V^\ast_{l_4k_1}U^\ast_{l_1k_2}V^\ast_{l_5k_3}V^\ast_{l_6k_4}V_{l_2k_5}V_{l_3k_6}
  + V^\ast_{l_4k_1}V^\ast_{l_5k_2}U^\ast_{l_1k_3}V^\ast_{l_6k_4}V_{l_2k_5}V_{l_3k_6} \notag \\*
  &- V^\ast_{l_4k_1}V^\ast_{l_5k_2}V^\ast_{l_6k_3}U^\ast_{l_1k_4}V_{l_2k_5}V_{l_3k_6}
  - U^\ast_{l_1k_1}U^\ast_{l_2k_2}V^\ast_{l_4k_3}V^\ast_{l_5k_4}U_{l_6k_5}V_{l_3k_6}
  + U^\ast_{l_1k_1}U^\ast_{l_2k_2}V^\ast_{l_4k_3}V^\ast_{l_5k_4}V_{l_3k_5}U_{l_6k_6} \notag \\*
  &+ U^\ast_{l_1k_1}V^\ast_{l_4k_2}U^\ast_{l_2k_3}V^\ast_{l_5k_4}U_{l_6k_5}V_{l_3k_6}
  - U^\ast_{l_1k_1}V^\ast_{l_4k_2}U^\ast_{l_2k_3}V^\ast_{l_5k_4}V_{l_3k_5}U_{l_6k_6}
  - U^\ast_{l_1k_1}V^\ast_{l_4k_2}V^\ast_{l_5k_3}U^\ast_{l_2k_4}U_{l_6k_5}V_{l_3k_6} \notag \\*
  &+ U^\ast_{l_1k_1}V^\ast_{l_4k_2}V^\ast_{l_5k_3}U^\ast_{l_2k_4}V_{l_3k_5}U_{l_6k_6}
  - V^\ast_{l_4k_1}U^\ast_{l_1k_2}U^\ast_{l_2k_3}V^\ast_{l_5k_4}U_{l_6k_5}V_{l_3k_6}
  + V^\ast_{l_4k_1}U^\ast_{l_1k_2}U^\ast_{l_2k_3}V^\ast_{l_5k_4}V_{l_3k_5}U_{l_6k_6} \notag \\*
  &+ V^\ast_{l_4k_1}U^\ast_{l_1k_2}V^\ast_{l_5k_3}U^\ast_{l_2k_4}U_{l_6k_5}V_{l_3k_6}
  - V^\ast_{l_4k_1}U^\ast_{l_1k_2}V^\ast_{l_5k_3}U^\ast_{l_2k_4}V_{l_3k_5}U_{l_6k_6}
  - V^\ast_{l_4k_1}V^\ast_{l_5k_2}U^\ast_{l_1k_3}U^\ast_{l_2k_4}U_{l_6k_5}V_{l_3k_6} \notag \\*
  &+ V^\ast_{l_4k_1}V^\ast_{l_5k_2}U^\ast_{l_1k_3}U^\ast_{l_2k_4}V_{l_3k_5}U_{l_6k_6}
  + U^\ast_{l_1k_1}U^\ast_{l_2k_2}U^\ast_{l_3k_3}V^\ast_{l_4k_4}U_{l_5k_5}U_{l_6k_6}
  - U^\ast_{l_1k_1}U^\ast_{l_2k_2}V^\ast_{l_4k_3}U^\ast_{l_3k_4}U_{l_5k_5}U_{l_6k_6} \notag \\*
  &+ U^\ast_{l_1k_1}V^\ast_{l_4k_2}U^\ast_{l_2k_3}U^\ast_{l_3k_4}U_{l_5k_5}U_{l_6k_6}
  - V^\ast_{l_4k_1}U^\ast_{l_1k_2}U^\ast_{l_2k_3}U^\ast_{l_3k_4}U_{l_5k_5}U_{l_6k_6} \left. \vphantom{U^\ast_{l_1k_1}} \right) \, , \\
  O^{33}_{k_1k_2k_3k_4k_5k_6}
  &= \sum_{l_1l_2l_3l_4l_5l_6} \Lambda^{33}_{l_1l_2l_3l_4l_5l_6} \left( \vphantom{U^\ast_{l_1k_1}} \right. \notag \\*
  &- V^\ast_{l_4k_1}V^\ast_{l_5k_2}V^\ast_{l_6k_3}V_{l_1k_4}V_{l_2k_5}V_{l_3k_6}
  + U^\ast_{l_1k_1}V^\ast_{l_4k_2}V^\ast_{l_5k_3}U_{l_6k_4}V_{l_2k_5}V_{l_3k_6}
  - U^\ast_{l_1k_1}V^\ast_{l_4k_2}V^\ast_{l_5k_3}V_{l_2k_4}U_{l_6k_5}V_{l_3k_6} \notag \\*
  &+ U^\ast_{l_1k_1}V^\ast_{l_4k_2}V^\ast_{l_5k_3}V_{l_2k_4}V_{l_3k_5}U_{l_6k_6}
  - V^\ast_{l_4k_1}U^\ast_{l_1k_2}V^\ast_{l_5k_3}U_{l_6k_4}V_{l_2k_5}V_{l_3k_6}
  + V^\ast_{l_4k_1}U^\ast_{l_1k_2}V^\ast_{l_5k_3}V_{l_2k_4}U_{l_6k_5}V_{l_3k_6} \notag \\*
  &- V^\ast_{l_4k_1}U^\ast_{l_1k_2}V^\ast_{l_5k_3}V_{l_2k_4}V_{l_3k_5}U_{l_6k_6}
  + V^\ast_{l_4k_1}V^\ast_{l_5k_2}U^\ast_{l_1k_3}U_{l_6k_4}V_{l_2k_5}V_{l_3k_6}
  - V^\ast_{l_4k_1}V^\ast_{l_5k_2}U^\ast_{l_1k_3}V_{l_2k_4}U_{l_6k_5}V_{l_3k_6} \notag \\*
  &+ V^\ast_{l_4k_1}V^\ast_{l_5k_2}U^\ast_{l_1k_3}V_{l_2k_4}V_{l_3k_5}U_{l_6k_6}
  - U^\ast_{l_1k_1}U^\ast_{l_2k_2}V^\ast_{l_4k_3}U_{l_5k_4}U_{l_6k_5}V_{l_3k_6}
  + U^\ast_{l_1k_1}U^\ast_{l_2k_2}V^\ast_{l_4k_3}U_{l_5k_4}V_{l_3k_5}U_{l_6k_6} \notag \\*
  &- U^\ast_{l_1k_1}U^\ast_{l_2k_2}V^\ast_{l_4k_3}V_{l_3k_4}U_{l_5k_5}U_{l_6k_6}
  + U^\ast_{l_1k_1}V^\ast_{l_4k_2}U^\ast_{l_2k_3}U_{l_5k_4}U_{l_6k_5}V_{l_3k_6}
  - U^\ast_{l_1k_1}V^\ast_{l_4k_2}U^\ast_{l_2k_3}U_{l_5k_4}V_{l_3k_5}U_{l_6k_6} \notag \\*
  &+ U^\ast_{l_1k_1}V^\ast_{l_4k_2}U^\ast_{l_2k_3}V_{l_3k_4}U_{l_5k_5}U_{l_6k_6}
  - V^\ast_{l_4k_1}U^\ast_{l_1k_2}U^\ast_{l_2k_3}U_{l_5k_4}U_{l_6k_5}V_{l_3k_6}
  + V^\ast_{l_4k_1}U^\ast_{l_1k_2}U^\ast_{l_2k_3}U_{l_5k_4}V_{l_3k_5}U_{l_6k_6} \notag \\*
  &- V^\ast_{l_4k_1}U^\ast_{l_1k_2}U^\ast_{l_2k_3}V_{l_3k_4}U_{l_5k_5}U_{l_6k_6}
  + U^\ast_{l_1k_1}U^\ast_{l_2k_2}U^\ast_{l_3k_3}U_{l_4k_4}U_{l_5k_5}U_{l_6k_6} \left. \vphantom{U^\ast_{l_1k_1}} \right) \, , \\
  O^{24}_{k_1k_2k_3k_4k_5k_6}
  &= \sum_{l_1l_2l_3l_4l_5l_6} \Lambda^{33}_{l_1l_2l_3l_4l_5l_6} \left( \vphantom{U^\ast_{l_1k_1}} \right. \notag \\*
  &+ V^\ast_{l_4k_1}V^\ast_{l_5k_2}U_{l_6k_3}V_{l_1k_4}V_{l_2k_5}V_{l_3k_6}
  - V^\ast_{l_4k_1}V^\ast_{l_5k_2}V_{l_1k_3}U_{l_6k_4}V_{l_2k_5}V_{l_3k_6}
  + V^\ast_{l_4k_1}V^\ast_{l_5k_2}V_{l_1k_3}V_{l_2k_4}U_{l_6k_5}V_{l_3k_6} \notag \\*
  &- V^\ast_{l_4k_1}V^\ast_{l_5k_2}V_{l_1k_3}V_{l_2k_4}V_{l_3k_5}U_{l_6k_6}
  - U^\ast_{l_1k_1}V^\ast_{l_4k_2}U_{l_5k_3}U_{l_6k_4}V_{l_2k_5}V_{l_3k_6}
  + U^\ast_{l_1k_1}V^\ast_{l_4k_2}U_{l_5k_3}V_{l_2k_4}U_{l_6k_5}V_{l_3k_6} \notag \\*
  &- U^\ast_{l_1k_1}V^\ast_{l_4k_2}U_{l_5k_3}V_{l_2k_4}V_{l_3k_5}U_{l_6k_6}
  - U^\ast_{l_1k_1}V^\ast_{l_4k_2}V_{l_2k_3}U_{l_5k_4}U_{l_6k_5}V_{l_3k_6}
  + U^\ast_{l_1k_1}V^\ast_{l_4k_2}V_{l_2k_3}U_{l_5k_4}V_{l_3k_5}U_{l_6k_6} \notag \\*
  &- U^\ast_{l_1k_1}V^\ast_{l_4k_2}V_{l_2k_3}V_{l_3k_4}U_{l_5k_5}U_{l_6k_6}
  + V^\ast_{l_4k_1}U^\ast_{l_1k_2}U_{l_5k_3}U_{l_6k_4}V_{l_2k_5}V_{l_3k_6}
  - V^\ast_{l_4k_1}U^\ast_{l_1k_2}U_{l_5k_3}V_{l_2k_4}U_{l_6k_5}V_{l_3k_6} \notag \\*
  &+ V^\ast_{l_4k_1}U^\ast_{l_1k_2}U_{l_5k_3}V_{l_2k_4}V_{l_3k_5}U_{l_6k_6}
  + V^\ast_{l_4k_1}U^\ast_{l_1k_2}V_{l_2k_3}U_{l_5k_4}U_{l_6k_5}V_{l_3k_6}
  - V^\ast_{l_4k_1}U^\ast_{l_1k_2}V_{l_2k_3}U_{l_5k_4}V_{l_3k_5}U_{l_6k_6} \notag \\*
  &+ V^\ast_{l_4k_1}U^\ast_{l_1k_2}V_{l_2k_3}V_{l_3k_4}U_{l_5k_5}U_{l_6k_6}
  + U^\ast_{l_1k_1}U^\ast_{l_2k_2}U_{l_4k_3}U_{l_5k_4}U_{l_6k_5}V_{l_3k_6}
  - U^\ast_{l_1k_1}U^\ast_{l_2k_2}U_{l_4k_3}U_{l_5k_4}V_{l_3k_5}U_{l_6k_6} \notag \\*
  &+ U^\ast_{l_1k_1}U^\ast_{l_2k_2}U_{l_4k_3}V_{l_3k_4}U_{l_5k_5}U_{l_6k_6}
  - U^\ast_{l_1k_1}U^\ast_{l_2k_2}V_{l_3k_3}U_{l_4k_4}U_{l_5k_5}U_{l_6k_6} \left. \vphantom{U^\ast_{l_1k_1}} \right) \, , \\
  O^{15}_{k_1k_2k_3k_4k_5k_6}
  &= \sum_{l_1l_2l_3l_4l_5l_6} \Lambda^{33}_{l_1l_2l_3l_4l_5l_6} \left( \vphantom{U^\ast_{l_1k_1}} \right. \notag \\*
  &+ V^\ast_{l_4k_1}U_{l_5k_2}U_{l_6k_3}V_{l_1k_4}V_{l_2k_5}V_{l_3k_6}
  - V^\ast_{l_4k_1}U_{l_5k_2}V_{l_1k_3}U_{l_6k_4}V_{l_2k_5}V_{l_3k_6}
  + V^\ast_{l_4k_1}U_{l_5k_2}V_{l_1k_3}V_{l_2k_4}U_{l_6k_5}V_{l_3k_6} \notag \\*
  &- V^\ast_{l_4k_1}U_{l_5k_2}V_{l_1k_3}V_{l_2k_4}V_{l_3k_5}U_{l_6k_6}
  + V^\ast_{l_4k_1}V_{l_1k_2}U_{l_5k_3}U_{l_6k_4}V_{l_2k_5}V_{l_3k_6}
  - V^\ast_{l_4k_1}V_{l_1k_2}U_{l_5k_3}V_{l_2k_4}U_{l_6k_5}V_{l_3k_6} \notag \\*
  &+ V^\ast_{l_4k_1}V_{l_1k_2}U_{l_5k_3}V_{l_2k_4}V_{l_3k_5}U_{l_6k_6}
  + V^\ast_{l_4k_1}V_{l_1k_2}V_{l_2k_3}U_{l_5k_4}U_{l_6k_5}V_{l_3k_6}
  - V^\ast_{l_4k_1}V_{l_1k_2}V_{l_2k_3}U_{l_5k_4}V_{l_3k_5}U_{l_6k_6} \notag \\*
  &+ V^\ast_{l_4k_1}V_{l_1k_2}V_{l_2k_3}V_{l_3k_4}U_{l_5k_5}U_{l_6k_6}
  - U^\ast_{l_1k_1}U_{l_4k_2}U_{l_5k_3}U_{l_6k_4}V_{l_2k_5}V_{l_3k_6}
  + U^\ast_{l_1k_1}U_{l_4k_2}U_{l_5k_3}V_{l_2k_4}U_{l_6k_5}V_{l_3k_6} \notag \\*
  &- U^\ast_{l_1k_1}U_{l_4k_2}U_{l_5k_3}V_{l_2k_4}V_{l_3k_5}U_{l_6k_6}
  - U^\ast_{l_1k_1}U_{l_4k_2}V_{l_2k_3}U_{l_5k_4}U_{l_6k_5}V_{l_3k_6}
  + U^\ast_{l_1k_1}U_{l_4k_2}V_{l_2k_3}U_{l_5k_4}V_{l_3k_5}U_{l_6k_6} \notag \\*
  &- U^\ast_{l_1k_1}U_{l_4k_2}V_{l_2k_3}V_{l_3k_4}U_{l_5k_5}U_{l_6k_6}
  + U^\ast_{l_1k_1}V_{l_2k_2}U_{l_4k_3}U_{l_5k_4}U_{l_6k_5}V_{l_3k_6}
  - U^\ast_{l_1k_1}V_{l_2k_2}U_{l_4k_3}U_{l_5k_4}V_{l_3k_5}U_{l_6k_6} \notag \\*
  &+ U^\ast_{l_1k_1}V_{l_2k_2}U_{l_4k_3}V_{l_3k_4}U_{l_5k_5}U_{l_6k_6}
  - U^\ast_{l_1k_1}V_{l_2k_2}V_{l_3k_3}U_{l_4k_4}U_{l_5k_5}U_{l_6k_6} \left. \vphantom{U^\ast_{l_1k_1}} \right) \, , \\
  O^{06}_{k_1k_2k_3k_4k_5k_6}
  &= \sum_{l_1l_2l_3l_4l_5l_6} \Lambda^{33}_{l_1l_2l_3l_4l_5l_6} \left( \vphantom{U^\ast_{l_1k_1}} \right. \notag \\*
  &- U_{l_4k_1}U_{l_5k_2}U_{l_6k_3}V_{l_1k_4}V_{l_2k_5}V_{l_3k_6}
  + U_{l_4k_1}U_{l_5k_2}V_{l_1k_3}U_{l_6k_4}V_{l_2k_5}V_{l_3k_6}
  - U_{l_4k_1}U_{l_5k_2}V_{l_1k_3}V_{l_2k_4}U_{l_6k_5}V_{l_3k_6} \notag \\*
  &+ U_{l_4k_1}U_{l_5k_2}V_{l_1k_3}V_{l_2k_4}V_{l_3k_5}U_{l_6k_6}
  - U_{l_4k_1}V_{l_1k_2}U_{l_5k_3}U_{l_6k_4}V_{l_2k_5}V_{l_3k_6}
  + U_{l_4k_1}V_{l_1k_2}U_{l_5k_3}V_{l_2k_4}U_{l_6k_5}V_{l_3k_6} \notag \\*
  &- U_{l_4k_1}V_{l_1k_2}U_{l_5k_3}V_{l_2k_4}V_{l_3k_5}U_{l_6k_6}
  - U_{l_4k_1}V_{l_1k_2}V_{l_2k_3}U_{l_5k_4}U_{l_6k_5}V_{l_3k_6}
  + U_{l_4k_1}V_{l_1k_2}V_{l_2k_3}U_{l_5k_4}V_{l_3k_5}U_{l_6k_6} \notag \\*
  &- U_{l_4k_1}V_{l_1k_2}V_{l_2k_3}V_{l_3k_4}U_{l_5k_5}U_{l_6k_6}
  + V_{l_1k_1}U_{l_4k_2}U_{l_5k_3}U_{l_6k_4}V_{l_2k_5}V_{l_3k_6}
  - V_{l_1k_1}U_{l_4k_2}U_{l_5k_3}V_{l_2k_4}U_{l_6k_5}V_{l_3k_6} \notag \\*
  &+ V_{l_1k_1}U_{l_4k_2}U_{l_5k_3}V_{l_2k_4}V_{l_3k_5}U_{l_6k_6}
  + V_{l_1k_1}U_{l_4k_2}V_{l_2k_3}U_{l_5k_4}U_{l_6k_5}V_{l_3k_6}
  - V_{l_1k_1}U_{l_4k_2}V_{l_2k_3}U_{l_5k_4}V_{l_3k_5}U_{l_6k_6} \notag \\*
  &+ V_{l_1k_1}U_{l_4k_2}V_{l_2k_3}V_{l_3k_4}U_{l_5k_5}U_{l_6k_6}
  - V_{l_1k_1}V_{l_2k_2}U_{l_4k_3}U_{l_5k_4}U_{l_6k_5}V_{l_3k_6}
  + V_{l_1k_1}V_{l_2k_2}U_{l_4k_3}U_{l_5k_4}V_{l_3k_5}U_{l_6k_6} \notag \\*
  &- V_{l_1k_1}V_{l_2k_2}U_{l_4k_3}V_{l_3k_4}U_{l_5k_5}U_{l_6k_6}
  + V_{l_1k_1}V_{l_2k_2}V_{l_3k_3}U_{l_4k_4}U_{l_5k_5}U_{l_6k_6} \left. \vphantom{U^\ast_{l_1k_1}} \right) \, .
\end{align}
\end{strip}
The graphical representation of the normal-ordered operator $O$ with respect to $|\Phi\rangle$, resp. $|\text{SD}\rangle$, and expressed in the quasi-particle basis, is given in Tab.~\ref{fig:no_qp_qpv_grid}, resp. Tab~\ref{fig:no_sdp_sd_grid}. One notices the presence of non-diagonal contributions in Tab.~\ref{fig:no_qp_qpv_grid} (as in Tab.~\ref{fig:no_sp_qpv_grid}) and in Tab~\ref{fig:no_sdp_sd_grid} (even if there is none in Tab.~\ref{fig:no_sp_sd_grid}).
This is due to the use of the quasi-particle algebra that is agnostic regarding the symmetry-breaking or the symmetry-preserving character of the vacuum.

\begin{table}[t!]
\centering
\setlength{\tabcolsep}{6.0pt}
\renewcommand{\arraystretch}{1.7}
\begin{tabular}{| c | c | c | c | c | c | c | c |}
\hline 
$\{\beta,\beta^\dagger\}, |\Phi \ra $	&	-6	&	-4 	& 	-2	&	0	&	+2	&	+4	&	+6 \\
\hline 
\hline 
$O^{[0]}$	&	&	&	&	$O^{00}$	&	&	&	\\
\hline 
$O^{[2]}$	&	&	&	$O^{02}$ &	$O^{11}$	&	$O^{20}$ &	&	\\
\hline 
$O^{[4]}$	&	&	$O^{14}$ &	$O^{13}$ &	$O^{22}$	&	$O^{31}$ &	$O^{40}$&	\\
\hline 
$O^{[6]}$	&	$O^{06}$ &	$O^{15}$&	$O^{24}$&	$O^{33}$	& $O^{42}$	& $O^{51}$	& $O^{60}$	\\
\hline
\end{tabular}
\caption{Contributions to the three-body operator $O$ in normal-ordered form with respect to the quasi-particle vacuum $|\Phi\rangle$ and expressed $\{\beta,\beta^\dagger\}$. The $O^{ij}$ contributions are sorted horizontally according to $i-j$ and vertically according to $i+j$.}
\label{fig:no_qp_qpv_grid}
\end{table}

\begin{table}[t!]
\centering
\setlength{\tabcolsep}{5.7pt}
\renewcommand{\arraystretch}{1.7}
\begin{tabular}{| c | c | c | c | c | c | c | c |}
\hline 
$\{\beta,\beta^\dagger\}, |\text{SD} \ra $	&	-6	&	-4 	& 	-2	&	0	&	+2	&	+4	&	+6 \\
\hline 
\hline 
$O^{[0]}$	&	&	&	&	$O^{00}$	&	&	&	\\
\hline 
$O^{[2]}$	&	&	&	$O^{02}$ &	$O^{11}$	&	$O^{20}$ &	&	\\
\hline 
$O^{[4]}$	&	&	$O^{14}$ &	$O^{13}$ &	$O^{22}$	&	$O^{31}$ &	$O^{40}$&	\\
\hline 
$O^{[6]}$	&	$O^{06}$ &	$O^{15}$&	$O^{24}$&	$O^{33}$	& $O^{42}$	& $O^{51}$	& $O^{60}$	\\
\hline
\end{tabular}
\caption{Contributions to the three-body operator $O$ in normal-ordered form with respect to the slater determinant $|\text{SD}\rangle$ and expressed in $\{\beta,\beta^\dagger\}$. The $O^{ij}$ contributions are sorted horizontally according to the value of $i-j$ and vertically according to the value of $i+j$.}
\label{fig:no_sdp_sd_grid}
\end{table}

%

\section{Particle-number-conserving NOkB approximation}
\label{AppPNOkB}

\subsection{Derivation}

For each $(i,j)$ such that $\max(i,j)\le k$ one has to prove that
  \begin{align}
    \label{eq:lambdaequality}
    \tilde{\Lambda}^{ij} &= \Lambda^{ij} + \breve{\Lambda}^{ij} \, .
  \end{align}
The n-body normal fields, for which $i=j=n$, are given by
  \begin{align}
    \label{eq:lambdatildedemo}
    \tilde{\Lambda}^{nn}_{l_1 \ldots l_{2n}}
    &= \tilde{\Lambda}^{nn(nn)}_{l_1 \ldots l_{2n}} + \sum_{m=n+1}^{k} \tilde{\Lambda}^{nn(mm)}_{l_1 \ldots l_{2n}} \notag \\
    &= \tilde{o}^{nn}_{l_1 \ldots l_{2n}} + \sum_{m=n+1}^{k} \tilde{\Lambda}^{nn(mm)}_{l_1 \ldots l_{2n}} \notag \\
    &= \Lambda^{nn}_{l_1 \ldots l_{2n}} \, ,
  \end{align}
  where the property $\tilde{\Lambda}^{nn(nn)} = \tilde{o}^{nn}$ and the definition of $\tilde{o}^{nn}$ in Eq.~\eqref{eq:otildedef} have been used. Consequently, $\breve{\Lambda}^{nn}=0$ \eqref{eq:lambdabrevedef3} as required.

Let us now focus on the n-body contributions to anomalous fields associated with the full operator $O$ and for which $n\ge \max(i,j)$ and, e.g., $i>j$. One can write
\begin{strip}
  \begin{align}
    \Lambda^{ij(nn)}_{l_1 \ldots l_{i+j}}
    &= \sum_{n_\kappa=\frac{i-j}{2}}^{\left\lfloor \frac{n-j}{2} \right\rfloor} \Lambda^{ij(nn)(n_\kappa)}_{l_1 \ldots l_{i+j}} \label{eq:lambdaijnn} \\
    &= \sum_{n_\kappa=0}^{\left\lfloor \frac{n-i}{2} \right\rfloor} \frac{n_\kappa!}{\left(n_\kappa + \frac{i-j}{2}\right)!} \left(\frac{1}{2}\right)^{\frac{i-j}{2}} \sum_{l_{i+j+1}\ldots l_{2i}} \Lambda^{ii(nn)(n_\kappa)}_{l_1 \ldots l_{2i}} \kappa_{l_{i+j+1}l_{i+j+2}} \ldots \kappa_{l_{2i-1} l_{2i}} \notag \\
    &= \frac{1}{\left(\frac{i-j}{2}\right)!}\left(\frac{1}{2}\right)^{\frac{i-j}{2}} \sum_{l_{i+j+1}\ldots l_{2i}} \left[ \Lambda^{ii(nn)}_{l_1 \ldots l_{2i}} - \sum_{n_\kappa=0}^{\left\lfloor \frac{n-i}{2} \right\rfloor} \left( 1 - \frac{1}{\binom{n_\kappa + \frac{i-j}{2}}{n_\kappa}}\right) \Lambda^{ii(nn)(n_\kappa)}_{l_1 \ldots l_{2i}} \right] \kappa_{l_{i+j+1}l_{i+j+2}} \ldots \kappa_{l_{2i-1} l_{2i}} \, ,
  \end{align}
where $\Lambda^{ij(nn)(n_\kappa)}$ denotes the contribution from $o^{nn}$ to $\Lambda^{ij}$ containing $n_\kappa$ anomalous contractions $\kappa$. Indeed, several contractions patterns associated with different numbers of normal and anomalous contractions can lead from $o^{nn}$ to $\Lambda^{ij}$, knowing that the minimal number of $\kappa$ contractions is $(i-j)/2$. The goal of the above rewriting is to factorize $(i-j)/2$ anomalous contractions $\kappa$ in order to express each contribution $\Lambda^{ij(nn)(n_\kappa)}$ in terms of its diagonal (i.e. $i=j$) partner $\Lambda^{ii(nn)(n_\kappa)}$. Summing \eqref{eq:lambdaijnn} over n-body contributions with $n=i,\ldots,N$, one obtains
  \begin{align}
    \Lambda^{ij}_{l_1 \ldots l_{i+j}}
    &= \frac{1}{\left(\frac{i-j}{2}\right)!}\left(\frac{1}{2}\right)^{\frac{i-j}{2}} \sum_{l_{i+j+1}\ldots l_{2i}} \left[ \Lambda^{ii}_{l_1 \ldots l_{2i}} - \sum_{n=i}^{N} \sum_{n_\kappa=0}^{\left\lfloor \frac{n-i}{2} \right\rfloor} \left( 1 - \frac{1}{\binom{n_\kappa + \frac{i-j}{2}}{n_\kappa}}\right) \Lambda^{ii(nn)(n_\kappa)}_{l_1 \ldots l_{2i}} \right] \kappa_{l_{i+j+1}l_{i+j+2}} \ldots \kappa_{l_{2i-1} l_{2i}} \, .
  \end{align}
A similar relation holds for $\tilde{\Lambda}^{ij}$ associated with $O^{\text{PNOkB}}$
  \begin{align}
    \tilde{\Lambda}^{ij}_{l_1 \ldots l_{i+j}}
    &= \frac{1}{\left(\frac{i-j}{2}\right)!}\left(\frac{1}{2}\right)^{\frac{i-j}{2}} \sum_{l_{i+j+1}\ldots l_{2i}} \left[ \tilde{\Lambda}^{ii}_{l_1 \ldots l_{2i}} - \sum_{n=i}^{k} \sum_{n_\kappa=0}^{\left\lfloor \frac{n-i}{2} \right\rfloor} \left( 1 - \frac{1}{\binom{n_\kappa + \frac{i-j}{2}}{n_\kappa}}\right) \tilde{\Lambda}^{ii(nn)(n_\kappa)}_{l_1 \ldots l_{2i}} \right] \kappa_{l_{i+j+1}l_{i+j+2}} \ldots \kappa_{l_{2i-1} l_{2i}} \, , 
  \end{align}
 the difference being that the sum over $n$ only extends up to $k$ instead of $N$ for $\Lambda^{ij}$ given that $\tilde{o}^{nn}$ is null for $n>k$ to begin with. Using Eq.~\eqref{eq:lambdatildedemo}, the combination of the two above identities allows one to relate both sets of fields through
\begin{align}
\tilde{\Lambda}^{ij}_{l_1 \ldots l_{i+j}}
    &=  \Lambda^{ij}_{l_1 \ldots l_{i+j}} + \breve{\Lambda}^{ij}_{l_1 \ldots l_{i+j}} \, ,
\end{align}
with
\begin{align}
\breve{\Lambda}^{ij}_{l_1 \ldots l_{i+j}}
    &\equiv \frac{1}{\left(\frac{i-j}{2}\right)!}\left(\frac{1}{2}\right)^{\frac{i-j}{2}} \sum_{n=i}^{N} \sum_{n_\kappa=0}^{\left\lfloor \frac{n-i}{2} \right\rfloor} \left(1 - \frac{1}{\binom{n_\kappa + \frac{i-j}{2}}{n_\kappa}}\right) \sum_{l_{i+j+1}\ldots l_{2i}} \left[ \Lambda^{ii(nn)(n_\kappa)}_{l_1 \ldots l_{2i}} - \tilde{\Lambda}^{ii(nn)(n_\kappa)}_{l_1 \ldots l_{2i}} \right] \kappa_{l_{i+j+1}l_{i+j+2}} \ldots \kappa_{l_{2i-1} l_{2i}} \, , \label{generallambdatilde}
  \end{align}
\end{strip}
where $\tilde{\Lambda}^{ii(nn)(n_\kappa)}$ is in fact zero for $n>k$. The case of anomalous field with $i<j$ is obtained through the same procedure by factorizing $\kappa^\ast$ contractions instead of $\kappa$ ones.

Of course, the standard \textit{NOkB approximation} is recovered from the \textit{PNOkB} one whenever the reference state is particle-number conserving, i.e. whenever $|\Phi\rangle$ reduces to a Slater determinant $|\text{SD}\rangle$.

\subsection{Examples}
\label{ExamplesapplicationPNOkB}

\subsubsection{PNO2B approximation of a three-body operator}

Let us exemplify the case of largest interest in current nuclear structure \textit{ab initio} calculations, i.e. the PNO2B approximation of the three-body operator
\begin{align}
  O &= o^{00} + o^{11} + o^{22} + o^{33} \, .
\end{align}

The normal-ordered form of $O$ with respect to $| \Phi \rangle$ reads in the single-particle basis as
\begin{align}
  O &= 
\phantom{+ \Lambda^{31}+} \Lambda^{00} \nonumber \\*
  &\phantom{=}+ \Lambda^{20}+ \Lambda^{11}+ \Lambda^{02} \nonumber \\*
  &\phantom{=} + \Lambda^{31}+ \Lambda^{22}+ \Lambda^{13} \nonumber \\*
  &\phantom{= + \Lambda^{42} }\,+ \Lambda^{33} \nonumber
\end{align}
and involves the set of normal and anomalous fields
  \begin{subequations}
    \begin{align}
      \Lambda^{33}
      &= o^{33}\\
      \Lambda^{31}
      &= \frac{1}{2}\tr[o^{33}\kappa] \, ,  \\
      \Lambda^{22}
      &= o^{22} + \tr[o^{33}\rho] \, ,  \\
      \Lambda^{20}
      &= \frac{1}{2}\tr[o^{22}\kappa] + \frac{1}{2}\tr[o^{33}\rho\kappa] \, ,  \\
      \Lambda^{11}
      &= o^{11} + \tr[o^{22}\rho] + \frac{1}{2}\tr[o^{33}\rho\rho] \nonumber \\
      &\phantom{=}  + \frac{1}{4}\tr[o^{33}\kappa^\ast\kappa] \, , \\
      \Lambda^{00}
      &= o^{00} + \tr[o^{11}\rho] + \frac{1}{2}\tr[o^{22}\rho\rho] \nonumber \\
      &\phantom{=}  + \frac{1}{4}\tr[o^{22}\kappa^\ast\kappa] + \frac{1}{6}\tr[o^{33}\rho\rho\rho] \nonumber \\
      &\phantom{=}  + \frac{1}{4}\tr[o^{33}\rho\kappa^\ast\kappa] \, .
    \end{align}
  \end{subequations}
The nNO2B approximation leads to just dropping $O^{[3]}=\Lambda^{33}$. The application of the PNO2B approximation is more involved and we now proceed to the construction of the corresponding operator
\begin{align}
  O^{\text{PNO2B}}
  &\equiv \tilde{o}^{00}+\tilde{o}^{11}+\tilde{o}^{22} \, , \label{PNOkBexample1bis}
\end{align}
where the different terms are to be obtained recursively on the basis of Eq.~\eqref{eq:otildedef}. In the present case, it leads to
  \begin{subequations}
    \begin{align}
      \tilde{o}^{22}
      &= \Lambda^{22} \notag \\
      &= o^{22} + \tr[o^{33}\rho] \, ,  \\
      \tilde{o}^{11}
      &= \Lambda^{11} - \tilde{\Lambda}^{11(22)} \notag \\
      &= o^{11} - \frac{1}{2} \tr[o^{33}\rho\rho] + \frac{1}{4} \tr[o^{33}\kappa^\ast\kappa] \, ,  \\
      \tilde{o}^{00}
      &= \Lambda^{00} - \tilde{\Lambda}^{00(11)} - \tilde{\Lambda}^{00(22)} \notag \\
      &= o^{00} + \frac{1}{6} \tr[o^{33}\rho\rho\rho] - \frac{1}{4} \tr[o^{33}\rho\kappa^\ast\kappa] \, ,
    \end{align}
  \end{subequations}
where the needed intermediate quantities, i.e. the diagonal normal-ordered fields originating from the successive contributions to $O^{\text{PNO2B}}$, are nothing but
  \begin{subequations}
    \begin{align}
      \tilde{\Lambda}^{11(22)}
      &= \tr[\tilde{o}^{22}\rho] \notag \\
      &= \tr[o^{22}\rho] + \tr[o^{33}\rho\rho] \, , \\
      \tilde{\Lambda}^{00(22)}
      &= \frac{1}{2}\tr[\tilde{o}^{22}\rho\rho] + \frac{1}{4}\tr[\tilde{o}^{22}\kappa^\ast\kappa] \notag \\
      &= \frac{1}{2}\tr[o^{22}\rho\rho] + \frac{1}{4}\tr[o^{22}\kappa^\ast\kappa] \nonumber \\
      &\phantom{=}   + \frac{1}{2}\tr[o^{33}\rho\rho\rho] + \frac{1}{4}\tr[o^{33}\rho\kappa^\ast\kappa] \, , \\
      \tilde{\Lambda}^{00(11)}
      &= \tr[\tilde{o}^{11}\rho] \notag \\
      &= \tr[o^{11}\rho] - \frac{1}{2}\tr[o^{33}\rho\rho\rho] \nonumber \\
      &\phantom{=}   + \frac{1}{4}\tr[o^{33}\rho\kappa^\ast\kappa]  \, .
    \end{align}
  \end{subequations}
The two above equations fully define the PNO2B approximation of the three-body operator. It is particle-number conserving by construction. Recent \textit{ab initio} BMBPT calculations~\cite{Tichai:2018mll} have been performed on the basis of this approximation although this was not explicited at the time.

It is useful to further characterize the operator by closely inspecting its normal-ordered contributions. Focusing for example on the normal field $\tilde{\Lambda}^{11}$, one obtains
  \begin{align}
    \tilde{\Lambda}^{11}
    &= \tilde{o}^{11} + \tr[\tilde{o}^{22}\rho] \notag \\
    &= o^{11} + \tr[o^{22}\rho] + \frac{1}{2} \tr[o^{33}\rho\rho] + \frac{1}{4} \tr[o^{33}\kappa^\ast\kappa] \notag \\
    &= \Lambda^{11} \, ,
  \end{align}
  which indeed satisfies the systematic property $\tilde{\Lambda}^{ii}=\Lambda^{ii}$. Focusing now on the anomalous field $\tilde{\Lambda}^{20}$, one obtains
  \begin{align}
    \tilde{\Lambda}^{20}
    &= \frac{1}{2}\tr[\tilde{o}^{22}\kappa] \notag \\
    &= \frac{1}{2}\tr[o^{22}\kappa] + \frac{1}{2}\tr[o^{33}\rho\kappa] \notag \\
    &= \Lambda^{20} \, ,
  \end{align}
  where again no extra term arises.
  
Eventually, the complete set of normal-ordered contributions to $O^{\text{PNO2B}}$ expressed in the single-particle basis relates to those of the original three-body operator through
  \begin{subequations}
    \begin{align}
      \tilde{\Lambda}^{33}
      &= 0 \, , \\
      \tilde{\Lambda}^{31}
      &= 0 \, , \\
      \tilde{\Lambda}^{22}
      &= \Lambda^{22} \, , \\
      \tilde{\Lambda}^{20}
      &= \Lambda^{20}\, ,  \\
      \tilde{\Lambda}^{11}
      &= \Lambda^{11} \, , \\
      \tilde{\Lambda}^{00}
      &= \Lambda^{00} \, .
    \end{align}
  \end{subequations}
One thus observes that the non-zero fields are strictly equal to those associated with the original operator, i.e. there is no so-called extra term when approximating a three-body operator.

\begin{figure*}[t!]
  \centering
  \includegraphics[width=\textwidth]{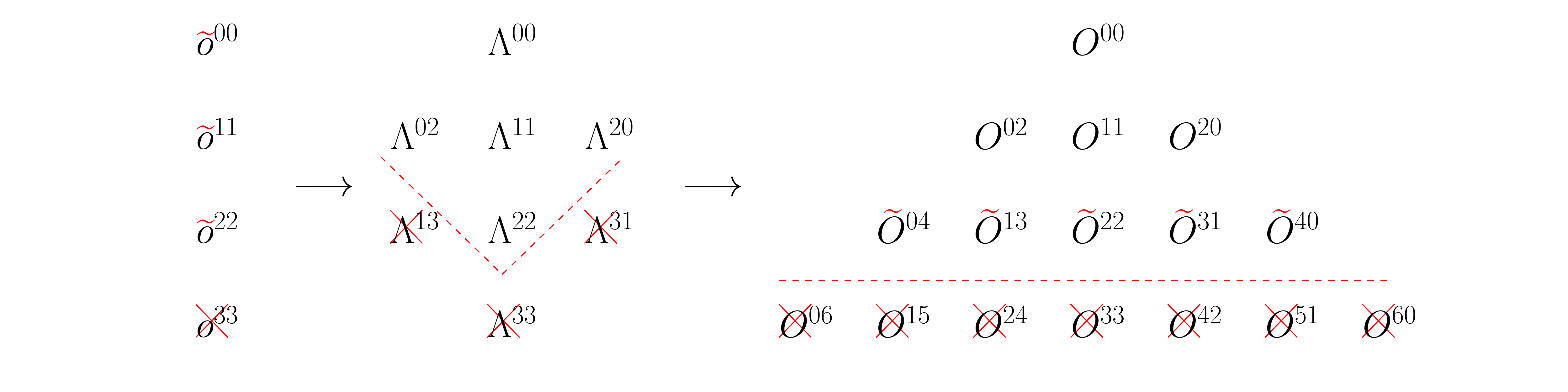}
  \caption{Representation of the \textit{PNO2B approximation} of a three-body operator $O$.
  Left column: Normal-ordered form with respect to the particle vacuum $|0\rangle$ expressed in $\{c,c^\dagger\}$.
  Middle column: Normal-ordered form with respect to the Bogoliubov vacuum $|\Phi\rangle$ expressed in $\{c,c^\dagger\}$.
  Right column: Normal-ordered form with respect to $|\Phi\rangle$ expressed in $\{\beta,\beta^\dagger\}$.
  The red tildes, crosses and dashed lines embody the effect of the PNO2B approximation, i.e. (1) red crosses indicate the suppressed terms, (2) red dashed lines separate suppressed terms from retained ones and (3) red tildes represent the retained terms that are modified.}
  \label{fig:pnokb}
\end{figure*}

A graphical representation of the PNO2B approximation of a three-body operator $O$ is given in Fig.~\ref{fig:pnokb}.
    
\subsubsection{PNO3B approximation of a four-body operator}

Let us now build the PNO3B approximation of the four-body operator
\begin{align}
  O &= o^{00} + o^{11} + o^{22} + o^{33} + o^{44} \, . \label{4body}
\end{align}

The normal ordering of $O$ with respect to $| \Phi \rangle$ reads in the single-particle basis as
\begin{align}
  O &= 
\phantom{+ \Lambda^{40}+ \Lambda^{31}+} \Lambda^{00} \nonumber \\*
  &\phantom{= + \Lambda^{40}}+ \Lambda^{20}+ \Lambda^{11}+ \Lambda^{02} \nonumber \\*
  &\phantom{=} + \Lambda^{40}+ \Lambda^{31}+ \Lambda^{22}+ \Lambda^{13}+ \Lambda^{04} \nonumber \\*
  &\phantom{= + \Lambda^{40}}+ \Lambda^{42}+ \Lambda^{33}+ \Lambda^{24} \nonumber \\*
&\phantom{= + \Lambda^{40}+ \Lambda^{31}} + \Lambda^{44} \, ,
\end{align}
and involves the following normal and anomalous fields
  \begin{subequations}
    \begin{align}
      \Lambda^{44}
      &= o^{44} \, , \\
      \Lambda^{42}
      &= \frac{1}{2}\tr[o^{44}\kappa] \, , \\
      \Lambda^{33}
      &= o^{33} + \tr[o^{44}\rho] \, , \\
      \Lambda^{40}
      &= \frac{1}{8}\tr[o^{44}\kappa\kappa] \, , \\
      \Lambda^{31}
      &= \frac{1}{2}\tr[o^{33}\kappa] + \frac{1}{2}\tr[o^{44}\rho\kappa] \, , \\
      \Lambda^{22}
      &= o^{22} + \tr[o^{33}\rho] + \frac{1}{2}\tr[o^{44}\rho\rho] + \frac{1}{4}\tr[o^{44}\kappa^\ast\kappa] \, , \\
      \Lambda^{20}
      &= \frac{1}{2}\tr[o^{22}\kappa] + \frac{1}{2}\tr[o^{33}\rho\kappa]  + \frac{1}{4}\tr[o^{44}\rho\rho\kappa] \nonumber \\
      &\phantom{=} + \frac{1}{16}\tr[o^{44}\kappa^\ast\kappa\kappa] \, , \\
      \Lambda^{11}
      &= o^{11} + \tr[o^{22}\rho] + \frac{1}{2}\tr[o^{33}\rho\rho] + \frac{1}{4}\tr[o^{33}\kappa^\ast\kappa] \nonumber \\
      &\phantom{=}  + \frac{1}{6}\tr[o^{44}\rho\rho\rho] + \frac{1}{4}\tr[o^{44}\rho\kappa^\ast\kappa] \, , \\
      \Lambda^{00}
      &= o^{00} + \tr[o^{11}\rho] + \frac{1}{2}\tr[o^{22}\rho\rho] + \frac{1}{4}\tr[o^{22}\kappa^\ast\kappa] \nonumber \\
      &\phantom{=} + \frac{1}{6}\tr[o^{33}\rho\rho\rho] + \frac{1}{4}\tr[o^{33}\rho\kappa^\ast\kappa] \notag \\
      &\hphantom{=} + \frac{1}{24}\tr[o^{44}\rho\rho\rho\rho] + \frac{1}{8}\tr[o^{44}\rho\rho\kappa^\ast\kappa] \nonumber \\
      &\phantom{=} + \frac{1}{64}\tr[o^{44}\kappa^\ast\kappa^\ast\kappa\kappa] \, .
    \end{align}
  \end{subequations}

The nNO3B approximation leads to just dropping $O^{[4]}=\Lambda^{44}$. The application of the PNO3B approximation is more involved and we now proceed to the construction of the corresponding operator
\begin{align}
  O^{\text{PNO3B}}
  &\equiv \tilde{o}^{00}+\tilde{o}^{11}+\tilde{o}^{22}+\tilde{o}^{33} \, , \label{PNOkBexample1}
\end{align}
where the different terms are to be obtained recursively on the basis of Eq.~\eqref{eq:otildedef}. In the present case, it leads to
  \begin{subequations}
    \begin{align}
      \tilde{o}^{33}
      &= \Lambda^{33} \notag \\
      &= o^{33} + \tr[o^{44}\rho] \, , \\
      \tilde{o}^{22}
      &= \Lambda^{22} - \tilde{\Lambda}^{22(33)} \notag \\
      &= o^{22} - \frac{1}{2} \tr[o^{44}\rho\rho] + \frac{1}{4} \tr[o^{44}\kappa^\ast\kappa] \, , \\
      \tilde{o}^{11}
      &= \Lambda^{11} - \tilde{\Lambda}^{11(22)} - \tilde{\Lambda}^{11(33)} \notag \\
      &= o^{11} + \frac{1}{6} \tr[o^{44}\rho\rho\rho] - \frac{1}{4} \tr[o^{44}\rho\kappa^\ast\kappa] \, , \\
      \tilde{o}^{00}
      &= \Lambda^{00} - \tilde{\Lambda}^{00(11)} - \tilde{\Lambda}^{00(22)} - \tilde{\Lambda}^{00(33)} \notag \\
      &= o^{00} - \frac{1}{24} \tr[o^{44}\rho\rho\rho\rho] + \frac{1}{8} \tr[o^{44}\rho\rho\kappa^\ast\kappa]  \nonumber \\
      &\phantom{=}  - \frac{3}{64} \tr[o^{44}\kappa^\ast\kappa^\ast\kappa\kappa] \, ,
    \end{align}
  \end{subequations}
where the needed intermediate quantities, i.e. the diagonal normal-ordered fields originating from the successive contributions to $O^{\text{PNO3B}}$, are
  \begin{subequations}
    \begin{align}
      \tilde{\Lambda}^{22(33)}
      &= \tr[\tilde{o}^{33}\rho] \notag \\
      &= \tr[o^{33}\rho] + \tr[o^{44}\rho\rho] \, , \\
      \tilde{\Lambda}^{11(22)}
      &= \tr[\tilde{o}^{22}\rho] \notag \\
      &= \tr[o^{22}\rho] - \frac{1}{2}\tr[o^{44}\rho\rho\rho]  + \frac{1}{4}\tr[o^{44}\rho\kappa^\ast\kappa] \, , \\
      \tilde{\Lambda}^{11(33)}
      &= \frac{1}{2}\tr[\tilde{o}^{33}\rho\rho] + \frac{1}{4}\tr[\tilde{o}^{33}\kappa^\ast\kappa] \notag \\
      &= \frac{1}{2}\tr[o^{33}\rho\rho] + \frac{1}{4}\tr[o^{33}\kappa^\ast\kappa] \nonumber \\
      &\phantom{=}  + \frac{1}{2}\tr[o^{44}\rho\rho\rho] + \frac{1}{4}\tr[o^{44}\rho\kappa^\ast\kappa] \, , \\
      \tilde{\Lambda}^{00(11)}
      &= \tr[\tilde{o}^{11}\rho] \notag \\
      &= \tr[o^{11}\rho] + \frac{1}{6}\tr[o^{44}\rho\rho\rho\rho]  - \frac{1}{4}\tr[o^{44}\rho\rho\kappa^\ast\kappa] \, ,  \\
      \tilde{\Lambda}^{00(22)}
      &= \frac{1}{2}\tr[\tilde{o}^{22}\rho\rho] + \frac{1}{4}\tr[\tilde{o}^{22}\kappa^\ast\kappa] \notag \\
      &= \frac{1}{2}\tr[o^{22}\rho\rho] + \frac{1}{4}\tr[o^{22}\kappa^\ast\kappa]  \nonumber \\
      &\phantom{=} - \frac{1}{4}\tr[o^{44}\rho\rho\rho\rho] + \frac{1}{16}\tr[o^{44}\kappa^\ast\kappa^\ast\kappa\kappa] \, , \\
      \tilde{\Lambda}^{00(33)}
      &= \frac{1}{6}\tr[\tilde{o}^{33}\rho\rho\rho] + \frac{1}{4}\tr[\tilde{o}^{33}\rho\kappa^\ast\kappa] \notag \\
      &= \frac{1}{6}\tr[o^{33}\rho\rho\rho] + \frac{1}{4}\tr[o^{33}\rho\kappa^\ast\kappa]  \nonumber \\
      &\phantom{=} + \frac{1}{6}\tr[o^{44}\rho\rho\rho\rho] + \frac{1}{4}\tr[o^{44}\rho\rho\kappa^\ast\kappa] \, .
    \end{align}
  \end{subequations}
The two above equations fully define the PNO3B approximation of the four-body operator. It is particle-number conserving by construction.  It is useful to further characterize the operator by closely inspecting its normal-ordered contributions. Focusing for example on the normal field $\tilde{\Lambda}^{22}$, one obtains
  \begin{align}
    \tilde{\Lambda}^{22}
    &= \tilde{o}^{22} + \tr[\tilde{o}^{33}\rho] \notag \\
    &= o^{22} + \tr[o^{33}\rho] + \frac{1}{2} \tr[o^{44}\rho\rho] + \frac{1}{4} \tr[o^{44}\kappa^\ast\kappa] \notag \\
    &= \Lambda^{22} \, ,
  \end{align}
  which indeed satisfies the systematic property $\tilde{\Lambda}^{ii}=\Lambda^{ii}$. Focusing now on the anomalous field $\tilde{\Lambda}^{20}$, one obtains
  \begin{align}
    \tilde{\Lambda}^{20}
    &= \frac{1}{2}\tr[\tilde{o}^{22}\kappa] + \frac{1}{2}\tr[\tilde{o}^{33}\rho\kappa] \notag \\
    &= \frac{1}{2}\tr[o^{22}\kappa] + \frac{1}{2}\tr[o^{33}\rho\kappa] + \frac{1}{4}\tr[o^{44}\rho\rho\kappa] + \frac{1}{8}\tr[o^{44}\kappa^\ast\kappa\kappa] \notag \\
    &= \left(\frac{1}{2}\tr[o^{22}\kappa] + \frac{1}{2}\tr[o^{33}\rho\kappa] + \frac{1}{4}\tr[o^{44}\rho\rho\kappa] \right. \nonumber \\
      &\phantom{=} \left. + \frac{1}{16}\tr[o^{44}\kappa^\ast\kappa\kappa]\right) + \frac{1}{16}\tr[o^{44}\kappa^\ast\kappa\kappa] \notag \\
    &= \Lambda^{20} + \frac{1}{16}\tr[o^{44}\kappa^\ast\kappa\kappa] \, , \label{exampleanomalous20}
  \end{align}
  where the extra term is nothing but $\breve{\Lambda}^{20}$. Four-body operators are thus the first for which such an extra term appears. 
  
In order to check the consistency of the derivation, the extra term can also be obtained via the application of Eq.~\eqref{generallambdatilde} to the present case of interest. This gives 
  \begin{align}
    \breve{\Lambda}^{20}
    &= \frac{1}{\left(\frac{2-0}{2}\right)!}\left(\frac{1}{2}\right)^{\frac{2-0}{2}} \sum_{n=2}^{4} \sum_{n_\kappa=0}^{\left\lfloor \frac{n-2}{2} \right\rfloor} \left(1 - \frac{1}{\binom{n_\kappa + \frac{2-0}{2}}{n_\kappa}}\right) \notag \\ &\times \sum_{l_{3} l_{4}} \left[ \Lambda^{22(nn)(n_\kappa)}_{l_1 \ldots l_{4}} - \tilde{\Lambda}^{22(nn)(n_\kappa)}_{l_1 \ldots l_{4}} \right] \kappa_{l_{3}l_{4}} \notag \\
    &= \frac{1}{4} \tr[\Lambda^{22(44)(n_\kappa=1)}\kappa] \notag \\
    &= \frac{1}{16} \tr[o^{44}\kappa^\ast\kappa\kappa] \, ,
  \end{align}
which is indeed consistent with Eq.~\eqref{exampleanomalous20}. Inspecting Eq.~\eqref{exampleanomalous20}, one realizes that the difference between $\tilde{\Lambda}^{20}$ and $\Lambda^{20}$ eventually boils down to using a prefactor $1/8$ instead of $1/16$ in the term $\tr[o^{44}\kappa^\ast\kappa\kappa]$. This originates from the non-equivalent combinatorial prefactor at play when generating  contributions of the type $\tr[o^{44}\kappa^\ast\kappa\kappa]$ with three anomalous contractions to $\tilde{\Lambda}^{20}$ and $\Lambda^{20}$ via the application of Wick's theorem to $O^{\text{PNO3B}}$ and $O$, respectively. The difference between the normal-ordered fields of the PNOkB operator and of the original operator, if any, can always be rephrased in terms of modified prefactors of certain contributions involving strings of anomalous contractions. 

Eventually, the complete set of normal-ordered contributions to $O^{\text{PNO3B}}$ expressed in the single-particle basis relates to those of the original four-body operator through
  \begin{subequations}
    \begin{align}
      \tilde{\Lambda}^{44}
      &= 0 \, , \\
      \tilde{\Lambda}^{42}
      &= 0 \, , \\
      \tilde{\Lambda}^{33}
      &= \Lambda^{33} \, ,  \\
      \tilde{\Lambda}^{40}
      &= 0 \, , \\
      \tilde{\Lambda}^{31}
      &= \Lambda^{31}\, ,  \\
      \tilde{\Lambda}^{22}
      &= \Lambda^{22} \, , \\
      \tilde{\Lambda}^{20}
      &= \Lambda^{20} + \frac{1}{16}\tr[o^{44}\kappa^\ast\kappa\kappa]\, , \\
      \tilde{\Lambda}^{11}
      &= \Lambda^{11} \, , \\
      \tilde{\Lambda}^{00}
      &= \Lambda^{00} \, .
    \end{align}
  \end{subequations}
Among the non-zero fields, one (plus its hermitian conjugate) contains an extra contribution, i.e. a term with a modified prefactor compared to the original operator $O$.

\subsubsection{PNO2B approximation of a four-body operator}

Let us now further approximate a four-body operator by building its PNO2B approximation. Starting from the operator defined in Eq.~\eqref{4body}, the nNO2B approximation now leads to dropping $O^{[3]}+O^{[4]}=\Lambda^{42}+ \Lambda^{33}+ \Lambda^{24} +\Lambda^{44}$. The application of the PNO2B approximation is more involved and we now proceed to the construction of the corresponding operator
\begin{align}
  O^{\text{PNO2B}}
  &\equiv \tilde{o}^{00}+\tilde{o}^{11}+\tilde{o}^{22} \, , \label{PNO2Bexample1}
\end{align}
where the different terms are to be obtained recursively on the basis of Eq.~\eqref{eq:otildedef}. In the present case, it leads to
  \begin{subequations}
    \begin{align}
      \tilde{o}^{22}
      &= \Lambda^{22} \notag \\
      &= o^{22} + \tr[o^{33}\rho] + \frac{1}{2}\tr[o^{44}\rho\rho] \nonumber \\
      &\phantom{=}  + \frac{1}{4}\tr[o^{44}\kappa^\ast\kappa] \, , \\
      \tilde{o}^{11}
      &= \Lambda^{11} - \tilde{\Lambda}^{11(22)} \notag \\
      &= o^{11} - \frac{1}{2}\tr[o^{33}\rho\rho] + \frac{1}{4}\tr[o^{33}\kappa^\ast\kappa]  \nonumber \\
      &\phantom{=} - \frac{1}{3}\tr[o^{44}\rho\rho\rho] \, , \\
      \tilde{o}^{00}
      &= \Lambda^{00} - \tilde{\Lambda}^{00(11)} - \tilde{\Lambda}^{00(22)} \notag \\
      &= o^{00} + \frac{1}{6}\tr[o^{33}\rho\rho\rho] - \frac{1}{4}\tr[o^{33}\rho\kappa^\ast\kappa] \notag \\
      &\hphantom{=} + \frac{1}{8}\tr[o^{44}\rho\rho\rho\rho] - \frac{1}{8}\tr[o^{44}\rho\rho\kappa^\ast\kappa] \nonumber \\
      &\phantom{=}  - \frac{3}{64}\tr[o^{44}\kappa^\ast\kappa^\ast\kappa\kappa] \, ,
    \end{align}
  \end{subequations}
where the needed intermediate quantities, i.e. the diagonal normal-ordered fields originating from the successive contributions to $O^{\text{PNO2B}}$, are
  \begin{subequations}
    \begin{align}
      \tilde{\Lambda}^{11(22)}
      &= \tr[\tilde{o}^{22}\rho] \notag \\
      &= \tr[o^{22}\rho] + \tr[o^{33}\rho\rho] + \frac{1}{2}\tr[o^{44}\rho\rho\rho] \nonumber \\
      &\phantom{=}  + \frac{1}{4}\tr[o^{44}\rho\kappa^\ast\kappa] \, ,  \\
      \tilde{\Lambda}^{00(11)}
      &= \tr[\tilde{o}^{11}\rho] \notag \\
      &= \tr[o^{11}\rho] - \frac{1}{2}\tr[o^{33}\rho\rho\rho] + \frac{1}{4}\tr[o^{33}\rho\kappa^\ast\kappa] \nonumber \\
      &\phantom{=}  - \frac{1}{3}\tr[o^{44}\rho\rho\rho\rho] \, , \\
      \tilde{\Lambda}^{00(22)}
      &= \frac{1}{2}\tr[\tilde{o}^{22}\rho\rho] + \frac{1}{4}\tr[\tilde{o}^{22}\kappa^\ast\kappa] \notag \\
      &= \frac{1}{2}\tr[o^{22}\rho\rho] + \frac{1}{4}\tr[o^{22}\kappa^\ast\kappa] + \frac{1}{2}\tr[o^{33}\rho\rho\rho] \nonumber \\
      &\phantom{=}  + \frac{1}{4}\tr[o^{33}\rho\kappa^\ast\kappa]  + \frac{1}{4}\tr[o^{44}\rho\rho\rho\rho] \nonumber \\
      &\phantom{=}  + \frac{1}{4}\tr[o^{44}\rho\rho\kappa^\ast\kappa] + \frac{1}{16}\tr[o^{44}\kappa^\ast\kappa^\ast\kappa\kappa] \, .
    \end{align}
  \end{subequations}
The two above equations fully define the PNO2B approximation of the four-body operator. It is particle-number conserving by construction. It is useful to further characterize the operator by closely inspecting its normal-ordered contributions. Focusing for example on the normal field $\tilde{\Lambda}^{11}$, one obtains
  \begin{align}
    \tilde{\Lambda}^{11}
    &= \tilde{o}^{11} + \tr[\tilde{o}^{22}\rho] \notag \\
    &= o^{11} + \tr[o^{22}\rho] + \frac{1}{2}\tr[o^{33}\rho\rho] \nonumber \\
      &\phantom{=}  + \frac{1}{4}\tr[o^{33}\kappa^\ast\kappa] + \frac{1}{6}\tr[o^{44}\rho\rho\rho] \nonumber \\
      &\phantom{=}  + \frac{1}{4}\tr[o^{44}\rho\kappa^\ast\kappa] \notag \\
    &= \Lambda^{11} \, ,
  \end{align}
  which indeed satisfies the systematic property $\tilde{\Lambda}^{ii}=\Lambda^{ii}$. Focusing now on the anomalous field $\tilde{\Lambda}^{20}$, one obtains
  \begin{align}
    \tilde{\Lambda}^{20}
    &= \frac{1}{2}\tr[\tilde{o}^{22}\kappa] \notag \\
    &= \frac{1}{2}\tr[o^{22}\kappa] + \frac{1}{2}\tr[o^{33}\rho\kappa] + \frac{1}{4}\tr[o^{44}\rho\rho\kappa] + \frac{1}{8}\tr[o^{44}\kappa^\ast\kappa\kappa] \notag \\
    &= \left(\frac{1}{2}\tr[o^{22}\kappa] + \frac{1}{2}\tr[o^{33}\rho\kappa] + \frac{1}{4}\tr[o^{44}\rho\rho\kappa] \right. \nonumber \\
      &\phantom{=} \left. + \frac{1}{16}\tr[o^{44}\kappa^\ast\kappa\kappa]\right) + \frac{1}{16}\tr[o^{44}\kappa^\ast\kappa\kappa] \notag \\
    &= \Lambda^{20} + \frac{1}{16}\tr[o^{44}\kappa^\ast\kappa\kappa] \, ,
  \end{align}
  where the extra term is nothing but $\breve{\Lambda}^{20}$. One notices that it is the same as in the PNO3B approximation of the four-body operator.

Eventually, the complete set of normal-ordered contributions to $O^{\text{PNO2B}}$ expressed in the single-particle basis relates to those of the original four-body operator through
  \begin{subequations}
    \begin{align}
      \tilde{\Lambda}^{44}
      &= 0 \, , \\
      \tilde{\Lambda}^{42}
      &= 0 \, , \\
      \tilde{\Lambda}^{40}
      &= 0 \, , \\
      \tilde{\Lambda}^{33}
      &= 0 \, , \\
      \tilde{\Lambda}^{31}
      &= 0 \, , \\
      \tilde{\Lambda}^{22}
      &= \Lambda^{22} \, , \\
      \tilde{\Lambda}^{20}
      &= \Lambda^{20} + \frac{1}{16}\tr[o^{44}\kappa^\ast\kappa\kappa] \, , \\
      \tilde{\Lambda}^{11}
      &= \Lambda^{11} \, , \\
      \tilde{\Lambda}^{00}
      &= \Lambda^{00} \, .
    \end{align}
  \end{subequations}
Among the non-zero fields, one (plus its hermitian conjugate) contains an extra contribution, i.e. a term with a modified prefactor compared to the original operator $O$.

\subsubsection{PNO3B and PNO4B approximations of a five-body operator}

We have also worked out the PNO3B and the PNO4B approximations to a five-body operator in full details. As the expressions become lengthy, they are not reported here. Still, it is interesting to note that the particular form of the extra terms depends on which PNOkB, e.g. PNO3B or PNO4B, approximation is performed starting from the same original five-body operator. This point can be illustrated by only reporting how the normal-ordered contributions to both  $O^{\text{PNO3B}}$ and  $O^{\text{PNO4B}}$ relate to those of the original five-body operator. One has
  \begin{subequations}
    \begin{align}
      \tilde{\Lambda}^{55}
      &= 0 \, , \\
      \tilde{\Lambda}^{53}
      &= 0 \, , \\
      \tilde{\Lambda}^{51}
      &= 0 \, , \\
      \tilde{\Lambda}^{44}
      &= 0 \, , \\
      \tilde{\Lambda}^{42}
      &= 0 \, , \\
      \tilde{\Lambda}^{40}
      &= 0 \, , \\
      \tilde{\Lambda}^{33}
      &= \Lambda^{33} \, ,  \\
      \tilde{\Lambda}^{31}
      &= \Lambda^{31} + \frac{1}{16}\tr[o^{55}\kappa^\ast\kappa\kappa] \, , \\
      \tilde{\Lambda}^{22}
      &= \Lambda^{22} \, , \\
      \tilde{\Lambda}^{20}
      &= \Lambda^{20} + \frac{1}{16}\tr[o^{44}\kappa^\ast\kappa\kappa]  + \frac{1}{16} \tr[o^{55}\rho\kappa^\ast\kappa\kappa] \, , \\
      \tilde{\Lambda}^{11}
      &= \Lambda^{11} \, , \\
      \tilde{\Lambda}^{00}
      &= \Lambda^{00} \, ,
    \end{align}
  \end{subequations}
 for the  PNO3B approximation and
  \begin{subequations}
    \begin{align}
      \tilde{\Lambda}^{55}
      &= 0 \, , \\
      \tilde{\Lambda}^{53}
      &= 0 \, , \\
      \tilde{\Lambda}^{51}
      &= 0 \, , \\
      \tilde{\Lambda}^{44}
      &= \Lambda^{44} \, , \\
      \tilde{\Lambda}^{42}
      &= \Lambda^{42} \, , \\
      \tilde{\Lambda}^{40}
      &= \Lambda^{40} \, , \\
      \tilde{\Lambda}^{33}
      &= \Lambda^{33} \, , \\
      \tilde{\Lambda}^{31}
      &= \Lambda^{31} + \frac{1}{16}\tr[o^{55}\kappa^\ast\kappa\kappa] \, , \\
      \tilde{\Lambda}^{22}
      &= \Lambda^{22} \, , \\
      \tilde{\Lambda}^{20}
      &= \Lambda^{20} \, , \\
      \tilde{\Lambda}^{11}
      &= \Lambda^{11} \, , \\
      \tilde{\Lambda}^{00}
      &= \Lambda^{00} \, ,
    \end{align}
  \end{subequations}
 for the PNO4B approximation. While in both cases the extra term entering $\tilde{\Lambda}^{31}$ is the same, one observes that $\tilde{\Lambda}^{20}$ does acquire an extra term in the PNO3B approximation but it does not in the PNO4B approximation.

\begin{figure*}[t!]
  \centering
  \includegraphics[width=\textwidth]{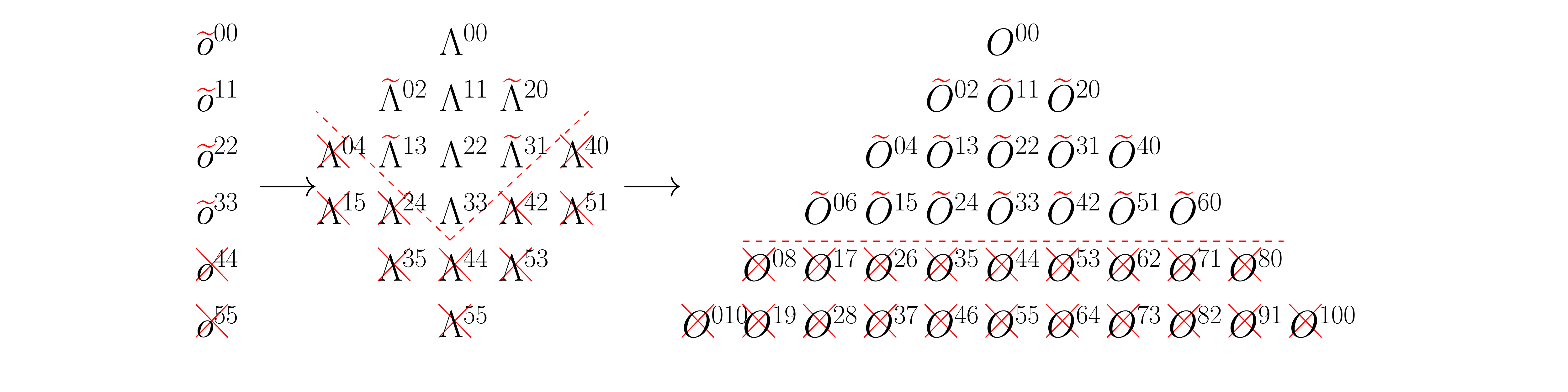}
  \caption{Representation of the \textit{PNO3B approximation} of a five-body operator $O$.
  Left column: Normal-ordered form with respect to the particle vacuum $|0\rangle$ expressed in $\{c,c^\dagger\}$.
  Middle column: Normal-ordered form with respect to the Bogoliubov vacuum $|\Phi\rangle$ expressed in $\{c,c^\dagger\}$.
  Right column: Normal-ordered form with respect to $|\Phi\rangle$ expressed in $\{\beta,\beta^\dagger\}$.
  The red tildes, crosses and dashed lines embody the effect of the PNO3B approximation, i.e. (1) red crosses indicate the suppressed terms, (2) red dashed lines separate suppressed terms from retained ones and (3) red tildes represent the retained terms that are modified.}
  \label{fig:pnokb_no3b_5body}
\end{figure*}

For illustration, a graphical representation of the  PNO3B approximation of the five-body operator $O$ is given in Fig.~\ref{fig:pnokb_no3b_5body}.
   
\subsection{Alternative approximation}
\label{alternativePNOkB}

The procedure to define the PNOkB approximation is not unique. Given the three objectives stated in Sec.~\ref{PNOkBprocedure},  two independent approaches cannot differ drastically in their philosophy and cannot produce very different approximate operators. In the present section, an alternative approximation procedure based on the quasi-normal ordering of Ref.~\cite{kong10a} is briefly investigated to highlight the similarities and differences.
  
\subsubsection{Quasi normal-ordering}

The quasi normal-ordering introduced in Ref.~\cite{kong10a} is stipulated through the set of equations 
\begin{subequations}
  \begin{align}
    \spc{l_1}\spa{l_2}
    &\equiv \kongno{\spc{l_1}\spa{l_2}} + \gamma_{l_2l_1} \, , \\
    \spc{l_1}\spc{l_2}\spa{l_4}\spa{l_3}
    &\equiv \kongno{\spc{l_1}\spc{l_2}\spa{l_4}\spa{l_3}} \\*
    &\phantom{\equiv} + \antisym{\kongno{\spc{l_1}\spa{l_3}}\gamma_{l_4l_2}}  + \antisym{\gamma_{l_3l_1}\gamma_{l_4l_2}} \nonumber  \\*
    &\hspace{5pt}\vdots  \, , \nonumber
  \end{align}
\end{subequations}
where \kongno{\ldots} denotes the quasi normal-ordering defined by the given of the elementary contraction $\gamma$. The antisymmetrization operator \antisym{\ldots} is the one defined in Ref.~\cite{shamasundar09a}.  At this point, the contraction is defined as any two-index tensor such that the above procedure is very general. It happens that such a definition is sufficient to have a generalized Wick's theorem governing the way the product of two quasi normal-ordered operators can be systematically decomposed into a sum of quasi normal-ordered operators weighted by a set of contractions~\cite{kong10a}.

In the present context, the quasi-normal ordering is only used as a systematic way to reshuffle as a large part as possible of an initial string of single-particle operators into operators of lower ranks in view of neglecting the quasi normal-ordered term having the same rank as the original string. Compared to the normal ordering with respect to the Bogoliubov vacuum associated with standard Wick's theorem, it is presently possible to only invoke a symmetry-conserving elementary contraction such that each individual term in the quasi normal-ordered form is a scalar under $U(1)$ transformations. For the rest, the expectation value of a quasi-normal-ordered operator defined in such a general way does not vanish \textit{a priori}, i.e.
\begin{subequations}
  \begin{align}
    \bogoexpect{\kongno{\spc{l_1}\spa{l_2}}}
    &\neq 0 \, , \\
    \bogoexpect{\kongno{\spc{l_1}\spc{l_2}\spa{l_4}\spa{l_3}}}
    &\neq 0  \, , \\
  &\vdots \, .  \nonumber
  \end{align}
\end{subequations}
As a result, the quasi normal ordering does not present the practicalities of more traditional Wick's theorems. 

Thus, the approach proceeds in two steps. First,  the original operator expressed in quasi normal-ordered form is truncated to produce the particle-number conserving quasi normal-ordered k-body (PQNOkB) approximation. Second, the resulting operator is brought into a normal-ordered form with respect to the Bogoliubov vacuum according to standard Wick's theorem in view of using it in the many-body formalism of interest.

\subsection{PQNO1B of a two-body operator}

Applying quasi normal-ordering to a particle-number-conserving two-body operator $O$ leads to
\begin{align}
  O
  &\equiv \sum_{l_1l_2} o^{11}_{l_1l_2} \spc{l_1}\spa{l_2} + \frac{1}{4}\sum_{l_1l_2l_3l_4} o^{22}_{l_1l_2l_3l_4} \spc{l_1}\spc{l_2}\spa{l_4}\spa{l_3} \notag \, , \\ 
  &= \Lambda^{00(\text{qno})} + \sum_{l_1l_2} \Lambda^{11(\text{qno})}_{l_1l_2} \kongno{\spc{l_1}\spa{l_2}} \nonumber \\
    &\phantom{=}  + \frac{1}{4}\sum_{l_1l_2l_3l_4} \Lambda^{22(\text{qno})}_{l_1l_2l_3l_4} \kongno{\spc{l_1}\spc{l_2}\spa{l_4}\spa{l_3}} \, ,
\end{align}
where the n-body fields $\Lambda^{nn(\text{qno})}$ are given by
\begin{subequations}
  \begin{align}
    \Lambda^{00(\text{qno})}
    &\equiv \sum_{l_1l_2} o^{11}_{l_1l_2} \gamma_{l_2l_1}  + \frac{1}{2} \sum_{l_1l_2l_3l_4} o^{22}_{l_1l_2l_3l_4} \gamma_{l_3l_1} \gamma_{l_4l_2} \, , \nonumber \\
    \Lambda^{11(\text{qno})}_{l_1l_2}
    &\equiv o^{11}_{l_1l_2} + \sum_{l_1l_2l_3l_4} o^{22}_{l_1l_3l_2l_4} \gamma_{l_4l_3} \, , \\
    \Lambda^{22(\text{qno})}_{l_1l_2l_3l_4}
    &\equiv o^{22}_{l_1l_2l_3l_4} \, .
  \end{align}
\end{subequations}

The PQNO1B approximation of $O$ is obtained by neglecting the effective two-body part $\Lambda^{22(\text{qno})}$
\begin{align}
  O^{\qnooneb}
  &\equiv \Lambda^{00(\text{qno})} + \sum_{l_1l_2} \Lambda^{11(\text{qno})}_{l_1l_2} \kongno{\spc{l_1}\spa{l_2}} \nonumber \\
  &\equiv o^{00(\text{qno})} + \sum_{l_1l_2} o^{11(\text{qno})}_{l_1l_2} \spc{l_1}\spa{l_2} \, ,
\end{align}
where
\begin{subequations}
  \begin{align}
    o^{00(\text{qno})}
    &= \Lambda^{00(\text{qno})} - \sum_{l_1l_2} \Lambda^{11(\text{qno})}_{l_1l_2} \gamma_{l_2l_1}  \\
    &= -\frac{1}{2} \sum_{l_1l_2l_3l_4} o^{22}_{l_1l_2l_3l_4} \gamma_{l_3l_1} \gamma_{l_4l_2} \, , \nonumber \\
    o^{11(\text{qno})}_{l_1l_2}
    &= \Lambda^{11(\text{qno})}_{l_1l_2} \\
    &= o^{11}_{l_1l_2} + \sum_{l_1l_2l_3l_4} o^{22}_{l_1l_3l_2l_4} \gamma_{l_4l_3} \, . \nonumber
  \end{align}
\end{subequations}

The above quantities are independent of the specific choice made for the elementary contraction $\gamma$. In the present case, $\gamma$ is taken as the normal density $\rho$ of the Bogoliubov vacuum. With this definition at hand, the PQNO1B operator is brought into a normal-ordered form with respect of the Bogoliubov state on the basis of standard Wick's theorem
\begin{align}
  O^{\qnooneb}
  &= \Lambda^{00} + \sum_{l_1l_2} \Lambda^{11}_{l_1l_2} \no{\spc{l_1}\spa{l_2}} \, ,
\end{align}
where
\begin{subequations}
  \label{eq:qno1bfields}
  \begin{align}
    \Lambda^{00}
    &\equiv o^{00(\text{qno})} + \sum_{l_1l_2} o^{11(\text{qno})}_{l_1l_2} \rho_{l_2l_1} \\
    &= \sum_{l_1l_2} o^{11}_{l_1l_2} \rho_{l_2l_1} + \frac{1}{2}\sum_{l_1l_2l_3l_4} o^{22}_{l_1l_2l_3l_4} \rho_{l_3l_1}\rho_{l_4l_2} \, , \nonumber\\
    \Lambda^{11}_{l_1l_2}
    &\equiv o^{11(\text{qno})}_{l_1l_2} \\
    &= o^{11}_{l_1l_2} + \sum_{l_1l_2l_3l_4} o^{22}_{l_1l_3l_2l_4} \rho_{l_4l_3} \, .\nonumber
  \end{align}
\end{subequations}
Comparing Eq.~\eqref{eq:qno1bfields} to Eq.~\eqref{lambdaijtwobody}, one observes that the contribution to $\Lambda^{00}$ originating from the two-body operator via two anomalous contractions is not present in the PQNO1B operator. The latter thus corresponds to a more drastic appoximation of the original operator $O$ than the PNO1B operator. Performing the PQNO2B approximation of a three-body operator, the PQNO2B operator similarly misses a three-body contribution to $\Lambda^{11}$ associated with two anomalous contractions that is present in the PNO2B operator.

\section{Double particle-number projection}
\label{appdoubleproj}

\subsection{Rotated Bogoliubov vacuum}

The rotated Bogoliubov state
\begin{align}
  \label{rotationvacuum}
  \langle \Phi (\varphi) \vert
  &\equiv \langle \Phi \vert R(\varphi)
  \equiv \mathcal{C} \displaystyle \langle 0 \vert \prod_{k} \bar{\beta}^\dagger_{k} \, ,
\end{align}
is a vacuum for the set of rotated quasiparticle operators defined through
\begin{align}
  \left(
  \begin{array} {c}
    \bar{\beta} \\
    \bar{\beta}^{\dagger}
  \end{array}
  \right) (\varphi)
  &\equiv 
  R^{-1}(\varphi)
  \left(
  \begin{array} {c}
    \beta \\
    \beta^{\dagger}
  \end{array}
  \right) 
  R(\varphi) \\
  &\equiv W^{\varphi \, \dagger} \left(
  \begin{array} {c}
    c \\
    c^{\dagger}
  \end{array}
  \right)  \, , \nonumber
\end{align}
where the associated Bogoliubov transformation reads as
\begin{align}
  W^{\varphi \, \dagger} &\equiv 
  \left(
  \begin{array} {cc}
    U^{\varphi \, \dagger} & V^{\varphi \, \dagger} \\
    V^{\varphi \, T} &  U^{\varphi \, T}
  \end{array}
  \right) \\
  &= 
  \left(
  \begin{array} {cc}
    U^{\dagger} e^{+i\varphi} & V^{\dagger} e^{-i\varphi} \\
    V^{T}  e^{+i\varphi} &  U^{T} e^{-i\varphi}
  \end{array}
  \right)\, . \nonumber
\end{align}
The rotated vacuum can also be expressed via a non-unitary Thouless transformation of $\vert \Phi \rangle$ according to
\begin{align}
  \label{thoulessbetweenbothvacua}
  \langle \Phi(\varphi) \vert
  &\equiv \langle \Phi(\varphi) \vert \Phi \rangle \langle \Phi \vert e^{Z(\varphi)} \, .
\end{align}
The Thouless operator appearing in Eq.~\eqref{thoulessbetweenbothvacua}
\begin{equation}
  \label{thoulessoprot}
  Z(\varphi) \equiv \frac{1}{2} \sum_{k_1k_2} Z^{02}_{k_1k_2} (\varphi) \beta_{k_2} \beta_{k_1}  \, ,
\end{equation}
involves the Thouless matrix
\begin{equation}
  \label{thoulessoprot}
  Z^{02} (\varphi) \equiv  B (\varphi) A^{-1} (\varphi)  \, ,
\end{equation}
which is expressed in terms of the Bogoliubov transformation connecting the quasiparticle operators annihilating $|  \Phi(\varphi) \rangle$ to those annihilating $|  \Phi \rangle$
\begin{align}
\left(
\begin{array} {c}
\bar{\beta} \\
\bar{\beta}^{\dagger}
\end{array}
\right) (\varphi)
&\equiv 
{\cal W}^{\dagger}(\varphi) \left(
\begin{array} {c}
\beta \\
\beta^{\dagger}
\end{array}
\right) \, ,
\end{align}
where
\begin{align}
{\cal W}^{\dagger}(\varphi) &\equiv 
\left(
\begin{array} {cc}
A^{\dagger}(\varphi) & B^{\dagger}(\varphi) \\
B^{T}(\varphi) &  A^{T}(\varphi)
\end{array}
\right)\label{transfobogotransition} \\
&=
\left(
\begin{array} {cc}
U^{\varphi \, \dagger}U+V^{\varphi \, \dagger}V & V^{\varphi \, \dagger}U^{\ast} + U^{\varphi \, \dagger}V^{\ast} \\
V^{\varphi \, T}U + U^{\varphi \, T}V &  U^{\varphi \, T}U^{\ast}+V^{\varphi \, T}V^{\ast}
\end{array}
\right)
\, . \nonumber
\end{align}

\subsection{Gauge-rotated contractions}
\label{Approtcontract}

Matrix elements of the doubly gauge-rotated contractions are defined through
\begin{subequations}
  \label{eq:doublygaugerotatedcontractions}
  \begin{align}
    \rho_{l_1l_2}(\varphi,\varphi^\pr)
    &\equiv \frac{\langle\Phi(\varphi)|c^\dagger_{l_2} c_{l_1}|\Phi(\varphi^\pr)\rangle}{\langle\Phi(\varphi)|\Phi(\varphi^\pr)\rangle} \, , \\*
    \kappa_{l_1l_2}(\varphi,\varphi^\pr)
    &\equiv \frac{\langle\Phi(\varphi)|c_{l_2} c_{l_1}|\Phi(\varphi^\pr)\rangle}{\langle\Phi(\varphi)|\Phi(\varphi^\pr)\rangle} \, , \\*
    \bar{\kappa}^\ast_{l_1l_2}(\varphi,\varphi^\pr)
    &\equiv \frac{\langle\Phi(\varphi)|c^\dagger_{l_1} c^\dagger_{l_2}|\Phi(\varphi^\pr)\rangle}{\langle\Phi(\varphi)|\Phi(\varphi^\pr)\rangle} \, ,
  \end{align}
\end{subequations}
such that singly-rotated ones are nothing but
\begin{subequations}
  \label{eq:singlygaugerotatedcontractions}
  \begin{align}
    \rho(\varphi)
    &\equiv  \rho(\varphi,0) \, , \\*
    \kappa(\varphi)
    &\equiv \kappa(\varphi,0) \, , \\*
    \bar{\kappa}^\ast(\varphi)
    &\equiv \bar{\kappa}^\ast(\varphi,0) \, .
  \end{align}
\end{subequations}
Doubly gauge-rotated contractions can be expressed in terms of the Bogoliubov transformation defining $| \Phi \rangle$ and of the gauge angles via~\cite{RiSc80}
\begin{subequations}
  \label{eq:doublygaugerotatedcontractions2}
  \begin{align}
    \rho(\varphi,\varphi^\pr)
    &= \phantom{e^{+2i\varphi^\pr}} \left(\rho + V^\ast Z^{02 \ast}(\varphi-\varphi^\pr) U^\dagger \right) \, , \\*
    \kappa(\varphi,\varphi^\pr)
    &= e^{-2i\varphi^\pr} \left(\kappa- V^\ast Z^{02 \ast}(\varphi-\varphi^\pr) V^\dagger \right) \, , \\*
    \bar{\kappa}^\ast(\varphi,\varphi^\pr)
    &= e^{+2i\varphi^\pr} \left(\kappa^\ast- U^\ast Z^{02 \ast}(\varphi-\varphi^\pr) U^\dagger \right) \, .
  \end{align}
\end{subequations}
Setting $\varphi^\pr = 0$ in Eq.~\eqref{eq:doublygaugerotatedcontractions2}, singly and doubly gauge-rotated contractions appear to be trivially related through
\begin{subequations}
  \label{eq:stodgaugerotatedcontractions}
  \begin{alignat}{3}
    \rho(\varphi,\varphi^\pr)
    &= &&\rho(\varphi-\varphi^\pr) \, , \\*
    \kappa(\varphi,\varphi^\pr)
    &= e^{-2i\varphi^\pr} &&\kappa(\varphi-\varphi^\pr) \, , \\*
    \bar{\kappa}^\ast(\varphi,\varphi^\pr)
    &= e^{+2i\varphi^\pr} &&\bar{\kappa}^\ast(\varphi-\varphi^\pr) \, .
  \end{alignat}
\end{subequations}

\subsection{Projection}
\label{Appprojsingdoub}

With Eq.~\eqref{eq:stodgaugerotatedcontractions} at hand, the singly- (left-) projected mean-field matrix element on the value A of the one-body operator $F$ reads as
\begin{align}
  \label{eq:singleprojectionratio}
  \langle\Phi|P^{\text{A}}F|\Phi\rangle
  &= \frac{1}{2\pi} \int_0^{2\pi} \!d\varphi e^{-i\varphi \text{A}} f^{(0)}(\varphi) \mathcal{N}^{(0)}(\varphi) \notag \\
  &= \sum_{l_1l_2} f^{11}_{l_1l_2} \frac{1}{2\pi} \int_{0}^{2\pi} \!d\varphi e^{-i\varphi \text{A}} \rho_{l_2l_1}(\varphi) \mathcal{N}^{(0)}(\varphi) \notag \\
  &\hphantom{=} + \frac{1}{2} \sum_{l_1l_2} f^{20}_{l_1l_2} \frac{1}{2\pi} \int_{0}^{2\pi} \!d\varphi e^{-i\varphi \text{A}} \bar{\kappa}^\ast_{l_1l_2}(\varphi) \mathcal{N}^{(0)}(\varphi) \notag \\
  &\hphantom{=} + \frac{1}{2} \sum_{l_1l_2} f^{02}_{l_1l_2} \frac{1}{2\pi} \int_{0}^{2\pi} \!d\varphi e^{-i\varphi \text{A}} \kappa_{l_1l_2}(\varphi) \mathcal{N}^{(0)}(\varphi) \notag \\
  &= \langle\Phi|P^{\text{A}}f^{11}|\Phi\rangle + \langle\Phi|P^{\text{A}}f^{20}|\Phi\rangle + \langle\Phi|P^{\text{A}}f^{02}|\Phi\rangle \, ,
\end{align}
while the doubly- (left- and right-) projected, on possibly two different values A and A', is
\begin{strip}
\begin{align}
  \label{eq:doubleprojectionratio}
  \langle\Phi|P^{\text{A}}FP^{\text{A}^\pr}|\Phi\rangle
  &= \frac{1}{2\pi} \int_0^{2\pi} \!d\varphi^\pr e^{+i\varphi^\pr \text{A}^\pr} \frac{1}{2\pi} \int_0^{2\pi} \!d\varphi e^{-i\varphi \text{A}} f^{(0)}(\varphi,\varphi^\pr) \mathcal{N}^{(0)}(\varphi - \varphi^\pr) \notag \\
  &= \frac{1}{2\pi} \int_0^{2\pi} \!d\varphi^\pr e^{+i\varphi^\pr (\text{A}^\pr-\text{A})} \frac{1}{2\pi} \int_0^{2\pi} \!d\varphi e^{-i(\varphi-\varphi^\pr) \text{A}} f^{(0)}(\varphi,\varphi^\pr) \mathcal{N}^{(0)}(\varphi - \varphi^\pr) \notag \\
  &= \frac{1}{2\pi} \int_0^{2\pi} \!d\varphi^\pr e^{+i\varphi^\pr (\text{A}^\pr-\text{A})} \frac{1}{2\pi} \int_{-\varphi^\pr}^{2\pi-\varphi^\pr} \!d\phi e^{-i\phi \text{A}} \mathcal{N}^{(0)}(\phi) \notag \\
  &\hphantom{=} \times \sum_{l_1l_2} f^{11}_{l_1l_2} \rho_{l_2l_1}(\phi) + \frac{1}{2} f^{20}_{l_1l_2} e^{+2i\varphi^\pr} \bar{\kappa}^\ast_{l_1l_2}(\phi) + \frac{1}{2} f^{02}_{l_1l_2} e^{-2i\varphi^\pr} \kappa_{l_1l_2}(\phi) \notag \\
  &= \sum_{l_1l_2} f^{11}_{l_1l_2} \frac{1}{2\pi} \int_{0}^{2\pi} \!d\phi e^{-i\phi \text{A}} \rho_{l_2l_1}(\phi) \mathcal{N}^{(0)}(\phi) \delta_{\text{A},\text{A}^\pr} \notag \\
  &\hphantom{=} + \frac{1}{2} \sum_{l_1l_2} f^{20}_{l_1l_2} \frac{1}{2\pi} \int_{0}^{2\pi} \!d\phi e^{-i\phi \text{A}} \bar{\kappa}^\ast_{l_1l_2}(\phi) \mathcal{N}^{(0)}(\phi) \delta_{\text{A},\text{A}^\pr+2} \notag \\
  &\hphantom{=} + \frac{1}{2} \sum_{l_1l_2} f^{02}_{l_1l_2} \frac{1}{2\pi} \int_{0}^{2\pi} \!d\phi e^{-i\phi \text{A}} \kappa_{l_1l_2}(\phi) \mathcal{N}^{(0)}(\phi) \delta_{\text{A},\text{A}^\pr-2} \notag \\
  &= \langle\Phi|P^{\text{A}}f^{11}|\Phi\rangle \delta_{\text{A},\text{A}^\pr} + \langle\Phi|P^{\text{A}}f^{20}|\Phi\rangle \delta_{\text{A},\text{A}^\pr+2} + \langle\Phi|P^{\text{A}}f^{02}|\Phi\rangle \delta_{\text{A},\text{A}^\pr-2} \, ,
\end{align}
\end{strip}
where $\mathcal{N}^{(0)}(\phi)$, $\rho(\phi)$, $\bar{\kappa}^\ast(\phi)$ and $\kappa(\phi)$ are $2\pi$-periodic functions.

\subsection{Particle-number variance}
\label{approxVARPART}

In Fig.~\ref{test2}, a small (non-zero) value was obtained from the nNO1B approximation of the particle-number variance operation the basis of the double particle-number projection. While a zero variance must be obtained for the exact particle-number variance operator, the fact that its nNO1B approximation systematically delivers a (almost) zero value is not immediately obvious. This feature, for which the double projection is essential, is now briefly analyzed.   

\subsubsection{Doubly-projected naive NO1B expectation value}

The doubly-projected expectation value of the nNO1B approximation of an operator $O$ can be systematically written as
\begin{align}
  \frac{\langle\Phi|P^{\text{A}}O^{\text{nNO1B}}P^{\text{A}}|\Phi\rangle}{\langle\Phi|P^{\text{A}}|\Phi\rangle}
  &= \tilde{o}^{00} + \frac{\langle\Phi|P^{\text{A}}\tilde{o}^{11}|\Phi\rangle}{\langle\Phi|P^{\text{A}}|\Phi\rangle} \label{doubleprojNO1BO} \\
  &= \tilde{o}^{00} + \tr{\tilde{o}^{11}\rho^{\text{proj}}} \notag \\
  &= \tilde{o}^{00} + \tr{\tilde{o}^{11}\rho} + \tr{\tilde{o}^{11} \Delta\rho^{\text{proj}}} \, , \notag
\end{align}
where
\begin{align}
  \rho^{\text{proj}}_{l_1l_2}
  &\equiv \frac{\langle\Phi|P^{\text{A}}c^\dagger_{l_2}c_{l_1}|\Phi\rangle}{\langle\Phi|P^{\text{A}}|\Phi\rangle} \, , \\
  \Delta\rho^{\text{proj}}_{l_1l_2}
  &\equiv \rho^{\text{proj}}_{l_1l_2} - \rho_{l_1l_2} \, ,
\end{align}
and where the explicit form of $\tilde{o}^{00}$ and $\tilde{o}^{11}$ are given in Eq.~\eqref{no1bpartbasisME}.

\subsubsection{Double projection and nNO1B approximation}

If $A^2$ is approximated at the nNO1B level, Eq.~\eqref{doubleprojNO1BO} specifies to
\begin{align}
  \frac{\langle\Phi|P^{\text{A}}(A^2)^{\text{nNO1B}}P^{\text{A}}|\Phi\rangle}{\langle\Phi|P^{\text{A}}|\Phi\rangle} 
  &=\tr{\tilde{(a^2)}^{11} \Delta\rho^{\text{proj}}} \notag \\
  &\phantom{=} +\tr{(a^2)^{11}\rho} + \frac{1}{2}\tr{(a^2)^{22}\rho\rho} \notag \\
  &\phantom{=} - \frac{1}{4}\tr{(a^2)^{22}\kappa^\ast\kappa} \notag \\
  &= \tr{\tilde{(a^2)}^{11} \Delta\rho^{\text{proj}}}  + (\tr{\rho})^2 \, ,
\end{align}
where the expressions
\begin{subequations}
  \begin{align}
    \tr{(a^2)^{11}\rho}
    &= \sum_{l_1l_2} \delta_{l_1l_2} \rho_{l_2l_1} \notag \\
    &= \tr{\rho} \, , \\
    \tr{(a^2)^{22}\rho\rho}
    &= \sum_{l_1l_2l_3l_4} 2 (\delta_{l_1l_3}\delta_{l_2l_4} - \delta_{l_1l_4}\delta_{l_2l_3}) \rho_{l_4l_2}\rho_{l_3l_1} \notag \\
    &= 2(\tr{\rho})^2 - 2\tr{\rho^2} \, , \\
    \tr{(a^2)^{22}\kappa^\ast\kappa}
    &= \sum_{l_1l_2l_3l_4} 2 (\delta_{l_1l_3}\delta_{l_2l_4} - \delta_{l_1l_4}\delta_{l_2l_3}) \kappa^\ast_{l_1l_2}\kappa_{l_3l_4} \notag \\
    &= -4\tr{\kappa^\ast\kappa} \, ,
  \end{align}
\end{subequations}
were used along with the identity
\begin{align}
  \rho^2 - \rho = \kappa\kappa^\ast \, , \notag
\end{align}
originating from the unitarity of the Bogoliubov transformation. Having constrained the average particle number in the Bogoliubov vacuum to equate the targeted particle number A, i.e.
\begin{align}
  \tr{\rho} &= \text{A} \, ,
\end{align}
the doubly-projected nNO1B approximation to the particle-number variance is eventually given by
\begin{align}
  \frac{\langle\Phi|P^{\text{A}}(A^2)^{\text{nNO1B}}P^{\text{A}}|\Phi\rangle}{\langle\Phi|P^{\text{A}}|\Phi\rangle} - A^2
  &= \tr{\tilde{(a^2)}^{11} \Delta\rho^{\text{proj}}} \, . \notag
\end{align}
The approximate particle-number variance is thus given in this case by the trace over $\tilde{(a^2)}^{11}$ and the difference between PHFB and HFB normal density matrices $\Delta\rho^{\text{proj}}$. The latter being expected to be small, the doubly-projected nNO1B particle-number variance is expected to be small as well.

\begin{figure}[t!]
  \centering
  \includegraphics[width=1.0\columnwidth]{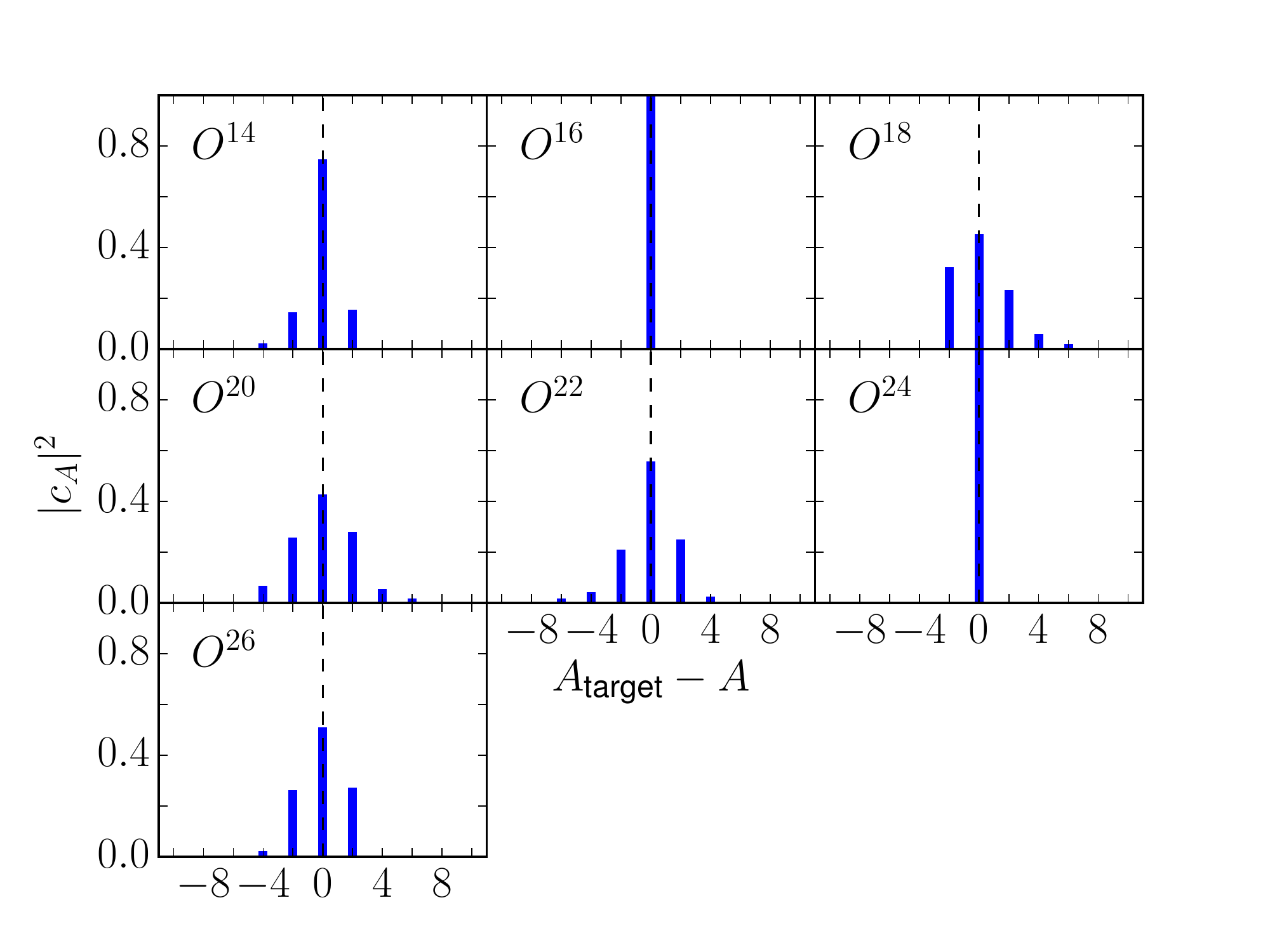}
  \caption{Distribution of good-particle components $c^2_{\text{A}}$ in each oxygen isotope. The dashed line denotes the average particle number in the underlying HFB vacuum.}
  \label{Fig:sumrulesA}
\end{figure}

\begin{figure}[t!]
  \centering
  \includegraphics[width=1.0\columnwidth]{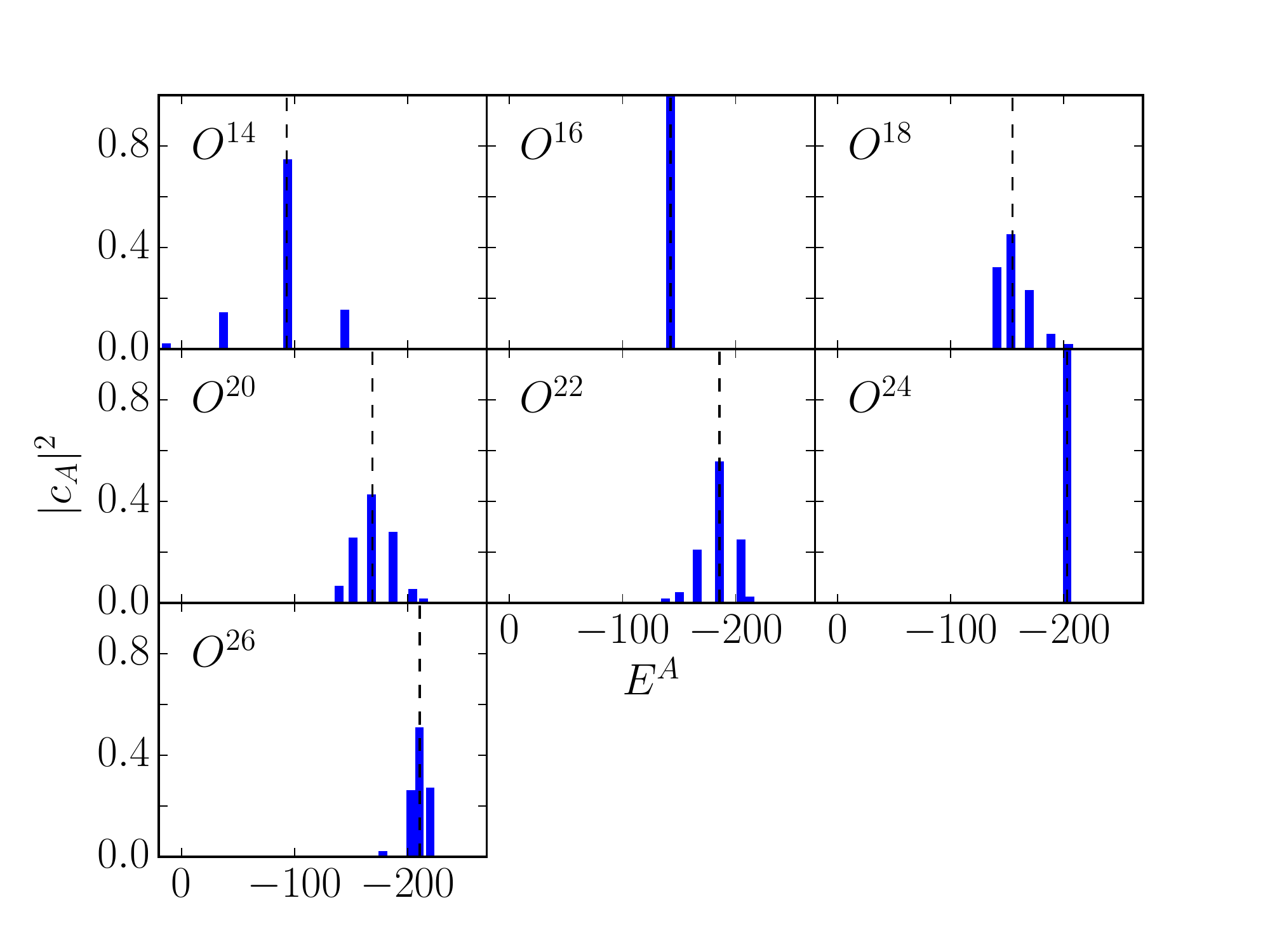}
  \caption{Good-particle components $c^2_{\text{A}}$ as a function of the projected energy $E^{\text{A}}$ in each oxygen isotope. The dashed line denotes the average, i.e. HFB, energy.}
  \label{Fig:sumrulesB}
\end{figure}

\section{Effect of particle number projection}
\label{sumrulesproj}

As seen in panel (b) of Fig.~\ref{Fig:energeticsO}, the particle-number projection provides further/no/lesser binding in $^{14,22}$O/$^{16,24}$O/$^{18,20,26}$O, knowing that the effect is essentially negligeable in $^{22,26}$O. 

In order to analyze this trend, the HFB state and energy can be decomposed into their particle-number projected components according to the sum rules\footnote{These sum rules are obtained from the Fourier expansion of uncorrelated singly-rotated off-diagonal kernels $\mathcal{N}^{(0)}(\varphi)=\langle\Phi(\varphi)|\Phi\rangle$ and $O^{(0)}(\varphi) \equiv o^{(0)}(\varphi) \mathcal{N}^{(0)}(\varphi)=\langle\Phi(\varphi)|O|\Phi\rangle$ computed at $\varphi = 0$.}
\begin{subequations}
\label{eq:sumrules}
\begin{align}
1 &= \sum_{\text{A}>0} c_{\text{A}}^2 \, , \label{eq:sumrulesA} \\
H^{00} &= \sum_{\text{A}>0} c_{\text{A}}^2 E^{\text{A}} \, , \label{eq:sumrulesB}
\end{align}
\end{subequations}
where the projected weights and energies are defined as
\begin{subequations}
\label{eq:defs}
\begin{align}
c^2_{\text{A}}  &\equiv \int_0^{2\pi} \!\frac{d\varphi}{2\pi} e^{-i\varphi \text{A}}  \mathcal{N}^{(0)}(\varphi) \, , \\
E^{\text{A}} &\equiv \int_0^{2\pi} \!\frac{d\varphi}{2\pi c^2_{\text{A}} } e^{-i\varphi \text{A}} h^{(0)}(\varphi) \mathcal{N}^{(0)}(\varphi) \, .
\end{align}
\end{subequations}

The decomposition underlying Eq.~\eqref{eq:sumrulesA} is displayed in Fig.~\ref{Fig:sumrulesA} along the sequence of oxygen isotopes. One first observes that $^{16}$O and $^{24}$O display a single component corresponding to their number of particles, i.e. the HFB vacuum reduces to a Slater determinant in these doubly closed-shell nuclei such that the subsequent particle number projection does not provide any static correlations. While in $^{14}$O the distribution is slightly skewed towards smaller particle numbers than the average value $14$, it is the opposite in $^{18,20}$O. As seen in Fig.~\ref{Fig:sumrulesB}, the skewness of the distribution eventually impacts the energy associated with each component obtained under the constraint that the energy sum rule (Eq.~\eqref{eq:sumrulesB}) must be fulfilled. In particular, the projected energy associated with the largest component, i.e. with the targeted particle number, ends up being more/less negative than the average, i.e. the HFB energy, in $^{14}$O/$^{18,20}$O.

\end{appendix}


\end{document}